\documentclass[
aps,
pre,
showpacs,
superscriptaddress,
preprintnumbers,
amsmath,
amssymb,
floatfix,
10pt,
reprint,
twocolumn]{revtex4-1}

\usepackage{hyperref}
\usepackage{graphicx}% Include figure files
\usepackage{dcolumn}% Align table columns on decimal point
\usepackage{bm}% bold math
\usepackage{multirow}
\usepackage{natbib}
\usepackage{braket}
\usepackage[caption=false]{subfig}
\usepackage[usenames,dvipsnames,svgnames,table]{xcolor}
\usepackage[normalem]{ulem}
\usepackage[T1]{fontenc}
\usepackage{txfonts}
\usepackage[utf8]{inputenc}
\usepackage{amsmath}
\usepackage{amssymb}
\usepackage{epstopdf}

\renewcommand{\paragraph}[1]{

\textit{#1}: }
\newcommand{\halfcolwidth}{0.48\columnwidth}

\newcommand{\pdf}{\textsc{pdf}}
\newcommand{\pdfs}{{\pdf}s}
\newcommand{\cdf}{\textsc{cdf}}

\newcommand{\ccdf}{\textsc{ccdf}}

\newcommand{\fire}{\textsc{fire}}

\newcommand{\matrixsym}[1]{\mathbf{#1}}

\renewcommand{\epsilon}{\varepsilon}

\newcommand{\for}{\textrm{for }}
\newcommand{\nplog}{N^2P\>\log_{10}(N)^{-0.7}}

\newcommand{\particlepair}{{ij}}
\newcommand{\upar}{\ensuremath{u_{\parallel, \particlepair}}}
\newcommand{\uperp}{\ensuremath{u_{\perp, \particlepair}}}

\newcommand{\mk}{\textrm{mk}}
\newcommand{\bk}{\textrm{bk}}
\newcommand{\cc}{\textrm{cc}}

\newcommand{\Pbk}{\ensuremath{\Pr(\bk)}}
\newcommand{\Pmk}{\ensuremath{\Pr(\mk)}}
\newcommand{\LR}{\textrm{LR}}
\newcommand{\dns}{\textrm{DNS}}

\newcommand{\gammaij}{\ensuremath{\gamma_{\particlepair}}}
\newcommand{\gammaijdagger}{\ensuremath{\gammaij^\dagger}}

\newcommand{\gammastar}{\ensuremath{\gamma_*}}
\newcommand{\gammastardns}{\ensuremath{\gamma_*^\dns}}
\newcommand{\gammastarlr}{\ensuremath{\gamma_*^\LR}}
\newcommand{\gammastarmklr}{\ensuremath{\gamma_{*,\mk}^\LR}}
\newcommand{\gammastarbklr}{\ensuremath{\gamma_{*,\mk}^\LR}}
\newcommand{\gammastarbk}{\ensuremath{\gamma_{*,\mk}}}

\newcommand{\gammacc}{\ensuremath{\gamma_\cc}}
\newcommand{\gammamk}{\ensuremath{\gamma_\mk}}
\newcommand{\gammabk}{\ensuremath{\gamma_\bk}}

\newcommand{\gammamklr}{\ensuremath{\gammamk^\LR}}
\newcommand{\gammabklr}{\ensuremath{\gammabk^\LR}}

\newcommand{\gammadist}{\ensuremath{\gamma_{\bk}^{\textrm{dist}}}}

\newcommand{\sigmacc}{\ensuremath{\sigma_{\textrm{cc}}}}

\begin{document}

\title{Contact Changes of Sheared Systems: Scaling, Correlations, and Mechanisms}

\author{Merlijn S. \surname{van Deen}}
\affiliation{Huygens-Kamerlingh Onnes Lab, Universiteit Leiden, Postbus
9504, 2300 RA Leiden, The Netherlands}
\affiliation{FOM Institute AMOLF, Science Park 104, 1098 XG Amsterdam, The Netherlands}

\author{Brian P. Tighe}
\affiliation{Process \& Energy Laboratory, Delft University of Technology, Leeghwaterstraat 39, 2628 CB Delft, The Netherlands}

\author{Martin \surname{van Hecke}}
%\email[]{mvhecke@physics.leidenuniv.nl}
\affiliation{Huygens-Kamerlingh Onnes Lab, Universiteit Leiden, Postbus 9504, 2300 RA
Leiden, The Netherlands}
\affiliation{FOM Institute AMOLF, Science Park 104, 1098 XG Amsterdam, The Netherlands}

\date{\today}

\begin{abstract}
We probe the onset and effect of contact changes in 2D soft harmonic particle packings which are sheared
quasistatically under controlled strain. First, we show that in the majority of cases, the first contact changes correspond to
the creation or breaking of contacts on a single particle, with contact breaking overwhelmingly likely for
low pressures and/or small systems, and contact making and breaking equally likely for large pressures and in the thermodynamic limit.
The statistics of the corresponding strains are near-Poissonian,
in particular for large enough systems. The mean characteristic strains exhibit scaling with the number of particles $N$ and pressure $P$, and reveal the existence of finite size effects akin to those seen for linear response quantities \cite{goodrich2012finitesizescaling,goodrich2013jamminginfinite}.
Second, we show that linear response accurately predicts the strains of the first contact changes, which allows us to accurately study the scaling of the characteristic strains of making and breaking contacts separately. Both of these show finite size scaling, and we formulate scaling arguments that are consistent with the observed behavior.
Third, we probe the effect of the first contact change on the shear modulus $G$, and show in detail how the variation of $G$ remains smooth and bounded in the large system size limit: even though contact changes occur then at vanishingly small strains, their cumulative effect, even at a fixed value of the strain, are limited, so that effectively, linear response remains well-defined. Fourth, we explore multiple contact changes under shear, and find strong and surprising correlations between alternating making and breaking events. Fifth, we show that by making a link with extremal statistics, our data
is consistent with a very slow crossover to self averaging with system size, so that the thermodynamic limit is reached much more slowly than expected based on finite size scaling of elastic quantities or contact breaking strains.
 %We close our paper with a discussion of related approaches to contact breaking, and the postulated role of anomalously weak contacts.
%TODO: * weird scaling uparl, uperp
\end{abstract}
%\pacs{Plasticity, 62.20.fq, deformation and plasticity, 62.20.fg, Rheo 83.10.Rs	 Computer simulation of molecular and %particle dynamics, 83.80.Fg	 Granular solids}
\maketitle
\cleardoublepage

How does a jammed system fail? Failure of amorphous systems under increasing driving generally leads to a complex chain
of events, where an initial linear response gets gradually eroded by
local micro events that lead to plasticity and
eventually organize in persistent flows \cite{ keim2013multipletransientmemories,maloney2006amorphoussystemsin,lemaitre2007plasticresponsetwo,salerno2012avalanchesinstrained,hentschel2010sizeplasticevents,chikkadi2013percolatingplasticfailure,olsson2007criticalscalingshear,tighe2010modelscalingstresses,chikkadi2015correlationsstrainand,woldhuis2015fluctuationsinflows}. For systems near the
critical jamming point, the question of failure is even more vexing, as the characteristic strain for the first deviations from linear response is vanishing, both with the number of particles in the system $N$, but also when the confining pressure $P$ is lowered towards the critical jamming point. Moreover, near the unjamming point
disordered solids are extremely fragile, and the tiniest of  perturbations can cause an intrinsically nonlinear response \cite{wyart2012marginalstabilityconstrains,olsson2007criticalscalingshear, tighe2010modelscalingstresses,schreck2011repulsivecontactinteractions, gomez2012uniformshockwaves,gomez2012shocksnearjamming,
wildenberg2013shockwavesin}.
Hence,
one may question the validity of linear response for  athermal amorphous solids, as the range of validity may vanish
\cite{combe2000strainversusstress,hentschel2010sizeplasticevents,schreck2011repulsivecontactinteractions,boschaninprep}.
Finally, the unjamming transition at vanishing $P$
bears hallmarks of a critical phase transition:
properties such
as the contact number and elastic moduli exhibit power law scaling
\cite{bolton1990rigiditylosstransition, durian1995foammechanicsat,lacasse1996modelelasticitycompressed,
ohern2002randompackingsfrictionless,
ohern2003jammingatzero,wyart2005rigidityamorphoussolids,wyart2008elasticityfloppyand,katgert2010jammingandgeometry,tighe2011relaxationsandrheology,boschaninprep},
time and length scales diverge \cite{ohern2003jammingatzero,
wyart2005geometricoriginexcess,wyart2005rigidityamorphoussolids,silbert2005vibrationsanddiverging,
ellenbroek2006criticalscalingin,tighe2011relaxationsandrheology,boschaninprep}, the material's response becomes singularly non-affine \cite{ellenbroek2006criticalscalingin, ellenbroek2009jammedfrictionlessdisks}
and finite size scaling governs the behavior for small numbers of particles $N$ and/or small $P$
\cite{dagoisbohy2012softspherepackings,goodrich2012finitesizescaling, goodrich2013jamminginfinite}.
The question we want to address is how, near jamming, when linear response vanishes and criticality dominates, a jammed system reacts and fails under increasing driving.

\begin{figure}[bt!]
\centering
\includegraphics[width=\halfcolwidth]{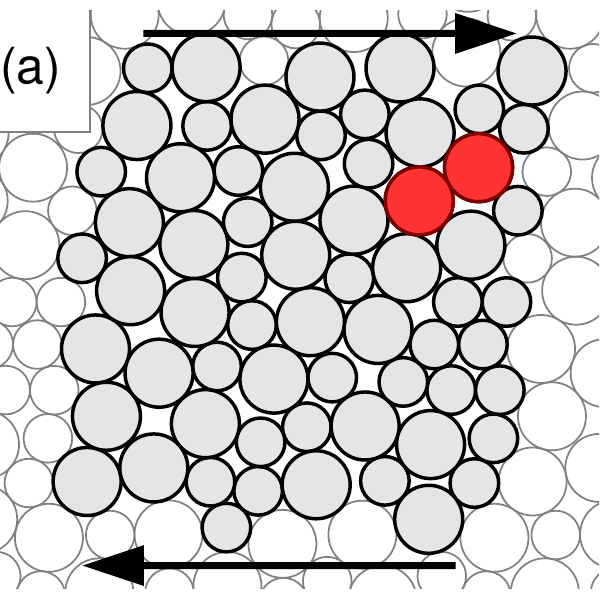}
\includegraphics[width=\halfcolwidth]{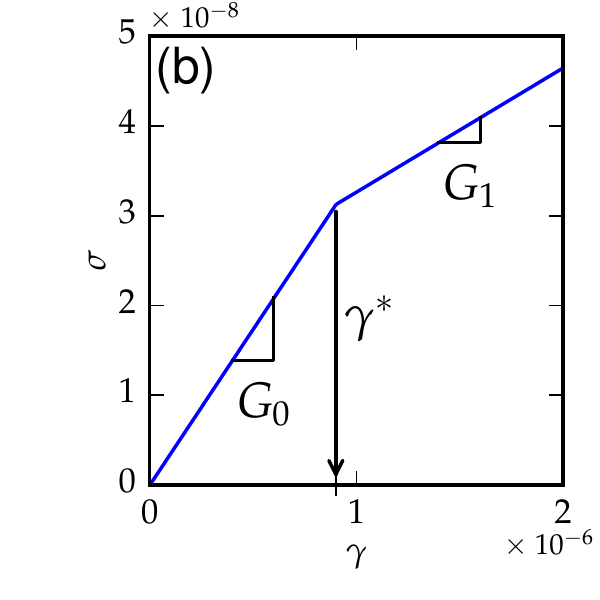}
\caption{(color online).
(a) The first contact change in a sheared packing  ($N=64$, $P=10^{-6}$) occurs
at a strain $\gammastar = 9.003851(2)\times 10^{-7}$, when the two marked particles lose their contact. (b) The corresponding stress-strain curve remains continuous but exhibits a
sharp kink; we define $G_0$ as the shear modulus of the undeformed packing, and $G_1$ as the shear modulus of the packing just above $\gammastar$.
\label{fig:pack-en-stresstrain}}
\end{figure}

Earlier work on contact changes has focused on vibrations \cite{schreck2011repulsivecontactinteractions,goodrich2013commentrepulsivecontact, schreck2013responsetocomment} or hard particle systems \cite{combe2000strainversusstress,lerner2013lowenergynon}. We instead focus on soft particle systems, as they are descriptive for a wider range of experimentally relevant systems, use experimentally relevant simple shear deformations, and
focus on the first unambiguous deviation from strict linear response:  contact changes under quasistatic shear (Fig.~\ref{fig:pack-en-stresstrain}a) \cite{deen2014contactchangesnear}.

We address the following questions:
{\em{(i)}} What is the nature of the first contact changes near jamming?
In systems far from jamming, rearrangements organize into avalanches: collective, plastic events in which multiple contacts are broken and formed and the stresses exhibit discontinuous drops \cite{maloney2006amorphoussystemsin, lemaitre2007plasticresponsetwo, salerno2012avalanchesinstrained,hentschel2010sizeplasticevents,manning2011vibrationalmodesidentify}.
For hard particles which represent a singular case where motion always involves unjamming,
even a single contact break may induce
a complete loss of rigidity \cite{combe2000strainversusstress,hentschel2010sizeplasticevents,karmakar2010predictingplasticflow,wyart2012marginalstabilityconstrains}.
In contrast, we find that {\em near} jamming the first events are the making or breaking of a
single contact, and that the stress remains continuous.
The probabilities for contact making and breaking are governed by finite size scaling, with making and breaking equally likely for $N^2P \gg 1$, but contact breaking dominant for $N^2P \ll 1$.

{\em (ii)} What is the mean strain $\gammacc$ at which the first
contact change arises? What are the mean strains of the first contact breaking $\gammabk$ or contact making $\gammamk$ events? We first show that we can use linear response calculations to accurately capture these
strains, and then show that all these characteristic strains
vanishes when either $N$ diverges or $P$ vanishes. All strains obey finite size scaling:
$\gammacc \sim \gammabk \sim P$ and $\gammamk \sim 1/N^2$
for small systems close to jamming ($N^2P \ll 1$), whereas
and $\gammacc \sim \gammabk \sim \gammamk \sim \sqrt{P}/N$
for $N^2P \gg 1$. As log-corrections to scaling are expected for jamming in 2D \cite{goodrich2013jamminginfinite,binder1985finitesizetests,goodrich2012finitesizescaling}, and in additional alternative corrections to scaling have been proposed \cite{wyart2012marginalstabilityconstrains,lerner2013lowenergynon}, we carefully study our data from this perspective, and find that our data is consistent with both - in 2D, extremely large systems are needed to distinguish between these different corrections.

{\em (iii)}
How do contact changes affect linear response? For finite systems close to jamming, even a single contact change can strongly affect the elastic response (Fig.~\ref{fig:pack-en-stresstrain}b). Clearly,
calculations based on the Hessian matrix of
the undeformed packing are then no longer strictly valid.
As a result, the relevance of the linear response scaling relations
are currently under dispute
for systems close to jamming, at finite temperature, or in the thermodynamic limit  \cite{schreck2011repulsivecontactinteractions, goodrich2013commentrepulsivecontact,schreck2013responsetocomment,ikeda2013dynamiccriticalityat,wang2013criticalscalingin,goodrich2014whendojammed}.
By comparing the shear modulus before ($G_0$) and after ($G_1$) the first contact change, we find that
their ratio again is governed by finite size scaling, and while the ratio
$G_1/G_0$ approaches 0.2 for small $N^2P$, for large $N^2P$,
$G_1/G_0 \rightarrow 1$. We also study the statistics of $G_1/G_0$ by the standard deviation  $\sigma$ of its distribution, and find three regimes: for small $N^2 P$, $\sigma \approx 0.3$, for $N^2P \approx 1$, the fluctuations are strongest and values of $G_1<0$ are most likely, whereas for large $N^2 P$, $\sigma$ scales roughly  as $[\nplog] ^{0.35}$. The latter scaling allows us to  estimate the cumulative effect of a diverging number of contact changes that occur when the strain is fixed and $N \rightarrow \infty$, and shows that this is limited: effective linear response, quantified by the shear modulus at finite strain, appears well-defined.

{\em (iv)}
We explore sequences of multiple contact changes under shear, and find strong correlations between alternating making and breaking events. A surprising effect is that while initial contact breakings drive the system precariously close to catastrophic failure (too few contacts to maintain rigidity), the subsequent sequence of contact making and breaking
extends the range before such failure sets in.

{\em (v)}
Fifth, we show that by making a link with extremal statistics, our data
is consistent with a very slow crossover to self averaging with system size, so that the thermodynamic limit is reached much more slowly than expected based on finite size scaling of elastic quantities or contact breaking strains.

Our work paints a clear and coherent picture of the role of contact changes near the critical jamming point.
While the range of {\em strict} validity of linear response vanishes for small $P$ and large $N$, macroscopic
quantities such as the shear modulus are relatively insensitive to contact changes
as long as $P \gg 1/N^2$. Hence, linear response quantities remain relevant for finite $P$ and large $N$, while for $P \ll 1/N^2$, a single contact change already changes the
packing significantly. The qualitative differences in
the nature of contact changes close to and far from jamming suggests that plasticity, creep, and flow \emph{near} jamming are controlled by fundamentally different mechanisms than plastic flows in systems \emph{far from} jamming \cite{tighe2010modelscalingstresses,salerno2012avalanchesinstrained,hentschel2010sizeplasticevents,maloney2006amorphoussystemsin,olsson2007criticalscalingshear,manning2011vibrationalmodesidentify,chikkadi2015correlationsstrainand,woldhuis2015fluctuationsinflows}.

\section{Method \& protocols}
\label{sec:shear-method-protocols}
We simulate bidisperse packings of massless, frictionless soft spheres in two dimensions \cite{ohern2002randompackingsfrictionless,ohern2003jammingatzero,durian1995foammechanicsat}.
Recently, it was shown that such finite packings are not guaranteed to have positive shear moduli, nor have zero residual stress \cite{donev2004commentjammingat,dagoisbohy2012softspherepackings,goodrich2013jamminginfinite}, which both could lead to problems when studying contact changes.
We therefore focus on so-called $\epsilon_\textrm{all}^+$ packings
that have positive moduli and zero residual shear stress as described in \cite{dagoisbohy2012softspherepackings,goodrich2013jamminginfinite}.
In Appendix \ref{apx:create-and-shear}, we describe in detail how to create and shear such packings, which
in particular necessitates the use of non-square unit cells   \cite{donev2004commentjammingat,dagoisbohy2012softspherepackings,goodrich2013jamminginfinite}. Here, we will focus on our algorithm to detect contact changes.

To find contact changes, we apply a strain (Eq.~A2)
\begin{equation}
\gamma = 10^{-9} \cdot 10^{\zeta}
\end{equation}
where we increase $\zeta=0,1,...$ until we detect a change in the contact network ($\delta_{ij} = 0 \leftrightarrow \delta_{ij} > 0$ for any pair $i,j$).
We then move back to the state before the contact change, and use bisection to
determine the strain at the contact change $\gammastar$ until $\Delta\gamma/\gammastar < 10^{-6}$.

\label{sec:linres-ccstrain-rattlers}
        \begin{figure}
        \includegraphics[width=\halfcolwidth]{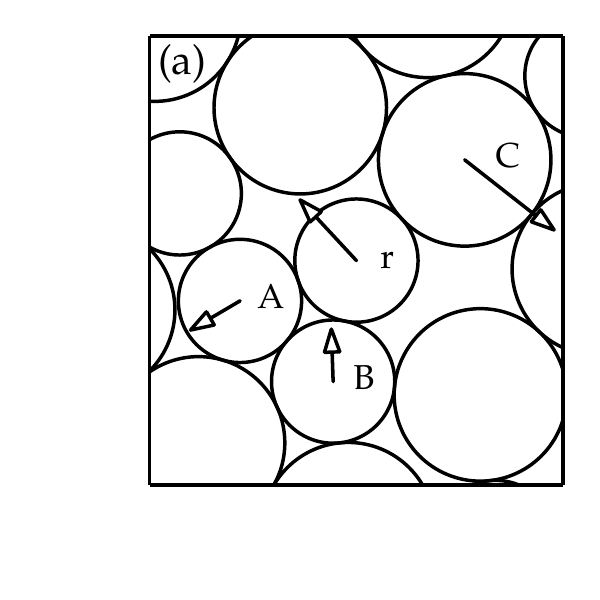}
        \includegraphics[width=\halfcolwidth]{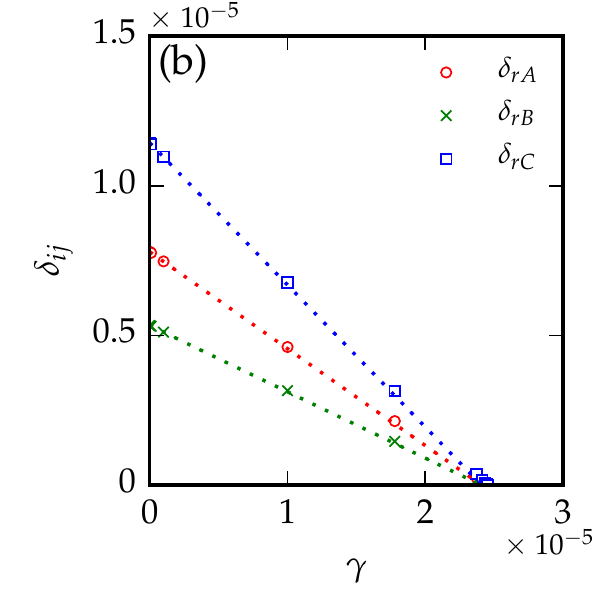}
        \subfloat{\label{fig:rattler-naar-nul-particles}}
        \subfloat{\label{fig:rattler-naar-nul}}
        \caption{(color online) (a) Zoom-in of a packing where particle $r$ becomes a rattler after the first contact change ($N=22$, $P=1.5\cdot 10^{-5}$). Neighboring particles $A$, $B$ and $C$ are indicated.
         Overlap of $r$ with the neighboring particles $A$, $B$ and $C$ as a function of strain $\gamma$. Markers are {\dns} simulation data points, lines indicate the linear response prediction. The {\dns} and {\LR} predictions for rattler creation are $\gammastardns=2.45\cdot 10^{-5}$ and $\gammastarlr=2.41\cdot 10^{-5}$.
        }
        \end{figure}
Rattlers require special attention: because they are free to move, their behavior is ill-defined. In our simulations, we encounter rattlers in two distinct types of events.
First, rattlers may become part of the load-bearing network. As the rattlers' position is ill-defined, the strain at which this occurs is algorithm dependent. We therefore exclude such contact making events in our analysis of the first contact change.
Second, a particle with three contacts can become a rattler, where force balance dictates that all three contacts go to zero overlap simultaneously. This is detected correctly in our simulations, and the event is recorded as a \emph{single} break event.
In linear response calculations (to be discussed below), the creation of a rattler is also well-defined. In Fig.~\ref{fig:rattler-naar-nul}, we show the overlap $\delta_{ri}$ of particle $r$ with its neighbours $A$, $B$ and $C$. In the simulations (symbols), we find the overlaps smoothly go to zero while approaching the contact change strain $\gammastar$. In linear response, calculated at $\gamma=0$, we find a slightly different contact change strain for each contact, but they are within $|\Delta\gamma/\gammastar| < 10^{-4}$.

\section{Numerical results}\label{sec:shear-numerical-results}
In this section, we discuss the results
of direct numerical simulations to determine the properties of the
strain $\gammastar$ at which the first contact change occurs. We first discuss the
relative prevalence of contact making and breaking events. We then study in detail the statistics
of $\gammastar$ at given $P$ and $N$, and finally discuss how the ensemble averages $\gammacc = \langle\gammastar\rangle$ scale with $N$ and $P$.

\subsection{The first contact change}
    \begin{figure}
    \subfloat{\label{fig:numres-CDFS/CDF-gammacc-N-256.pdf}}
    \includegraphics[width=\halfcolwidth]{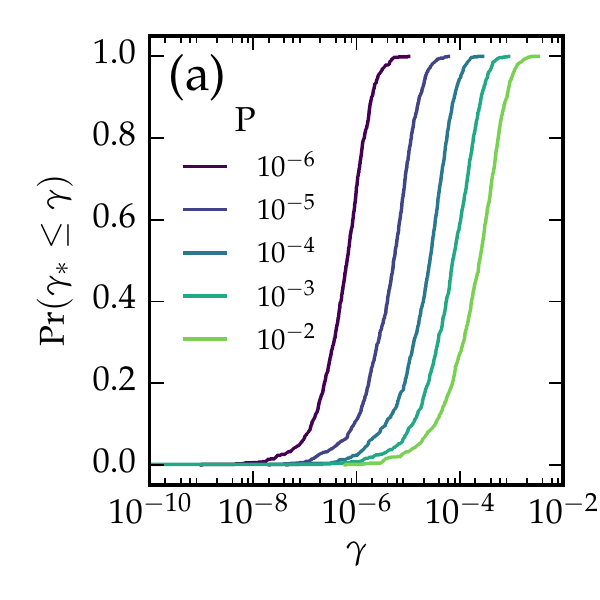}
    \subfloat{\label{fig:mk-bk-mixed-probabilities}}
    \includegraphics[width=\halfcolwidth]{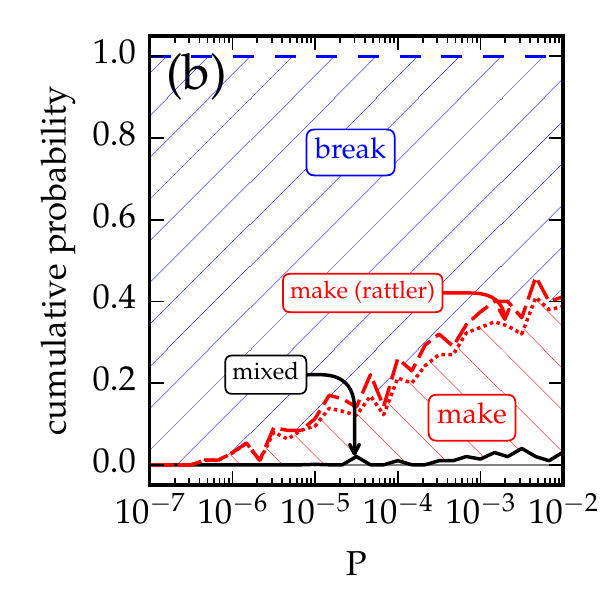}

        \caption{(color online)
            (a) Cumulative distribution functions $\Pr(\gammastar < \gamma)$ of the contact change strain $\gammastar$ for $N=256$, $P=10^{-6}$ (left) $ \ldots 10^{-2}$ (right).
            (b) Stacked probabilities for the first contact change being a break event (blue striped), a make event (red striped), a make event involving a rattler (red) and a mixed event, where contacts are both broken and created (black), for $N=256$ ensembles.
        }
    \end{figure}

For each packing in an $(N,P)$ ensemble, we determine the strain of the first contact change $\gammastar$, as described in Sec.~\ref{sec:simple-shear-cc-method}. In Fig.~\ref{fig:numres-CDFS/CDF-gammacc-N-256.pdf} we show the cumulative distribution function (\cdf) of $\gammastar$ for $N=256$ ensembles at various pressures. We observe that, first, the typical scale of the strain $\gammastar$ increases with pressure $P$, and secondly that their shape is mostly independent of $P$.

In Fig.~\ref{fig:mk-bk-mixed-probabilities}, we show a stacked probability graph of the different contact change types. We distinguish events where one or more contacts are broken (\emph{break}), events where one or more contacts are created (\emph{make}) and events where contacts are both broken and created (\emph{mixed}).
The number of mixed events increases with pressure, but is less than $5\%$, independent of $N$.
Within the \emph{make} class, we can distinguish events where a particle which originally was a rattler now becomes part of the contact network (\emph{make (rattler)}). Of all \emph{make} events, $5-15\%$ involve rattlers. This is consistent between ensembles, with no clear dependence on either $N$ or $P$.
At low pressures, we find that the vast majority of events consists of contacts being broken. At large pressures, we find that roughly half of the events create a new contact. In Sec.~\ref{sec:lr-ensemble-averages}, we will show how these probabilities vary as a function of $N^2P$.
In the remainder of this paper, we will focus on the simple \emph{make} and \emph{break} cases.

\subsection{Strain distributions}
\label{sec:strain-distributions}
    \begin{figure}
        \subfloat{\label{fig:numres-CDFS/CCDF-gammacc-N-256.pdf}}
        \subfloat{\label{fig:numres-CDFS/weibulla}}
        \subfloat{\label{fig:anderson-tests/p-005.pdf}}
        \includegraphics[width=\columnwidth]{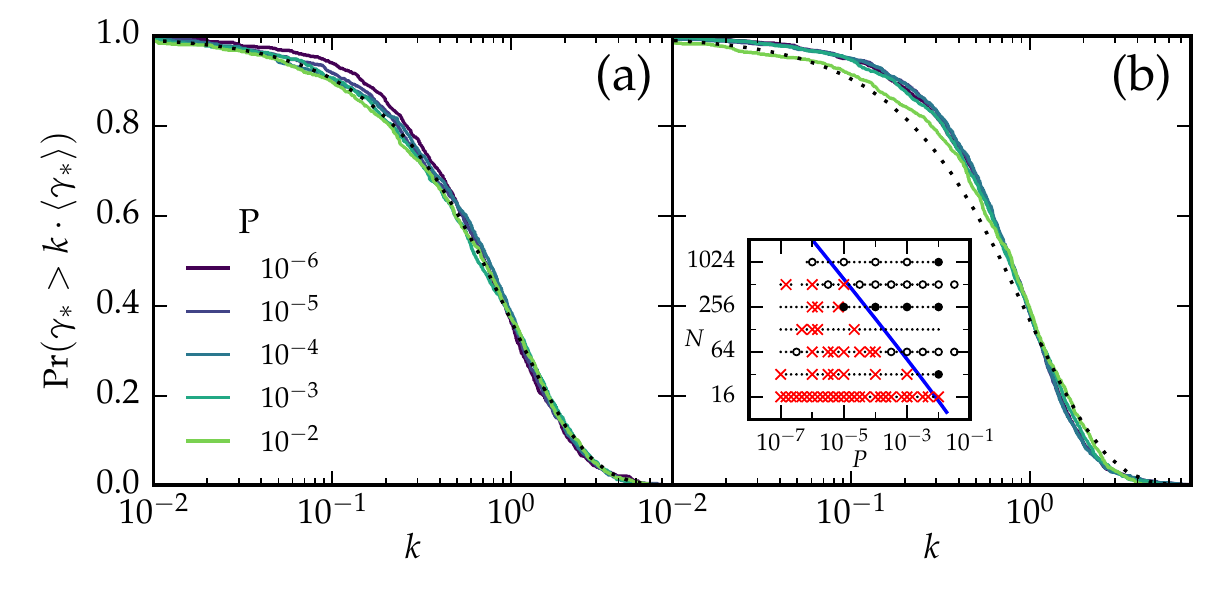}

         \caption{(color online) (a) Rescaled complementary cumulative distribution functions (\ccdf),  $N=256$, $P=10^{-6}\ldots 10^{-2}$ (highest pressures have lowest values for $k=0.1$). The dotted line gives the \ccdf{} for an exponential distribution.
        (b) Same, for $N=16$ systems at various pressures.
         (inset) Result of the Anderson-Darling test. Ensembles that fail the test are indicated with a red cross. Other ensembles are indicated with a dot ($\leq 100$ samples), open circle ($100 \sim 1000$ samples) or filled circle ($\approx 1000$ samples).
         The blue line indicates the finite size threshold $\nplog=1$ (see Sec.~\ref{sec:scaling}).
         }
\end{figure}

We now take a more detailed look at the distributions of $\gammastar$ and show that contact changes can essentially be described as a Poisson process, as the \cdf{} close resembles an exponential distribution, with $\Pr(\gammastar \leq \gamma) = 1-e^{-\gamma/\beta}$.
In Fig.~\ref{fig:numres-CDFS/CCDF-gammacc-N-256.pdf} we show $\Pr(\gammastar > k \langle\gammastar\rangle)$, i.e. the complimentary \cdf{} of $\gammastar$, rescaled by the ensemble mean $\langle\gammastar\rangle$. If $\gammastar$ is exponentially distributed, the \ccdf{} is a simple exponential: $\Pr(\gammastar > k\cdot\langle\gammastar\rangle) = e^{-k} \hspace{1em} (k \ge 0)$, and as Fig.~\ref{fig:numres-CDFS/CCDF-gammacc-N-256.pdf} shows, our distributions for $N=256$ are close to exponential. This is consistent with a Poisson process, where contact changes are independent of each other.

To check conformance to an exponential distribution as a function of $N$ and $P$, we use the \emph{Anderson-Darling test} \cite{NIST2014ehandbookstatistical}, with which we test the hypothesis "these values of $\gammastar$  were drawn from an exponential distribution". We use a 5\% confidence interval, i.e., there is a 5\% probability we reject the hypothesis for samples that \emph{were} drawn from an exponential distribution. In Fig.~\ref{fig:anderson-tests/p-005.pdf}, we show the results of this test.
We observe deviations from exponential behavior for small systems and low pressures. The boundary between rejection and non-rejection corresponds with the transition between systems for which finite size effects dominate and large systems, at $N^2P \approx 1$ \cite{goodrich2012finitesizescaling,goodrich2013jamminginfinite,deen2014contactchangesnear}. This suggests that for large systems ($N^2P \gg 1$), contact changes are uncorrelated, while for small systems, correlations build up.

How do distributions for systems in the finite size regime deviate from exponential? In Fig.~\ref{fig:numres-CDFS/weibulla}, we show rescaled {\ccdf}s for $N=16$ systems at various pressures. The most significant deviation is at low $k$, where we find $\Pr(\gammastar > k\cdot\langle\gammastar\rangle)$ is larger than expected for an exponential distribution. As $\Pr(\gammastar > k\cdot\langle\gammastar\rangle)$ is the \emph{survival probability}, this indicates a lack of events at small strain, which means that, in small systems, events are antibunched. Notwithstanding this deviation from exponential behavior, the mean remains well-defined, as is further evidenced by recent work which shows the number of contact changes scales linearly with strain \cite{boschaninprep}.

\subsection{Scaling}
\label{sec:scaling}
\begin{figure}[t!]
\includegraphics[width=\columnwidth]{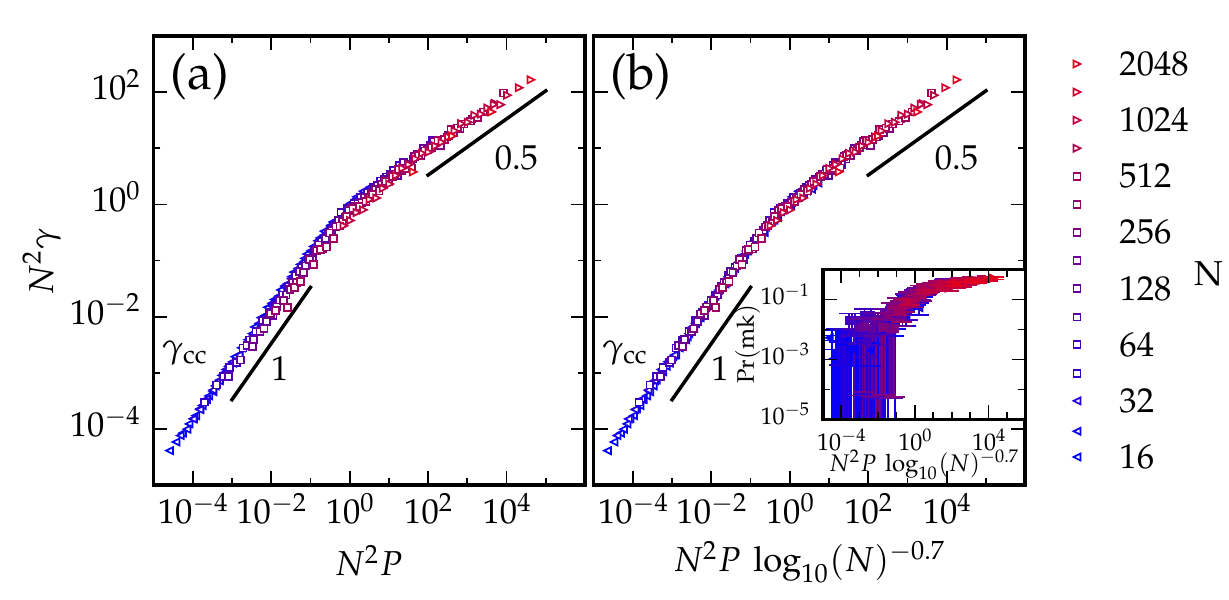}
\subfloat{\label{fig:gammamed-collapse}}
\subfloat{\label{fig:gammamed-collapse-log}}
\caption{(color online) (a) Scaling of the strain at first contact change $\gammacc$ as function of $N$ and $P$. Symbols and shades (colors) indicate packing sizes. Lines indicate power law functions with exponent $1$ (lower branch) and $0.5$ (upper branch). (b) Log corrections improve the  collapse.
(inset) Probability of the first contact change creating a new contact. At high $\nplog$, $\Pmk \approx \Pbk \approx 0.5$, but at low $\nplog$, breaking strongly dominates. \label{fig:gammamed-collapse-both}}
\end{figure}

We now discuss the variation of the mean contact change strain $\gammacc=\langle\gammastar\rangle$ with $N$ and $P$. As discussed in \cite{deen2014contactchangesnear}, we can obtain data collapse for $\gammacc$ when we plot $N^2\gammacc$ as a function of $N^2P$.
As shown in Fig.~\ref{fig:gammamed-collapse}, this results in a good (but not great) data collapse. It has been suggested that the upper critical dimension for jamming is two, which implies logarithmic corrections to scaling \cite{goodrich2013jamminginfinite}.
Using the form suggested in  \cite{goodrich2013jamminginfinite}, we find a very good data collapse (Fig.~\ref{fig:gammamed-collapse-log}).

How do we think about these strains? As we will show later, strictly linear response captures the deformations well up to the first contact change.
It is thus useful to consider, on the one hand, the overlaps and ``underlaps'' between pairs of particles in (near) contact and, on the other hand, the relative motion of such pairs. The former are set by the packing, and in particular, the overlaps scale trivially with the pressure. The latter follow
from the full linear response via the set of eigenmodes that characterize the system (Appendix C). This way of thinking strongly suggests that we should consider the behavior for $N^2P$ smaller or larger than one separately. For  $N^2 P \ll 1$, the number of contacts is constant, and the eigenmodes are essentially independent of $P$ (Appendix C). Hence, here the main variation with $P$ is in the overlaps, which vanish when $P \rightarrow 0$. Therefore, we expect breaking to happen at much smaller strains than making, and hence that the
contact change strain is simply linear in $P$ --- consistent with the data in Fig.~\ref{fig:gammamed-collapse-both}. Moreover, this simple picture suggests that the amount of shear stress at the first contact change is proportional to $P$.

The situation for $N^2P \gg 1$ is more complex, because here the eigenmode spectrum changes with $P$, and indeed, it is known that the relative motions normal and transverse to a contact pair's center-to-center line scale as $u_\parallel \sim P^{1/4} \gamma$ and $u_\perp \sim \gamma/P^{1/4}$, respectively \cite{ellenbroek2009jammedfrictionlessdisks}. As the transverse motion diverges near jamming (for large $N^2 P$), it dominates the change $\delta \ell \sim u_\perp^2/\ell$ in the center-to-center distance $\ell$. A na\"ive argument for the pressure dependence of the breaking strain can then be constructed by balancing $\delta \ell$ with the typical overlap in the initial condition, $\delta \ell(\gamma_{\rm bk}) \sim \delta$, yielding the prediction $\gamma_{\rm bk} \sim P^{3/4}$. Indeed, a strain proportional to $P^{3/4}$ also arises in a recent scaling theory of the jamming transition \cite{goodrich2015scalingtheoryjamming}. While this argument correctly predicts nontrivial $P$-dependence in the characteristic strains for $N^2 P \gg 1$,  the $3/4$ exponent is inconsistent  with our data shown in Fig.~\ref{fig:gammamed-collapse-both}. We believe the essence of this discrepancy is that the assumption that the first broken contact is typical of all contacts is incorrect. We note, in passing, that studies that consider
``typical'' contacts, do find a scaling consistent with a 3/4 exponent \cite{schreck2011repulsivecontactinteractions}.

After this introduction, we
we now discuss two distinct arguments that both lead to a scaling relation which is consistent with our data:
an argument for \emph{compressive} strain, and a stress-based argument for shear strain.

        \begin{figure}
        \includegraphics[width=\halfcolwidth]{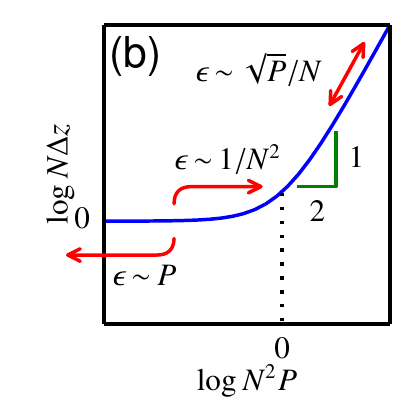}
        \caption{(color online)\label{fig:PREr-graphs-F2b-dZscaling}Excess number of contacts $N\Delta z/2$ as a function of $N^2P$ (blue curve, based on \cite{dagoisbohy2012softspherepackings,
        goodrich2012finitesizescaling,
        goodrich2013jamminginfinite}). Arrows indicate volumetric strains corresponding to a single contact change.}
        \end{figure}

\paragraph{Compression}
We start with a compressional argument, based on estimating the strain scale for making and breaking a contact under compression. There is a clear relationship between compression and the number of contacts: we gain contacts if we compress the system and we lose contacts if we expand the system. The scaling relation that relates the excess contact number $N_\textrm{exc} = N\Delta z/2$ to $N$ and $P$ is well-known from earlier work \cite{dagoisbohy2012softspherepackings,
goodrich2012finitesizescaling,
goodrich2013jamminginfinite}, and is shown in Fig.~\ref{fig:PREr-graphs-F2b-dZscaling}. There are two branches: a plateau $N_\textrm{exc} \sim 1$ at low pressures and a square root pressure dependence $N_\textrm{exc} \sim \sqrt{N^2P}$ at higher pressures.

How far do we need to expand or compress a system at given $N$ and $P$ to induce a contact change?
In the high-pressure regime, the derivative $\pm\frac{\partial}{\partial P}\left(\sqrt{N^2P}\right) \sim \pm N/\sqrt{P}$ gives the number of contacts changed due to unit pressure change. Its inverse $\delta P \sim \pm\sqrt{P}/N$, then gives the pressure change needed for a single contact change. The compressional strain is the pressure change divided by the bulk modulus $K$: $\epsilon_{\cc} \sim \pm\delta P/K$. As $K$ is independent of $N$ and $P$ \cite{ohern2003jammingatzero}, we simply find $\epsilon_{\cc} \sim \pm\sqrt{P}/N$.

In the low-pressure finite size regime, the number of contacts is independent of pressure. Nevertheless, the plateau has a finite length. On the one hand, the plateau ends at $P=0$, as we unjam our system and lose all contacts. On the other hand, the plateau ends when we enter the large system size regime at $N^2P \sim 1$ and gain one new contact.

The scales for making and breaking a contact are thus no longer the same in the finite size regime: To break a contact, we unjam the system by reducing the pressure with $\delta P \sim P$, and we find $\epsilon_{\bk} \sim -P$. To create a contact, we increase pressure up to the beginning of the large system regime, at $P_\textrm{target} =1/N^2$. As we are initially in the small system regime, the current pressure $P\ll 1/N^2$ and can be neglected, and the pressure change $\delta P = P_\textrm{target} - P \approx -1/N^2$. We thus need to apply a strain $\epsilon_{\mk} \sim -1/N^2$. The contact \emph{change} strain, independent of direction, will be given by the minimum of the absolute making and breaking strains. As $P\ll 1/N^2$, we thus expect $\epsilon^{\cc} \sim P$.

Summarized, this argument leads to these characteristic strains for contact changes under compression:
\begin{equation}\begin{array}{rcccr}
~                  & \epsilon_{\bk} & \epsilon_{\mk} & \epsilon_{\cc}\\
\multirow{2}{*}{\ensuremath{\epsilon \sim\ \bigg\{}}
                   & -P           & 1/N^2       & P & \for N^2P \ll 1,\\
                   & -\sqrt{P}/N  & \sqrt{P}/N  & \sqrt{P}/N & \for N^2P \gg 1.
\end{array}\label{eqn:predicted-scalings-compression}\end{equation}
As we will see, arguments based on shear, as well as our our results, find the same scaling for these strains.
5248860101717435

\paragraph{Shear}
\label{sec:shear-scaling-argument}
        \begin{figure}
        \begin{center}
        \includegraphics[width=\columnwidth]{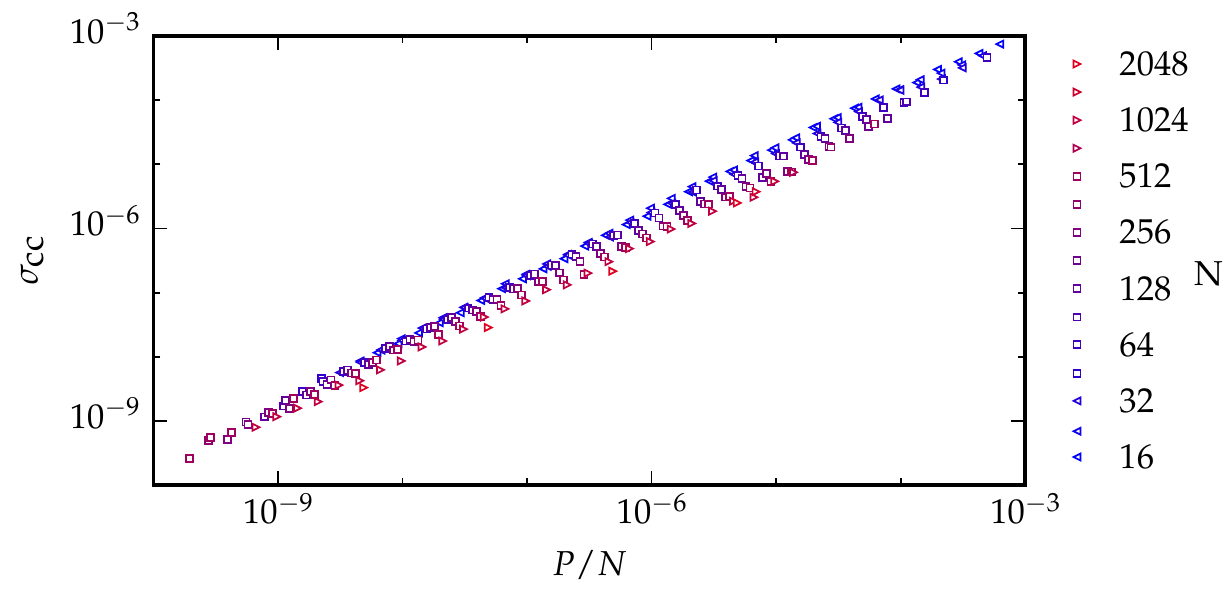}
        \end{center}
        \caption{(color online) $\sigmacc$ as a function of $P/N$. Symbols and shades (colors) indicate system size. The data supports an overall scaling $\sigma \sim P/N$, but the lack of a good collapse suggests this does not describe the entire behavior --- larger $N$ tend to have lower $\sigma_{cc}$. \label{fig:scaling-sigma}}
        \end{figure}
We can also formulate an argument for the scaling of $\gammacc$ under \emph{shear} from dimensional analysis. Other than taking $\gammacc$ constant, there is no clear strain scale, so we will construct the argument using stress instead. We will start by determining the typical stress scale $\sigmacc$.

There are three stress scales in the system: the confining pressure $P$, the bulk modulus $K$ and the shear modulus $G$.
As we are describing shear, it seems unlikely that $K$ is relevant. If the stress scale $\sigmacc$ were to scale with $G$, we would end up with a constant strain, and we have already seen that $\gammacc$ is not constant. This suggests that the only relevant stress scale is the confining pressure $P$, which we already have argued to govern the behavior for $N^2 P \ll 1$; we now assume it also to govern the large system limit, and take $\sigmacc \sim P$. The stress scale must also depend on the system size. Say we have a packing with $N$ particles, which has a contact change at $\sigma=\sigmacc$. If we duplicate this system, we will have $2N$ particles, and two contact changes will have happened at the same stress $\sigmacc$, so that we expect that $\sigmacc \sim 1/N$.
Combining these two scalings leads to the following suggested scaling:
\begin{equation}\sigmacc \sim P/N~. \label{PN}\end{equation}~

We  determine the strain scale $\gammacc$ via the shear modulus $G = \sigma/\gamma$.
From earlier work \cite{durian1995foammechanicsat,dagoisbohy2012softspherepackings,goodrich2012finitesizescaling}, we know $G$ scales as
\begin{equation}
G \sim \begin{cases}
\sqrt{P} & \for N^2P \gg 1, \\
1/N & \for N^2P \ll 1,
\end{cases}\end{equation}
which, combined with the stress scaling we derived, suggests the following scaling for $\gammacc$:
\begin{equation}\gammacc \sim \sigmacc/G \sim \begin{cases}
\makebox[15ex][l]{$(P/N) / \sqrt{P}$} \sim \sqrt{P}/N & \for N^2P \gg 1, \\
\makebox[15ex][l]{$(P/N) / (1/N)$} \sim P & \for N^2P \ll 1~,
\end{cases}\end{equation}
consistent with the scaling proposed in Eq.~(\ref{eqn:predicted-scalings-compression})

Finally, we note that Eq.~(\ref{PN}) suggests to plot the stress at the first contact change, $\sigma_{cc}$ as a function of $P/N$. As shown in
Fig.~\ref{fig:scaling-sigma}, this gives a reasonable, but not excellent, data collapse. Nevertheless, the quality of the scaling collapse
of $\gammacc$ shows that ultimately the proposed scaling is correct, despite the hand waving nature of the underlying arguments to derive it.

\section{Linear response}\label{sec:shear-linear-response}
We now show and utilize that many properties of the first contact change can be deduced from the initial state at $\gamma=0$ using linear response. The idea is to estimate the trajectories of (non-rattler) particles from their linear elastic response: $\vec{x_i}(\gamma) = \vec{x_i}(0) + \vec{u_i}(0) \cdot \gamma$, where $\vec{u_i}(0) = [\partial\vec{x_i}/\partial\gamma](0)$ is calculated at the initial state. From the linear trajectories, we extract the variation of all overlaps (contacts) and underlaps (gaps between particles) with strain. Contact changes then correspond to sign changes of the overlaps and underlaps. As we will see, this strategy not only allows us to accurately obtain the strain for the first contact change, but also gives us insight into the microscopic mechanisms. In particular, linear response allows us to probe the closing of contacts in detail, which is difficult in direct numerical simulations (\dns) since, at low $N^2P$, it becomes exceedingly rare for the first contact change to be a closing event (Fig.~\ref{fig:mk-bk-mixed-probabilities}). In this picture, the contact changes stem from a combination of geometric and linear response properties not explicitly considered before.

 In this section, we show that the response remains \emph{essentially} linear up to the first contact change: the nonlinear behavior of jammed packings under deformations arises mainly due to the cumulative effects of many contact changes. First, the stess-strain response is essentially linear between contact changes (\ref{sec:lr-stress-response}). Then, we show that linear response predicts the contact change strains with surprising accuracy (\ref{sec:lr-contact-change-strains}): Linear response predicts its own demise.
Finally, we investigate the first breaking and first closing events according to linear response (\ref{sec:lr-ensemble-averages}).

\subsection{Stress response}
\label{sec:lr-stress-response}
        \begin{figure}
        \includegraphics[width=\columnwidth]{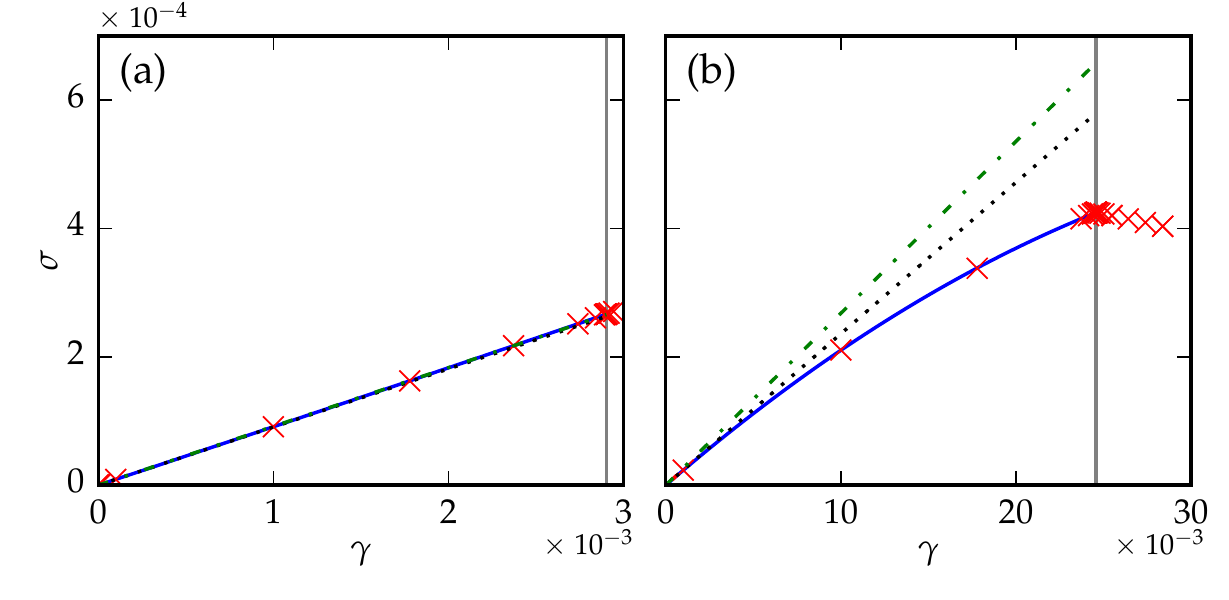}
        \subfloat{\label{fig:linres-stress-response-example-typical}}
        \subfloat{\label{fig:linres-stress-response-example-extreme}}
        \caption{\label{fig:linres-stress-response-example-both}(color online) Stress response for (a) a packing with typical $Q =0.014$ and (b) a packing with a very strong nonlinearity ($Q=0.267$; both $N=16, P=10^{-2}$). %packing $113$  en $9109$
        The simulation data (red $\times$) is fitted with the second-order polynomial (blue solid curves) $\sigma = G_2\gamma + \lambda\gamma^2$.
        The black dotted curves are the linear contribution $\sigma = G_2 \gamma$; the green dash-dotted curves are the linear response predictions $\sigma = G_\LR\gamma$.
        The gray vertical lines indicate the strain at the first contact change $\gammastar$.
        }
        \end{figure}
                \begin{figure}
                \includegraphics[width=\columnwidth]{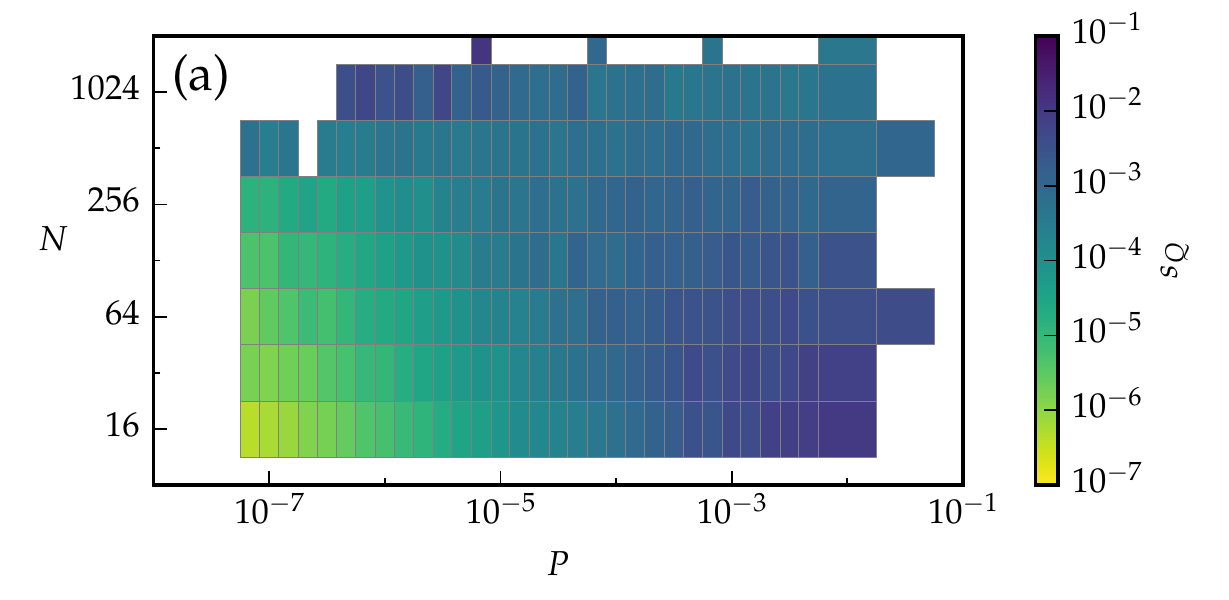}
                \includegraphics[width=\columnwidth]{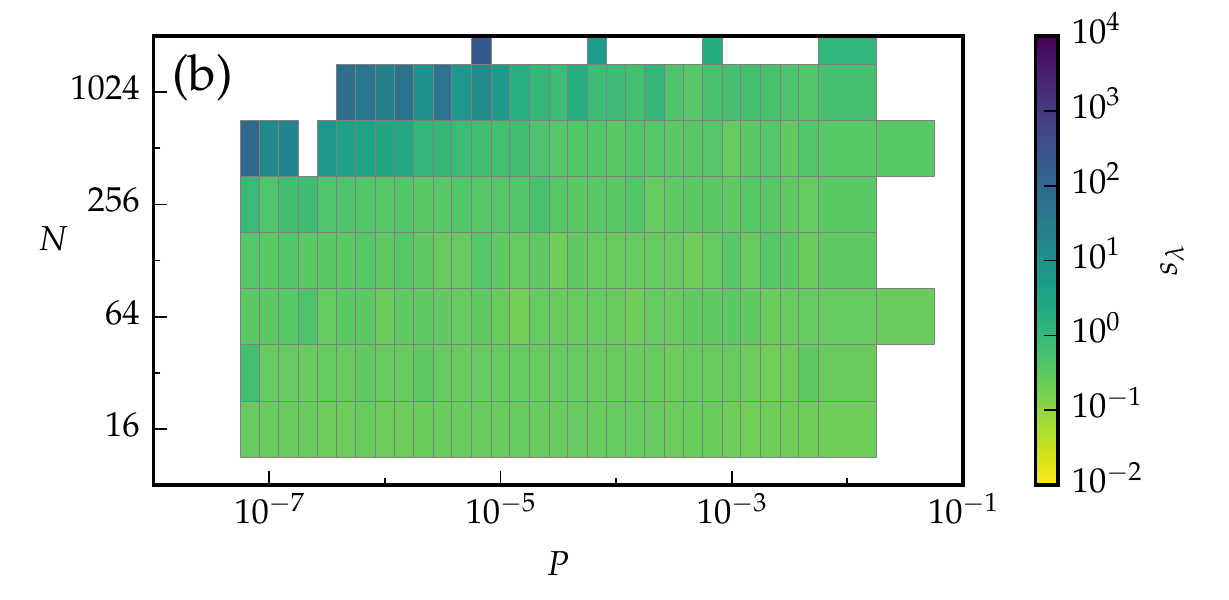}
                \caption{\label{fig:linres-stress-response-scaling-N-P} (color online) (a) 
                Width of the distribution of $Q$, the relative deviation from linear response at the first contact change for different ensembles. Clearly, the stresses are very well described by linear response for small systems at low pressures. More significant deviations occur for small systems at high pressures and large systems at low pressures.
                (b) \label{fig:linres-stress-response-scaling-lambdasq-N-P} Width of the distribution of $\lambda$. The quadratic component is on the order of $10^{-1}$ in most cases, but grows large for large systems at low pressures (i.e. close to jamming).             }
                \end{figure}
First, we will show that the stress-strain response of our systems is essentially linear in the {\dns} simulations up to the first contact change. From the simulations, we obtain the shear stress $\sigma(\gamma)$ at various strains before the first contact change (Figs.~\ref{fig:linres-stress-response-example-both}a-b).
We fit this response with a second-order polynomial $\sigma = G_2\gamma + \lambda\gamma^2$, and quantify the relative contribution of the quadratic component as the ratio between the quadratic and linear contributions at $\gammastar$:
\begin{equation}
Q = \frac{\lambda\gammastar^{2}}{G_2\gammastar} = \frac{\lambda\gammastar}{G_2}.
\end{equation}

For a given $N$ and $P$, the fluctuations in $Q$ are much larger than the mean. Hence, the relative importance of the nonlinearities is given by
the width of the distribution $P(Q)$. Our data indicates that these distributions exhibits fat tails, i.e. decays significantly slower than exponential, and that
the second moment of $Q$ is ill-defined. Therefore, we characterize the width of $P(Q)$ by halve of
the 16\%-84\% width that we denote $S_Q$ --- for Gaussian distributions this corresponds to one
standard deviation. We have checked that $S_Q$ allows to collapse the CDFs of $Q$ (such integrals over $P(Q)$ are more robust to small sample fluctuations than PDFs), such that $S_Q$ presents a robust measure of the fluctuations and magnitude of $Q$.

In Fig.~\ref{fig:linres-stress-response-scaling-N-P}, we plot $S_Q$ as function of $N$ and $P$.
The most important observation is that $S_Q$ remains small in the vast majority of cases, and the strongly nonlinear case shown in Fig.~\ref{fig:linres-stress-response-example-extreme} is truly exceptional. The two regions where $S_Q$ appears to be largest are for small $N$ and large $P$, and for large $N$ and small $P$. The origins for these deviations are different. For large systems at low pressure, the larger deviation is caused by inherent nonlinearities in the system. Small systems at high pressures exhibit also significant deviations from linear response, as the characteristic strains at the first contact change become large when $N$ is small and $P$ is large. Nevertheless, the quadratic contribution to the stress, $\lambda \gamma^2$  is  small compared to the linear contribution $G_2 \gamma$, and we therefore
expect to be able to predict the response of the system directly from linear response.

\subsection{Contact change strains}
\label{sec:lr-contact-change-strains}
In this section we describe how to calculate the contact change strains from linear response, and compare these values to the results from direct numerical simulations. First, for each particle pair $i,j$, we determine the contact change strain $\gammaij$, defined as the strain where the particles, assuming linear trajectories, break contact or make a new contact. By minimizing over all these strains, we calculate the strain at which the first new contact is made $\gammastarmklr$, the strain at which the first contact breaks $\gammastarbklr$, and their minimum gives the strain at the first contact change $\gammastarlr$. We then, for each packing, compare these values to their counterparts obtained by simulations.

\paragraph{Calculating $\gammastarlr$}
For each particle pair $i,j$, we determine the center-to-center distance $\vec{r_{ij}}$, and use linear response at $\gamma=0$ to determine $\vec{u_i} = \partial\vec{x_i}/\partial\gamma$. The inter-particle velocities are then given by $\vec{u_{ij}} = \vec{u_i} - \vec{u_j} - n_{y,ij} L_{yy} \hat{x}$, where the last term incorporates the velocity between the copies of the periodic box.
Combining these, we can solve $|\vec{r_{ij}} + \gammaij \vec{u_{ij}}| = R_i + R_j$ for $\gammaij$ to determine when the overlap $\delta_{ij} = 0$.

We determine the first broken and closed contact independently:
\begin{align}
\gammastarbklr \equiv & \min_{i,j\textrm{ in contact}} \gammaij,\label{eqn:gamma-bk-min-over-systeem}\\
\gammastarmklr \equiv & \min_{i,j\textrm{ not in contact}} \gammaij.\label{eqn:gamma-mk-min-over-systeem}
\end{align}
which allows us to study opening and closing events directly and independently, which is impossible in {\dns} simulations. The first contact change for the entire system is then determined by taking the minimum of the strain over \emph{all} particle pairs $i,j$:
\begin{align}
\gammastarlr \equiv & \min (\gammaij).
\end{align}

\paragraph{Comparison with {\dns} simulations}
        \begin{figure}
        \includegraphics[width=\columnwidth]{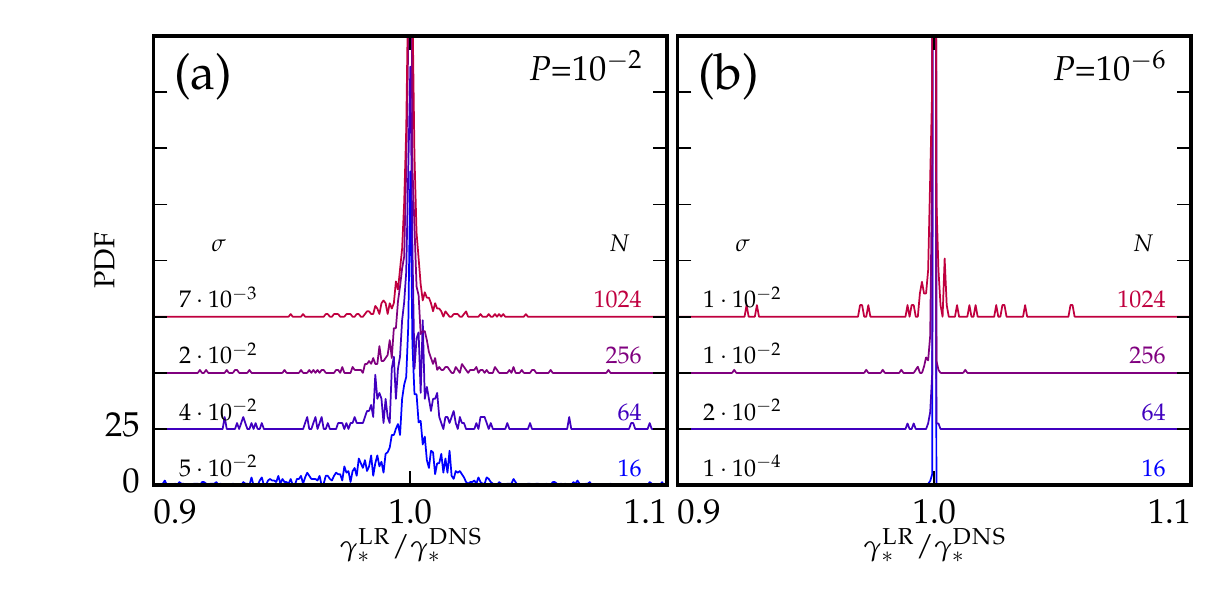}
        \caption{(color online) PDFs of $\gammastarlr / \gammastardns$ for various system sizes as indicated at (a) $P=10^{-2}$ and (b) $P=10^{-6}$. For each PDF, the standard deviation $\sigma$ is indicated. \label{fig:gammalr_vs_gammamd_pdf}}
        \end{figure}
We now show that linear response accurately predicts the contact change strain. For each individual system, we compare the linear response values $\gammastarlr$ to the corresponding strain $\gammastardns$ from the {\dns} simulations. In Fig.~\ref{fig:gammalr_vs_gammamd_pdf}, we plot \textsc{pdf}s of $\gammastarlr/\gammastardns$ to quantify the relative deviation from the simulation.
We observe that $\gammastarlr$ is a good predictor for $\gammastardns$.
First, these distributions are peaked around $1$, which shows the mean strain found in linear response matches that of the simulations very well.
Secondly, the standard deviation of the distributions, $\sigma$ is of the order of $5\%$ for small systems and $1\%$ for large systems. At $P=10^{-2}$, the largest packings have a standard deviation of $7\cdot 10^{-3}$, which increases to $5\cdot10^{-2}$ for small systems. The largest standard deviation is obtained for very small systems ($N=16$) at high P ($10^{-2}$). We find a large dependency on pressure: for $P=10^{-6}$, the distributions become very narrow around $1$. The standard deviation remains on the order of $10^{-2}$ due to outliers. We conclude that for all parameters considered, the differences between the strains obtained by linear response and direct numerical simulation are small.
In addition to determining the right contact change \emph{strain}, we found that in over $90\%$ of cases linear response also correctly identifies the contact $i,j$ where the first contact change takes place.

In conclusion, linear response provides us with a powerful tool to predict the behavior of packings. It allows us to predict the correct first contact change, as well as determining microscopic properties unavailable in the {\dns} simulations. We  note in passing that the correct prediction of contact changes suggests that shearing jammed packings might be modeled in terms of a \emph{discrete event simulation}, where, instead of slowly stepping through strain space, we immediately jump from contact change to contact change.

\subsection{Scaling of ensemble averages obtained in linear response}
\label{sec:lr-ensemble-averages}
\begin{figure}
\includegraphics[width=\columnwidth]{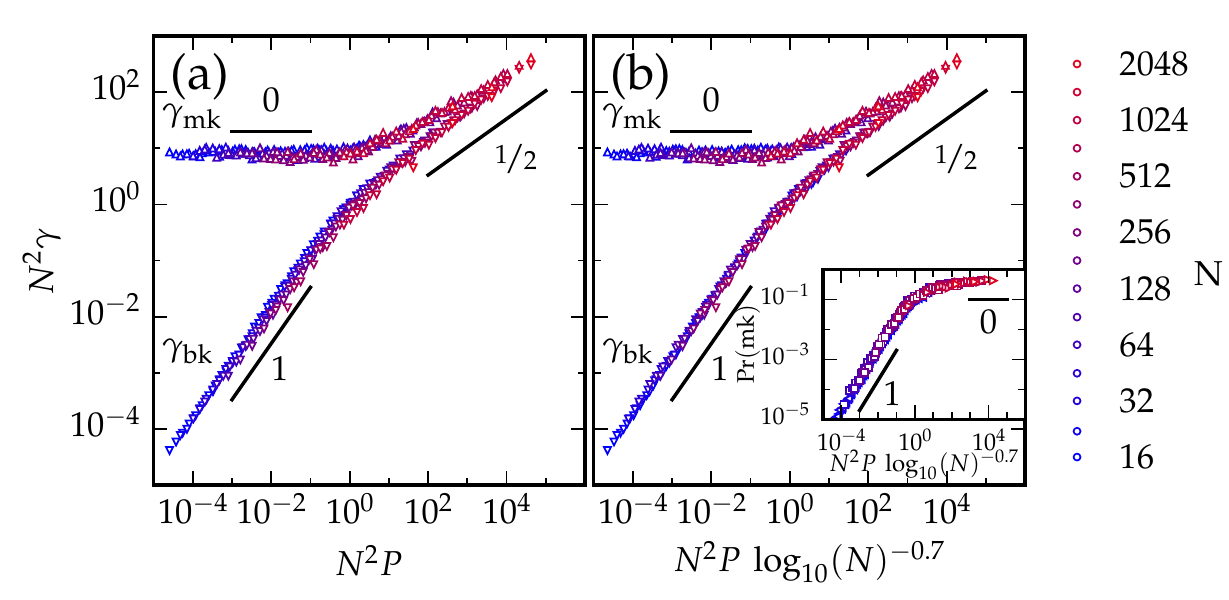}
\subfloat{\label{fig:lr-ensemble-averages-N2P}}
\subfloat{\label{fig:lr-ensemble-averages-N2Plog}}
\caption{ (color online) (a) Scaling of ensemble averaged breaking ($\bigtriangledown$) and making ($\bigtriangleup$) strains $\gammabklr = \langle\gammastarbklr\rangle$ and $\gammamklr = \langle\gammastarmklr\rangle$ from linear response. (b) As in Fig.~\ref{fig:gammamed-collapse-both}, log corrections significantly improve the quality of the collapse.
(inset) $\Pmk^\LR = 1/(1+\gammamk/\gammabk)$ is approximately $0.5$ for high $\nplog$, and scales as $\gammabk$ for small $\nplog$.}
\end{figure}

We now use linear response to study the strains at which contacts are broken or created in detail. Based on Eq.~\ref{eqn:predicted-scalings-compression}, we expect three scaling regimes for the contact change strains: for low $N^2P$, $\gammamk \sim 1/N^2$ and $\gammabk\sim P$, while for high $N^2P$, both $\gammabk$ and $\gammamk$ are expected to scale as $\sqrt{P/N^2}$. As before, these scalings suggest scaling collapse if we plot $N^2\gamma$ as a function of $N^2P$:

\begin{figure*}[t]
        \includegraphics[width=\textwidth]{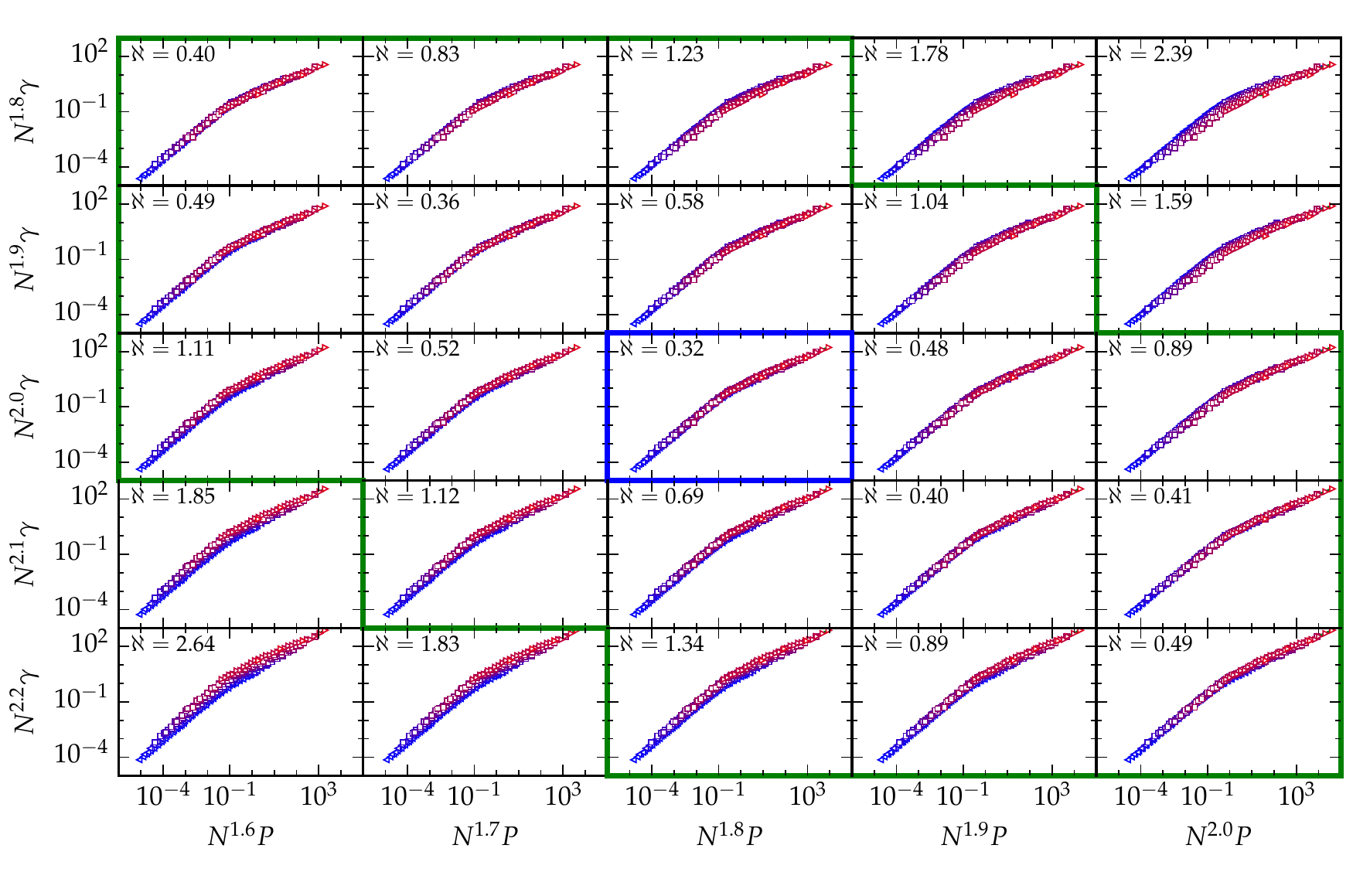}  %{../figs/fig00530.pdf}
\caption{Alternate scaling plots for $N^q \gamma \sim F(N^r P)$, with $q = 1.8\ldots2.2$ (vertical) and $r = 1.6\ldots 2.0$ (horizontal). $\aleph$ is a measure for collapse quality (see text), and the best collapse is found for $N^2 \gamma \sim F(N^{1.8} P)$ (blue border), and all collapses with $0 \leq r-q \leq 0.4$ are reasonable (green border).  \label{fig:alternate-scaling-collapses}}
\end{figure*}

In Fig.~\ref{fig:lr-ensemble-averages-N2P}, we plot our linear response data using this rescaling. As in Sec.~\ref{sec:scaling}, applying log corrections \cite{goodrich2013jamminginfinite} improves the collapse.
For low $N^2P$, we find that the data is well described by the expected power laws $\gammamk \sim (N^2P)^0$ and $\gammabk \sim (N^2P)^1$. For high $N^2P$, we find that neither branch cleanly scales as $\gamma \sim \sqrt{N^2P}$. Nevertheless, the linear response data support the scaling arguments, and in particular reveal the plateau for $\gammamk$ that cannot be obtained in {\dns}. We expect that for larger systems  the clean square root scaling will be recovered for both branches, as here both $N^2P$ can be large while $P$ remains small.

We have seen that linear response provides us with a powerful tool to understand what happens in the simulations. We not only predict the first contact change with surprising accuracy, we can also capture the prevalence of different types of events.

\subsection{Log-corrections versus freely adjustable exponents}

Here we investigate how accurately we can determine the power laws via scaling collapse of our data, and compare the log corrections we applied in Sec.~\ref{sec:scaling} to power law corrections.
In Sec.~\ref{sec:scaling}, we provided three arguments that predict the following scaling for the first contact change strain $\gammacc$:
\begin{equation}
N^2\gammacc \sim F\left(N^2P\right)
\end{equation}
where $F(x) \sim x$ for small $N^2P$ and $F(x) \sim x^{0.5}$ for large $N^2P$. In the same section, we have seen the results from the simulation collapse when plotted in this way. Furthermore, we have seen that by adding the log correction
\begin{equation}
N^2\gammacc \sim F\left(N^2P\log_{10}(N)^{-0.7}\right),
\end{equation}
with the same $F(x)$ the collapse improves.

First, we investigate for which exponents in $N$ the collapse, without the log correction, is satisfactory, i.e., for what values of $q$ and $r$ does
\begin{equation}
N^q\gammacc \sim F\left(N^r P\right)
\end{equation}
give an acceptable collapse? To make this quantitative, we measure the running maximum (starting at low $N^r P$) and the running minimum (starting at high $N^r P$), and calculate the effective area between the curves
\begin{equation}
\aleph = \int \left[ \log_{10}(M(N^r P)) - \log_{10}(m(N^r P)) \right] d \log_{10}(N^r P),
\end{equation}
where
\begin{align}
M(x) =& \max(N^q\gammacc | N^r P \leq x), \\
m(x) =& \min(N^q\gammacc | N^r P > x).
\end{align}
In Fig.~\ref{fig:alternate-scaling-collapses}, we show collapse plots for $q=1.8\ldots 2.2$ and $r=1.6\ldots 2.0$. We observe that all plots with
\begin{equation}
r \le q \le r + 0.4
\label{eqn:reasonable-collapse}
\end{equation}
are reasonable ($\aleph \lessapprox 1$), and that $N^2\gamma \sim F\left(N^{1.8} P\right)$ has the best overall scaling collapse ($\aleph = 0.32$). Our log-corrected collapse is very close to this, with $\aleph = 0.37$.

\begin{figure}
        \includegraphics[width=\halfcolwidth]{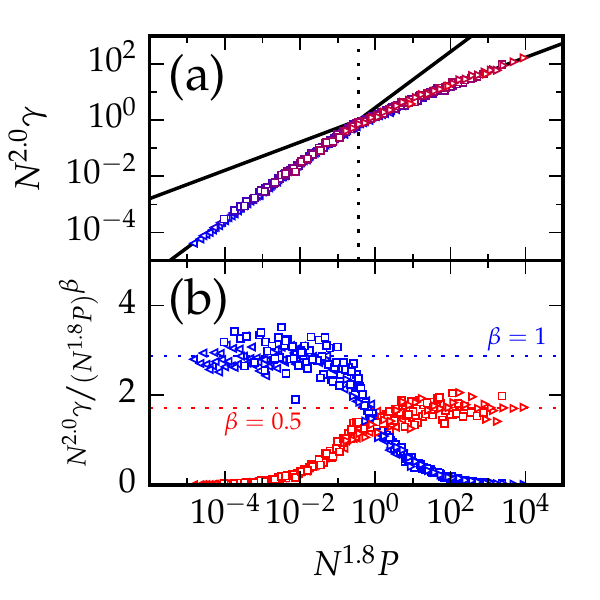}
        \includegraphics[width=\halfcolwidth]{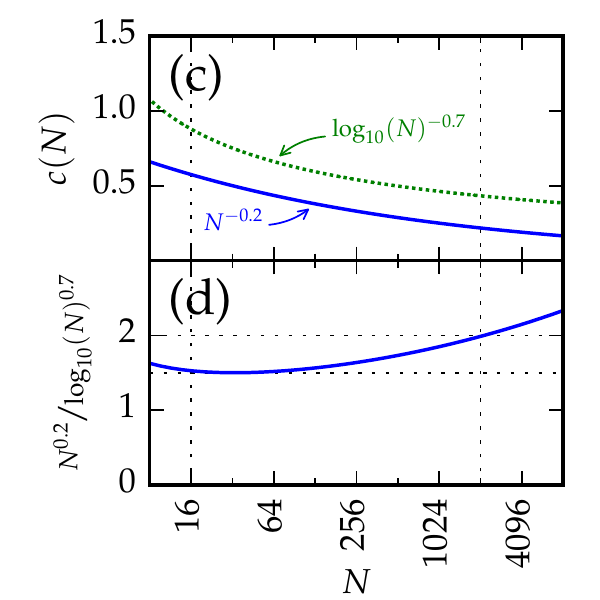}
        \subfloat{\label{fig:gammacc-vs-N2P-fitting-power-law}}
        \subfloat{\label{fig:gammacc-vs-N2P-residue}}
        \subfloat{\label{fig:gammacc-vs-N2P-Npwr-vs-logNpwr-corrections}}
        \subfloat{\label{fig:gammacc-vs-N2P-Npwr-vs-logNpwr-ratio}}
\caption{(color online) (a) Asymptotical behavior of $F(x)$. Black lines show the result from the power law fit: $F(x) \sim x^{1.0}$ for low x and $F(x) \sim x^{0.5}$ for high x. The crossover between the two regimes is at $x=0.4$.
(b) Residual plot $F(x)/x^{1.0}$ (dark/blue) and $F(x)/x^{0.5}$ (light/red) show the fitted power laws match the behavior very well in their respective regimes, as they scatter around a constant value.
(c) Log-correction $c(N)=\log_{10}(N)^{-0.7}$ and power law correction $c(N)=N^2 / N^{1.8} = N^{-0.2}$ as function of system size $N$. Both vary roughly by a factor of two in the range of $N$ we probe. (d) The ratio of the two varies by less than 35\%.}
\end{figure}
Secondly, we can wonder about the correct asymptotical behavior of $F(x)$. To find this behavior, we fit $F(x) = C \cdot x^\beta$ separately for both the upper ($N^{1.8}P > 10$) and lower ($N^{1.8}P < 0.1$) branches (Fig.~\ref{fig:gammacc-vs-N2P-fitting-power-law}). Here, we find
\begin{equation}
F(x) = \begin{cases}
(1.7 \pm 0.1) \cdot x^{0.50\pm0.01} & (x \ll 1) \\
(2.7 \pm 0.3) \cdot x^{1.00\pm0.01} & (x \gg 1)
\end{cases},
\end{equation}
which means that the best overall scaling of $\gamma$ becomes
\begin{equation}
\label{eqn:scaling-fit-two-regimes}
\gamma = \begin{cases}
(1.7 \pm 0.1) \cdot P^{0.5} N^{-1.1} & (N^2P \ll 1) \\
(2.7 \pm 0.3) \cdot P^{1} N^{-0.2} & (N^2P \gg 1)
\end{cases}.
\end{equation}
The error bars are given by the variation of the parameters when the fit range is increased or decreased by a decade. When $p$ and $q$ are varied within the collapse region, the exponents vary by $\sim \pm 0.05$.

When we compare the power laws to our expected scaling, we find the scaling of $\gamma$ with $P$ is as expected, but note two differences from the expected scaling of $\gamma$ with $N$. First, we observe $\gamma$ decreases as $N^{-0.17}$ for small systems, instead of the independence of $N$ our scaling model predicted. Secondly, for large systems, we observe $\gammacc$ scales as $N^{-1.1}$ instead of $N^{-1}$.

\paragraph{Comparison between power law and log corrections}
We can interpret the $1.8$ exponent in $N$ as a correction to the predicted $N^2P$ scaling: $N^{1.8}P =  N^{-0.2} (N^2P)$. In Fig.~\ref{fig:gammacc-vs-N2P-Npwr-vs-logNpwr-corrections}, we compare this correction to the log correction described in Sec.~\ref{sec:scaling}. We observe the corrections produce largely the same effect in the range of $N$ that our simulations cover. When we plot the ratio of the two (Fig.~\ref{fig:gammacc-vs-N2P-Npwr-vs-logNpwr-ratio}), we observe that the  deviations between both corrections are less than $35\%$, over a range where $N^2$ changes by three orders of magnitude.

To achieve a measurable difference of a factor three, systems of at least $60000$ particles are required. Alternatively, simulations can be performed in three dimensions, in which case the log corrections disappear \cite{goodrich2013jamminginfinite}. As the variation in the quality of the collapse is small, caution is warranted.

To conclude, we find our deviations from the expected scaling can be described by both a log correction and a power law correction. Much larger or three-dimensional simulations are required to fully distinguish the two corrections.

\section{Multiple contact changes}\label{sec:shear-more-than-one}
In this section, we discuss the behavior of our systems when they are strained beyond their first contact change, focussing on the the implications of contact changes for continuum elasticity, and reveal intriguing patterns of subsequent make and break events.

% and will take a look at the effects of switching from free boundaries in the relaxed case to fixed %boundaries in the strained systems.

\subsection{Shear modulus}
\begin{figure}
\subfloat{\label{fig:G1-G0-histograms}}
\subfloat{\label{fig:G1-lt-0-fraction}}
\subfloat{\label{fig:G1-G0-stddev}}
\includegraphics[width=\halfcolwidth]{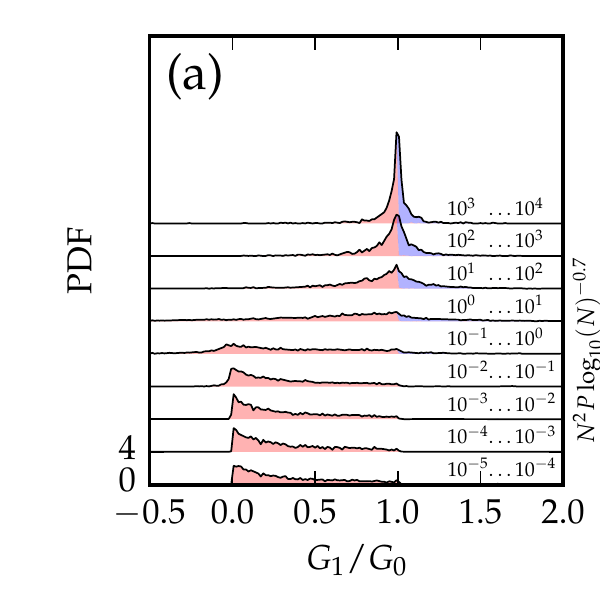}
\includegraphics[width=\halfcolwidth]{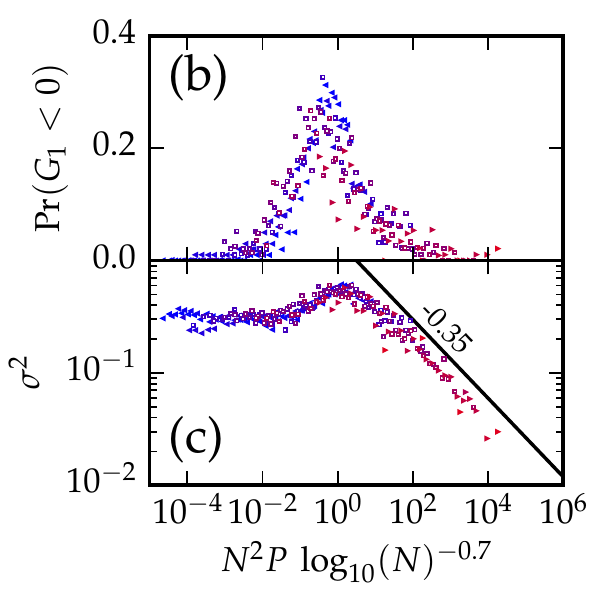}
\caption{(color online)
(a) Probability distribution functions for $G_1/G_0$, the relative shear modulus after the first contact change. For small systems at low pressures (bottom), we find $0 \leq G_1/G_0 \leq 1$; for intermediate system we find $G_1/G_0$ is typically smaller than $1$, but can become negative (indicating an unstable system). For large systems at high pressures (top), we find $G_1/G_0 \approx 1$. The creation of contacts (dark/blue) correlates with an increase in $G$, while the breaking of contacts (light/red) correlates with a decrease in $G$.
(b) The fraction of events where $G_1 < 0$ peaks around $\nplog\approx 1$.
(c) The standard deviation of $G_1/G_0$. For small systems at low pressures, $\sigma \approx 0.3$, whereas for large systems $\sigma \sim (N^2P)^{-\beta}$ with $\beta=0.35\pm0.01$.
}
\end{figure}
As we have seen, the first contact change happens at lower and lower strains as systems get larger. \citet{schreck2011repulsivecontactinteractions} suggested that this implies that linear response is no longer valid for disordered systems at large $N$.
It is clear that changing a single contact can have a large effect on small systems, but one would expect the effect to vanish in larger systems:
in the thermodynamic limit, systems are expected to behave increasingly like an elastic solid,  and this apparent paradox lead to a lively debate \cite{goodrich2013commentrepulsivecontact,schreck2013responsetocomment,goodrich2014whendojammed}.

Here we show how the effect of a single contact change on the shear modulus becomes smaller and smaller when the system size is increased. We note that, as long as the shear modulus does not change significantly, we can consider the system to have an \emph{effective} linear response, even though it is no longer \emph{strictly} linear.
To quantify the effect of a single contact change, we calculated the shear modulus before ($G_0$) and after ($G_1$) the first contact change using Eq.~\ref{eqn:elastic-modulus}. For each value of $N$ and $P$, we have calculated the probability distributions $\rho(G_1/G_0)$, and from these determine in particular $\rho(G_1 <0)$, and the width of these distributions (Fig.~\ref{fig:G1-G0-histograms}).
We find that the shape of these distributions varies strongly and that
we can organize our data using the finite size parameter
$\nplog$, and as function of this parameter we distinguish three regimes.

{\em(i) $\nplog \ll 1$:} In the small system size limit, we find that $\rho(G_1/G_0)$ is a strongly asymmetric distribution, with most weight around zero. We find that the mean $\langle G_1/G_0 \rangle \approx 0.2$, and that   $ 0< G_1 < G_0$. To understand this, we note that in this regime, the first contact change is a breaking event, which weakens the system. We find that $G_1$ is significantly smaller than $G_0$ because, in this regime, there is typically only a single excess contact ($N_c - 2N = 1$). Surprisingly, the system does not unjam immediately, for reasons we will discuss in Sec.~\ref{sec:whoop-whoop-whoop}.

{\em(ii)  $\nplog \approx 1$:} In the intermediate regime,
the number of excess contacts remains small, contact changes are predominantly contact breaking events, and we observe that $G_1 < G_0$. However, the probability that $G_1<0$ becomes finite, inc contrast to the behavior in regime {\em(i)}. This follows from the variation of prestress: without
prestress, $G$ has to be non-negative \cite{wyart2005geometricoriginexcess,goodrich2013jamminginfinite}, but as $P$ increases in regime {\em (ii)}
there is  sufficient prestress to allow for negative values of $G_1$, in up to $35\%$ of cases (Fig.~\ref{fig:G1-lt-0-fraction}).

{\em(iii)  $\nplog \gg 1$:} For large systems, we enter the
continuum regime, where  the distribution $\rho(G_1/G_0)$ peaks around one and becomes increasingly symmetric and narrow. Hence $G_1 \approx G_0$, and this is the essence of the solution of the apparent paradox. The symmetry of the distribution is consistent with out observation that contact creation and contact breaking becomes equally likely in this regime.

A simple scaling argument for the width of this distribution can be obtained from combining the scaling of $G$ with $P$,  $G \sim \Delta z \sim \sqrt{P}$ with the observation that making and breaking of contacts is equally likely. As a single contact change modifies $\Delta z$ by $\pm 1/N$, we thus expect $G_1^\pm \sim \Delta z_0 \pm 1/N$. The width of this distribution  scales as
\begin{equation}
\sigma \sim \frac{G_1^+ - G_1^-}{G_0} \sim \frac{1/N}{\Delta z_0} \sim \frac{1/N}{\sqrt{P}} = \frac{1}{\sqrt{N^2P}}.
\end{equation}

We measured the width of this distribution using the standard deviation $\sigma$, and observe that it vanishes as $(\nplog)^{-\beta}$ with $\beta = 0.35 \pm 0.01$ (Fig.~\ref{fig:G1-G0-stddev}).
We suggest that the contacts changed under a shear deformation have a relatively large impact on the shear modulus - a relatively small number of contacts contribute disproportionally to the elastic moduli \cite{goodrich2015principleindependentbond}.

Nevertheless, the observed diminishing of the width of the distribution $\rho(G_1/G_0)$ is sufficiently strong to be consistent with an effective linear response picture. We call a material \emph{effectively linear} if, for a small fixed deformation $\gamma_t$, the standard deviation of $G(\gamma_t)$ vanishes for $N \rightarrow \infty$. In terms of contact changes, we thus need to establish how the number of contact changes experienced up to $\gamma_t$ grows with $N$, and how the effect of single contact changes decreases with $N$.
We estimate the number of contact changes between $\gamma = 0$ and the test strain $\gamma_t$ as \begin{equation}
n = \gamma_t / \gammacc = \gamma_t / (\sqrt{P} / N).
\end{equation}
We then assume that all contact changes are independent of each other, and assume each contact change causes a change in $G$ drawn from the distribution $\rho(G_1/G_0)$ with standard deviation $\sigma \sim (N^2P)^{-\beta}$. The central limit theorem then states the standard deviation after $n$ contact changes is given by
\begin{equation}
\sigma_n \sim \sqrt{n} (N^2P)^{-\beta}.
\end{equation}
Combining these, we find that the standard deviation after a strain $\gamma_t$ is given by
\begin{equation}
\sigma_{\gamma_t} \sim \sqrt{\gamma_t / (\sqrt{P} / N)} \left(N^2P\right)^{-\beta}
\sim \sqrt{\gamma_t} \cdot N^{\frac{1}{2}-2\beta} P^{-\frac{1}{4}-\beta},
\end{equation}
which vanishes for large $N$ as long as $\frac{1}{2} - 2\beta < 0$, or
\begin{equation}
\beta > 1/4.
\end{equation}

Clearly, $0.35 > 1/4$, so, for $N \rightarrow \infty$, our systems approach the continuum limit.
Significant correlations between subsequent values of $G_{i+1/G_i}$ could in principle lead to a more problematic approach to the continuum limit. However, recent work by Boschan et al. \cite{boschaninprep} found that the ensemble-averaged stress-strain curve is linear with a slope compatible with $\langle G_0 \rangle$ up to a strain of order $P$. Though not a definitive test, on the basis of these results we consider that strong correlations are unlikely to be present. Our data is thus consistent with the picture where, for large $N$, the effective value of $G$ depends on the applied shear $\gamma$ rather than the number of contact changes $n$ \cite{dagoisbohy2014oscillatoryrheologynear, boschaninprep}.

\subsection{Alternating contact changes}
\label{sec:whoop-whoop-whoop}
\begin{figure}
\includegraphics[width=\columnwidth]{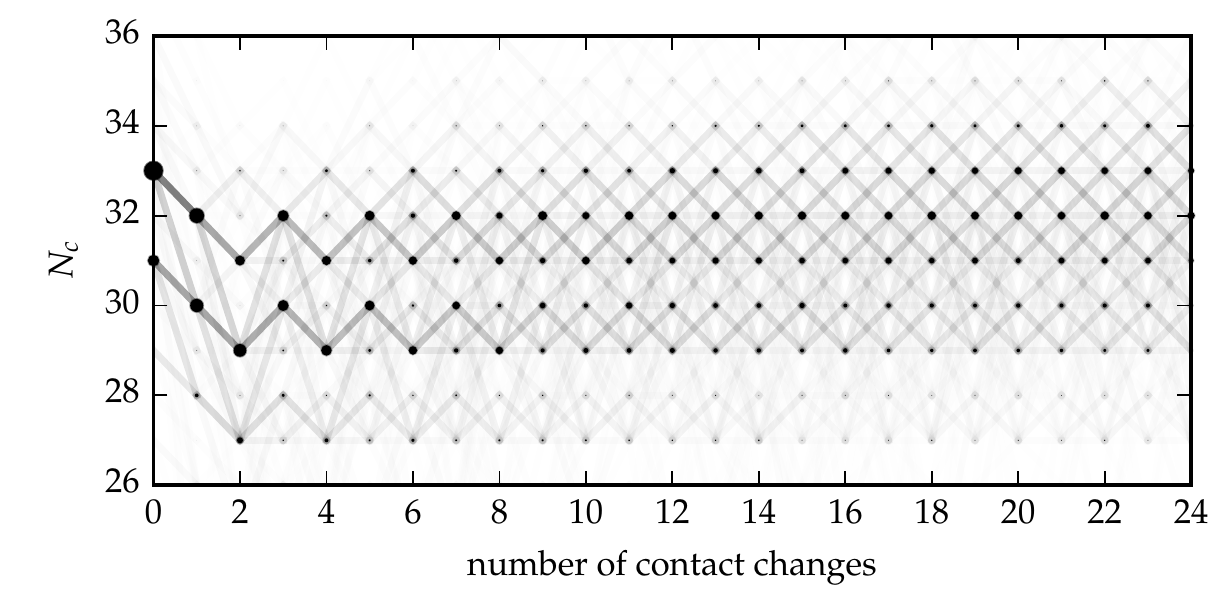}
\caption{The number of contacts, $N_c$, for systems with $N=16$ particles at $P=10^{-6}$ as function of the cumulative number of contact changes. The circle area represents the fraction of systems with a given number of contacts; the thickness of the lines represent transition probabilities.
Initially, the systems start off with the minimum number of contacts $2N+1 = 33$ ($31$ or $29$ when there are one or two rattlers, respectively). In the first and second contact change, the system loses one contact (three when a rattler is created). In the following events, the system alternately gains and loses a contact. \label{fig:excess-contacts-N16-lowP}}
\end{figure}

Here, we investigate correlations between consecutive contact changes, focussing on the $N^2 P \ll 1$ regime. In Fig.~\ref{fig:excess-contacts-N16-lowP}, we show the number of contacts in the system, $N_c$, as a function of the number of contact changes for systems with $N=16$ particles, at $P=10^{-6}$. Before shearing, $N_c$ reflects the number of rattlers, with $N_c=33,31,29$  corresponding
to zero, one and two rattlers, respectively. The presence of these rattlers accounts for the  parallel tracks in the dominant transition pathways ( Fig.~\ref{fig:excess-contacts-N16-lowP}). For definiteness, let us focus on the case where the initial packing has no rattlers $(N_c=33)$. When the system is sheared, rattlers occasionally  form, and $N_c$ is then seen to drop by three. In roughly one in ten packings, shearing cause three successive breaking events, causing  the system to unjam.
In most cases, however, we find that there are first two breaking events, followed by a series of alternating making and breaking events; clearly, throughout this process the pressure remains finite and the system remains jammed. This alternating behavior stays apparent at least until the 10$^\textrm{th}$ contact change. This evidences correlations between subsequent events. We note that for larger pressures $(N^2 P>1)$, such correlations are absent.

To interpret the values of  $N_c$, we recall that the initial condition of these simulations are
$\epsilon_\textrm{all}^+$ packings
that have positive moduli and zero residual shear stress; for these packings it is well known that the minimal number of contacts
equals $2N+1$, consistent with the initial values of $33,31,\dots$
observed here \cite{dagoisbohy2012softspherepackings,goodrich2013jamminginfinite}.
The reason the system under shear remains jammed for lower contact numbers, is that the boundary conditions during shear, and during initial
equilibration are different. Once the system is equilibrated, the box shape parameters $\alpha$ and $\delta$ (see Appendix A) are fixed, the system has two degrees of freedom less, and can remain jammed down to $N_c=2N-1$ \cite{dagoisbohy2012softspherepackings,goodrich2013jamminginfinite}. The situation is somewhat subtle though. We have observed that whether we fix the simulation box volume (as shown in Fig.~\ref{fig:excess-contacts-N16-lowP}) or fix the pressure does not change the minimal contact number during shear. However, if we fix the deviatoric (pure shear) stress $\tau = (1/2)\left(\sigma_{xx} - \sigma_{yy}\right)$ instead of the pure shear strain
$\delta = (L_{yy} - L)/L = \sqrt{L_{yy}/L_{xx}} - 1~$, we find that the minimal contact number is $2N$ instead of $2N-1$.

We note that  the same contact is often involved in multiple contact changes, although typically not in \emph{subsequent} contact changes.
This is an example of the intriguing correlations in the spatiotemporal patterns of contact changes that invite further studies.
We already discussed one aspect of the boundary conditions. In the constant volume protocol, the pressure increases with shear due to dilatancy --- for the example shown in Fig.~\ref{fig:excess-contacts-N16-lowP}, the pressure becomes of order $10^{-4}$ in the strain interval leading to the first contact creation event. How simulations at constant pressure, and/or constant shear stress influence this phenomenology is an open question.

\section{Extremal value scaling}

 Second, we will approach the problem from a statistical perspective. Starting from the distribution of $\gammastar$ of all contacts in all packings, we apply extreme value analysis to find the expected mean first contact change. We find that this does not yield a good prediction for the measured value, and determine that this cannot be explained by a few weak contacts, but rather  points to strong correlations involving the whole system --- i.e., the statistics of the first $n$ changes in the system are different from the statistics of the \emph{first} contact change in $n$ systems.

\label{sec:extremal-value-scaling}
In this section we probe whether we can predict the scaling of $\gammacc$ and distribution of $\gammastar$
based on the distribution of all contact change strains $\rho(\gamma_{ij})$ for a given ensemble $(N,P)$. Note that before (Secs.~\ref{sec:shear-numerical-results} and \ref{sec:shear-linear-response}), we have determined the scaling of $\gammacc$ by determining $\gamma_*$ {\em for each packing}, and averaging over those values. We have found that the distribution of $\gamma_*$ is close to a exponential distribution. Assuming that large enough packings are statistically similar, it should be possible to predict $\gammacc$ from the distribution of  $\rho(\gamma_{ij})$ using extremal statistics. In particular, one might expect that $\rho(\gamma_{ij})$ takes on a simple form for sufficiently large $N$, possibly even amenable to a theoretical description.
Deviations from this picture may point to lack of self-averaging or other subtleties, and as such provide important information for developing a deeper theoretical understanding for the characteristic strains of the first contact change.
Before starting, we note that for contact creation, it is difficult to establish which potential contacts should be considered, and we therefore focus on the breaking of contacts only, using $\gammastarbklr$ from linear response. We will also limit our discussion to contacts that break for shear in the positive direction, i.e., $\gamma > 0$.
\begin{figure}
        \includegraphics[width=\halfcolwidth]{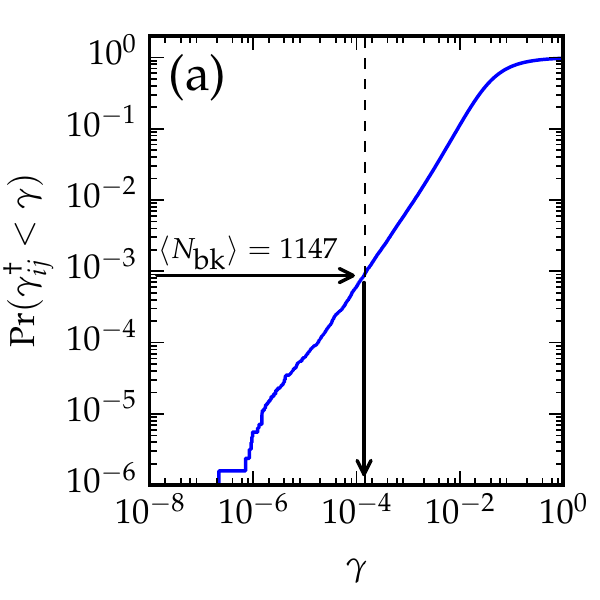}
        \includegraphics[width=\halfcolwidth]{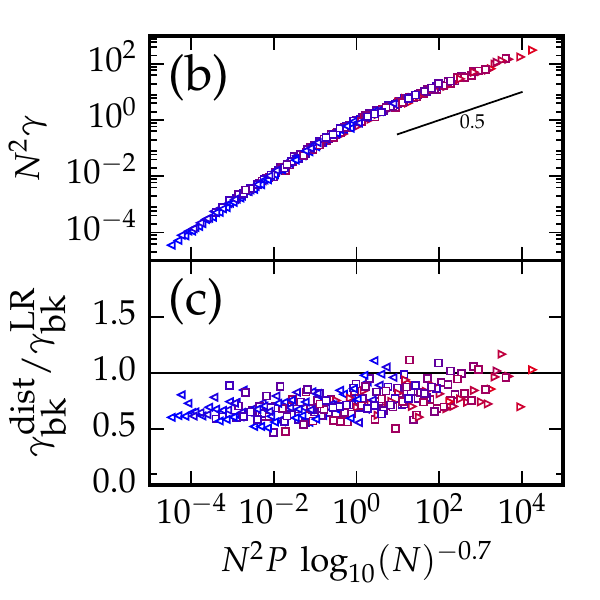}
        \subfloat{\label{fig:cdf-extremal}}\subfloat{\label{fig:scaling-extremal-via-cdf}}\subfloat{\label{fig:scaling-extremal-via-cdf-ratio}}
\caption{(color online) (a) {\cdf} of $\gamma_{ij}^\dagger$ for every contact for every packing in the $N=1024$, $ P=10^{-2}$ ensemble. The strain at which $\Pr(\gamma_{ij}^\dagger < \gammadist) = 1/\langle N_c\rangle$ is the expected contact breaking strain for this ensemble: $\gammadist=1.4\times 10^{-4}$. The mean breaking strain from linear response is $\gammabklr=1.5\times 10^{-4}$, and is indicated with the dashed line. (b) Colored symbols: resulting scaling of $\gammadist$. Gray background: scaling of $\gammabklr$, as in Fig.~\ref{fig:lr-ensemble-averages-N2Plog}. (c) The ratio $\gammadist/\gammabklr$ varies slowly with $\nplog$, from $\gammadist/\gammabklr \approx 0.5$ to $\gammadist/\gammabklr \approx 1.0$.}
\end{figure}
As a first probe of the usefulness of extremal value statistics for contact breaking,
we compare the results of two distinct methods to calculate
the mean contact breaking strain.
First, we define $\gammabklr=\langle\gammastarbklr\rangle $, the mean of the contact breaking strains determined for an ensemble of packings, as we have done in Sec.~\ref{sec:shear-linear-response}. Second,
we determine $\gammadist$ from the distribution of positive contact change strains $\rho(\gamma_{ij}^\dagger)$
by solving
\begin{equation}
\frac{1}{\langle N_\bk \rangle} = \int_0^{\gammadist} \rho(\gammaijdagger) d\gammaijdagger~.\label{eqn:CDF-extremal-value}
\end{equation}
To implement this, we first compute the numerical \cdf  $\Pr(\gammaijdagger < \gamma)$ based on the breaking strain $\gammaij$ for \emph{every} contact in \emph{every} packing in the ensemble  and then solve
\begin{equation}
\Pr(\gamma_{ij}^\dagger < \gammadist) = 1/\langle N_\bk \rangle~,\label{eqn:numerical-extremal-value}
\end{equation}
where $\langle N_\bk \rangle \approx 0.5 \langle N_c \rangle$ is the mean number of contacts that break under positive strain, for which we take the numerical ensemble average. This procedure is illustrated in Fig.~\ref{fig:cdf-extremal} for the $N=1024$, $P=10^{-2}$ ensemble, where $\langle N_\bk \rangle = 1147$.
For this particular example we find that $\gammabklr=1.5\times 10^{-4}$ whereas
$\gammadist=1.4\times 10^{-4}$. These values are close but distinct ($\gammadist/\gammabklr = 0.93$) --- as we will show below, there are systematic deviations between these numbers which provide insight into the statistics of contact breaking.

We can repeat this procedure for a synthetic ensemble of uncorrelated systems. From the frequentist distribution of contact breaking strains $\rho(\gammaijdagger)$ of the $N=1024$, $P=10^{-2}$ ensemble, we draw $N_\bk = 1147$ contacts for each of $N_s = 1000$ systems (bootstrapping). For each system, we calculate the minimum strain $\gammastar$. We then compare the mean breaking strain $\gammabk = \langle \gammastar \rangle = 1.34(4)\times 10^{-4}$ to $\gammadist=1.4\times 10^{-4}$. Here, we find $\gammadist / \gammabk = 1.05 \pm 0.04 > 1$. Values below $1$ thus indicate significant deviations from uncorrelated systems.

\begin{figure}
\includegraphics[width=\columnwidth]{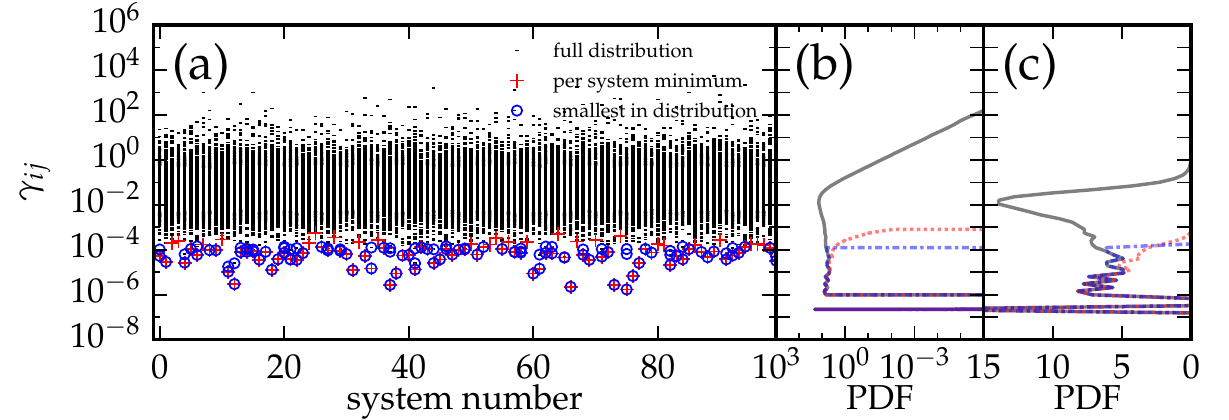}
\includegraphics[width=\columnwidth]{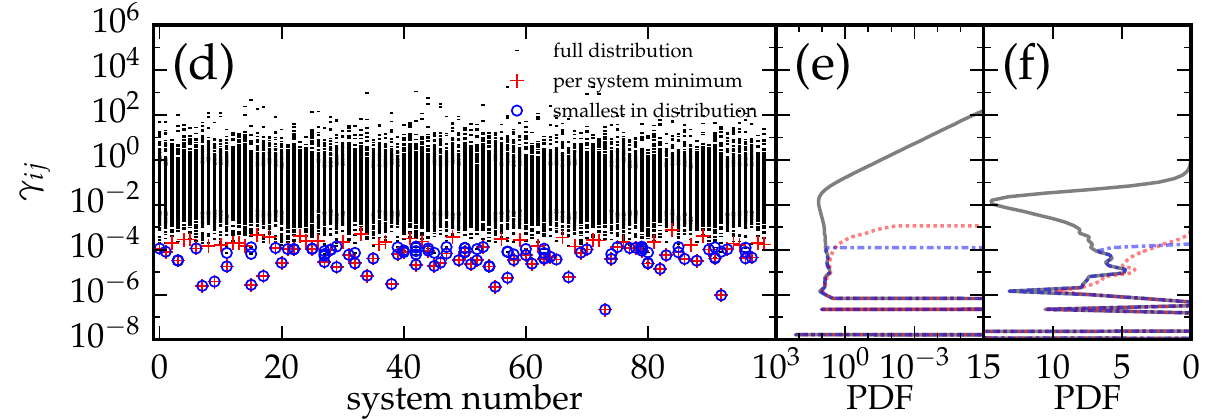}
\includegraphics[width=\columnwidth]{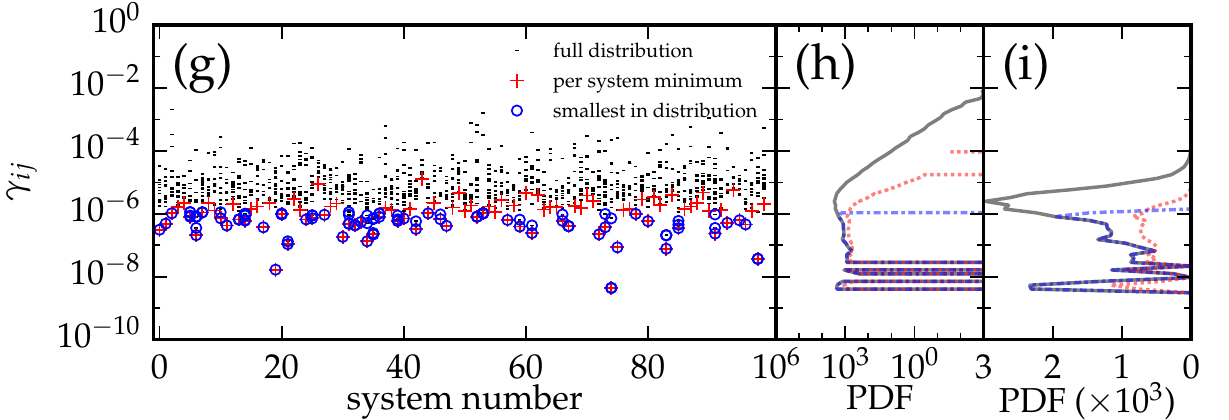}
\subfloat{\label{fig:lognormal-synth-scatter}}\subfloat{\label{fig:lognormal-synth-pdf}}{\subfloat{}}
{\subfloat{}}{\subfloat{}}{\subfloat{}}
\subfloat{\label{fig:N16-gammas-scatter}}
\caption{\label{fig:weakest-two-distriplot} (color online) (a) Scatter plot of each positive contact breaking strain $\gammaijdagger$ for $100$ synthetic systems drawn (bootstrapped) from the distribution $\rho(\gammaijdagger)$ for $N=1024$, $P=10^{-2}$ (black dots). For each system, $\gammastar \equiv \min \gammaijdagger$ is indicated with a red $+$. All values below the $1/N_c$ percentile are indicated with a blue $\circ$.
(b) The \pdf $\rho(x_{ij})$ (black). The distribution of per system minima ($\rho(\gammastar)/M$, red dashed) and values below the $1/N_c$ percentile ($\rho(\gamma_<)/M$, blue dash-dotted) as part of the whole are indicated.
(c) Same as (b), but with a linear \pdf axis.
(d,e,f) Same as (a,b,c), with numerical data from the $N=1024$, $P=10^{-2}$ ensemble.
(g,h,i) Same, with numerical data from the $N=16$, $P=10^{-6}$ ensemble.
}
\end{figure}

\paragraph{Distribution of strains}
We now probe the {\em distribution} of strains of first contact breaks.
Consider an ensemble of $M$ packings of $N$ particles, each with $N_\bk(m)$ contacts for which we calculate the breaking strains $\gamma_{ij}^\dagger$.  This yields a total of $\Sigma_{m=1}^M N_\bk(m) \equiv M \langle N_\bk\rangle$ samples (values of $\gamma_{ij}^\dagger$), as illustrated
in Fig.~\ref{fig:weakest-two-distriplot} for a synthetic data set, as well as for two data sets at fixed $P$ and $N$.
First, we can collect all breaking strains in a distribution $\rho( \gamma_{ij}^\dagger)$ (black curves in panels b,e,h).
As illustrated in Fig.~\ref{fig:weakest-two-distriplot} there are now two operations we can perform. Equivalent to what we
do to determine $\gammabklr$ in linear response, we can determine the minimum breaking strain for each of the $M$
packings, obtaining $M$ breaking strains (red crosses in panels a,d,g)
and the corresponding distribution $\rho(\gammastarbklr)$ (shown as red curves in panels b,e,h, as a fraction of $\rho(\gammaijdagger)$). Alternatively, we
may also consider the $M$ smallest values out of $M \langle N_\bk\rangle$  samples taken out of
the distribution
$\rho(\gammaijdagger)$ (blue circles), which yields the distribution $\rho (\gamma_<):= \rho( \gamma | \gamma \leq \gammadist)$ (blue curve).
The mean values considered above are related to these distributions as follows: $\gammabklr$ is the mean of the $\rho(\gammastarbklr)$, whereas
$\gammadist$ is the {\em maximum} value of $\gamma_<$ in $\rho (\gamma_<)$. Clearly, the distributions $\rho(\gammastarbklr)$ and $\rho (\gamma_<)$ in general will be different, but if the different packings are statistically indistinguishable and
large enough to allow for self-averaging, so that $\gammabklr \approx \gammadist$,
these distributions are directly related (see below), which yields a statistical test on the nature of the contact breaking strains.

\paragraph{Results}
We have determined $\gammabk$ and $\gammadist$ for all $(N,P)$ ensembles. In Fig.~\ref{fig:scaling-extremal-via-cdf}
we plot  $N^2 \gammadist$ vs $\nplog$, and in Fig.~\ref{fig:scaling-extremal-via-cdf-ratio} we plot the ratio
$\gammadist/ \gammabklr$ vs $\nplog$.
At low $\nplog$, we find that $\gammadist$ and $\gammabklr$ exhibit similar scaling with $\nplog$, but that their ratio $\gammadist/\gammabklr \approx 0.6 < 1.05 \pm 0.05$
points to deviations from self-averaging. At very high $\nplog$, $\gammadist$ increases faster than $\gammabklr$ and appears to reach equality for the highest values of $N^2P$ --- we suggest that here the packings are large enough to be self-averaging.

\begin{figure}
\includegraphics[width=\columnwidth]{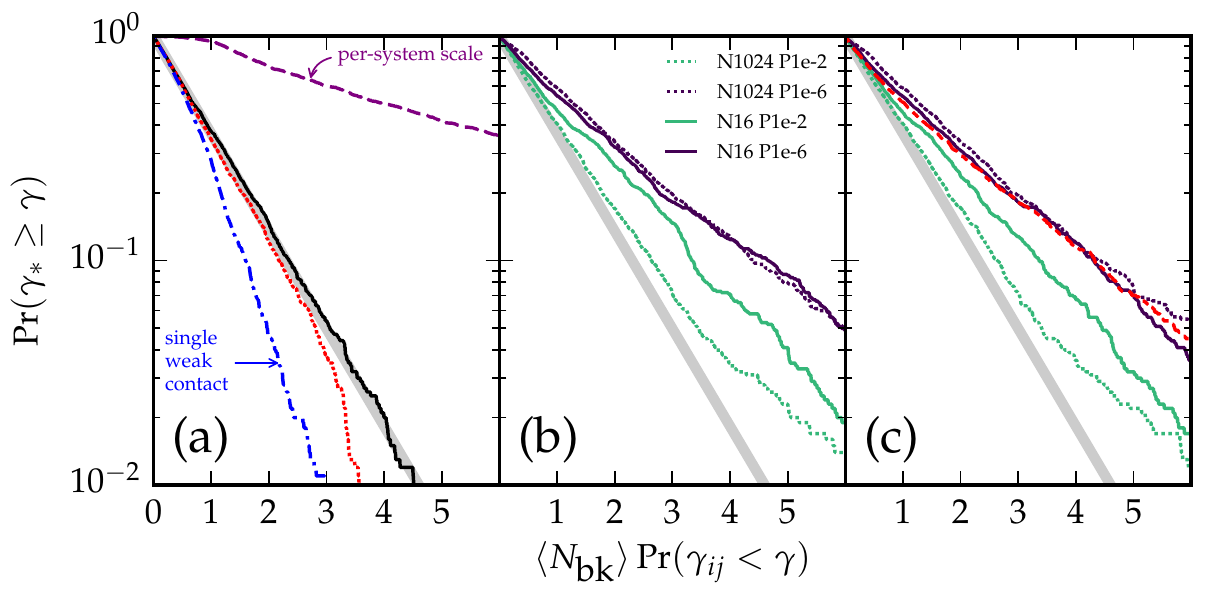}
\subfloat{\label{fig:weakest-two-simu}}
\subfloat{\label{fig:weakest-two-data}}
\subfloat{\label{fig:weakest-two-rescale}}
\caption{\label{fig:weakest-two} (color online) $\Pr(\gammastar \geq \gamma)$ as a function of $\langle N_\bk \rangle \Pr(\gammaij < \gamma)$ (see text). (a) Solid black: Synthetic data, drawn from $\rho(\gammaijdagger)$ in the $N=1024$, $P=10^{-2}$ ensemble ($\langle N_\bk \rangle = 1147$). For the same ensemble, data with a single value from a distribution with lower mean (dot-dashed blue) and for systems with an overall per-system scale (dashed purple) are also shown. Dotted red: Synthetic data, from $\rho(\gammaijdagger)$ in the $N=16$, $P=10^{-6}$ ensemble ($\langle N_\bk \rangle = 16$). The gray line indicates $\Pr(\gammastar \geq \gamma) = \exp(-\langle N_\bk \rangle \Pr(\gammaij < \gamma))$.
(b) Data from our simulations. We observe the curves decay \emph{slower} than exponential, indicating correlations between contacts. Curves from top to bottom: $N=1024$, $P=10^{-6}$; $N=16$, $P=10^{-6}$; $N=16$, $P=10^{-2}$; $N=1024$, $P=10^{-2}$;
(c) Data from (b), but with all strains rescaled to the mean of strains within one system. This reduces the effect of a per-system scale (dot-dashed red), but does not completely negate it. The behavior for the packing-derived data is unchanged as compared to (b). }
\end{figure}

To further characterize the origins of this breakdown of self averaging in small systems, we take a closer look at the distributions
$\rho(\gammastarbk)$ and $\rho (\gamma_<)$ in Figs.~\ref{fig:weakest-two-distriplot} and \ref{fig:weakest-two}.
In Fig.~\ref{fig:weakest-two-distriplot}(a--c), we plot each value of $\gammaijdagger$ for the first $100$ systems in the synthetic ensemble described above.
When we compare the \pdfs of the per system $\rho(\gammastarbk)$ (red curves in panel b) and distribution minima $\rho (\gamma_<)$ (blue curves in panel b), we note they are similar for small values of $\gammaijdagger$, but different for larger values of $\gammaijdagger$.

In Fig.~\ref{fig:weakest-two-simu} we compare the \cdf of the per system minima to the \cdf of the whole distribution. In the synthetic data, we can deduce that the inverse \cdf of minima $\Pr(\gammastar \geq \gamma)$ relates to the \cdf of the distribution $\Pr(\gammaij < \gamma)$ as
\begin{align}
\Pr(\gammastar \geq \gamma) &= (1 - \Pr(\gammaij < \gamma))^{\langle N_\bk \rangle} \nonumber\\
& = \left[1 - \frac{\#_{\gammaij < \gamma}}{N_s \langle N_\bk\rangle}\right]^{\langle N_\bk \rangle} \nonumber \\
& \approx \exp(- \frac{\#_{\gammaij < \gamma}}{N_s}) \nonumber \\
& = \exp(-\langle N_\bk \rangle \Pr(\gammaij < \gamma))~,\label{eqn:synth-cdf-comparison}
\end{align}
for large enough $\langle N_\bk \rangle$ for a given $\langle N_\bk \rangle \Pr(\gammaij < \gamma)$.
In Fig.~\ref{fig:weakest-two-simu}, we plot $\Pr(\gammastar \geq \gamma)$ as a function of $\langle N_\bk \rangle \Pr(\gammaij < \gamma)$ for both the synthetic distribution described above, as well as for a synthetic distribution with small $\langle N_\bk \rangle$. We observe the exponential scaling predicted in Eq.~\ref{eqn:synth-cdf-comparison} for both. Hence, one expects $63\%$ of the $N_s$ per-system minima $\gammastar$ to be present in the set of $N_s$ global minima $\gamma_<$.

In Fig.~\ref{fig:weakest-two-distriplot}(d--f), we plot each value of $\gammaijdagger$ for the first $100$ systems, taken from the $N=1024$, $P=10^{-2}$ ensemble. The relation between the \pdfs
of the per system $\rho(\gammabklr)$ (red curves in panel e) and distribution minima  $\rho(\gamma_<)$ (blue curves in panel e) are similar to those of the synthetic data, and
$\gammadist=1.4\times 10^{-4}$ and $\gammabklr=1.5\times 10^{-4}$ are quite similar. Consistent with this, a plot of
 $\Pr(\gammastar \geq \gamma)$ as a function of $\langle N_\bk \rangle \Pr(\gammaij < \gamma)$
is approximately exponential, although slight deviations can be seen in the tails of these distributions (Fig.~\ref{fig:weakest-two-data}).

In Fig.~\ref{fig:weakest-two-distriplot}(g--i), we plot each value of $\gamma_{ij}$ for the first $100$ systems, taken from the $N=16$, $P=10^{-6}$ ensemble. The differences between the \pdfs
of the per system $\rho(\gammabk)$ (red curves in panel h) and distribution minima  $\rho_< (\gammadist)$ (blue curves in panel h) are more significant, and
$\gammabk = 1.6\times 10^{-6} $ and $ \gammadist = 1.1\times 10^{-6}$ are quite distinct. Consistent with this, a plot of
 $\Pr(\gammastar \geq \gamma)$ as a function of $\langle N_\bk \rangle \Pr(\gammaij < \gamma)$
deviates significantly from an exponential
(Fig.~ \ref{fig:weakest-two-data}).
This deviation points to a lack of self-averaging in small systems.

\paragraph{Interpretation}
We now discuss two possible scenarios to explain the deviations for small $\nplog$. First, each finite packing could have a different distribution of $\gammaij$, but between packings these distributions are related by an overall scale factor. The data shown in Fig.~\ref{fig:N16-gammas-scatter} suggests that this is possible. To understand the effect of such `overall scale factor' for the statistics, we draw an overall system scale from a uniform distribution $\mathcal{U}(0,1)$ for each of the synthetic systems, and multiply the strains for each system with this scale factor.
The resulting behavior is shown in Fig.~\ref{fig:weakest-two-simu}, where we see the decay is much slower than for uncorrelated systems. The reason for this is that packings with a low minimum will typically come from a system which contains other low strains. This saturates the low strain region of the overall distribution with strains that are not system minima.
The data extracted from our direct simulations (Fig.~\ref{fig:weakest-two-data}) show a similar decay, slower than exponential, with slower decays for lower pressures.
To directly check whether a per-system scale \emph{can} explain the behavior, we divide all strains by the mean strain for each system, and show the results in Fig.~\ref{fig:weakest-two-rescale}. In the case of a simple scale incorporated in synthetic data, this brings the behavior closer to the simple exponential (dashed purple line). The behavior is still not purely exponential due to the subtle effects we induce with this normalization step. Nevertheless, we note that the rescaling has very little effect on the contact change strains shown in Fig.~\ref{fig:weakest-two-data}. We therefore conclude the correlations cannot be simply explained by an overall system scale.

Second, inspired by \citet{lerner2012towardmicroscopicdescription}, we now investigate whether we can recover the behavior of $\gammabklr$ using extremal value statistics by assuming that most contacts are drawn from a distribution with mean $k$, but a limited number of 'weak' contacts are drawn from a distribution with mean $k' \ll k$.
In the case of one extraordinarily weak contact in each packing, we expect most of the $k$ system minima to show up in the lowest $k$ values of the entire set of strains. We have simulated this by dividing one strain in each of the synthetic packings by $10^{3}$. As we see in Fig.~\ref{fig:weakest-two-simu}, $\Pr(\gammastar \geq \gamma)$ decreases much more rapidly than exponential, and drops to $\Pr(\gammastar \geq \gamma)=0$ around $\langle N_\bk \rangle \Pr(\gammaij < \gamma) \approx 3$ --- in other words, the $k$ minima are all found in the lowest $3k$ values of the full set. The exact point of intersection depends on how weak the contact is, and on how many weak samples are in the packing. However, our data for actual packings shows a slower than exponential decay, thus discounting
the 'weak contact' hypothesis as source for the correlations in our systems.

Hence, in conclusion: for sufficiently large systems, packings are self averaging, and
extremal value statistics may be sufficient to determine the mean value and distribution for the first contact break strains.

\section{Conclusion}

We have presented a systematic analysis of the first contact changes in soft spheres sheared quasistatically under controlled strain. There are several important conclusions. To begin, contact changes are strongly sensitive to both the system size and the distance to jamming, and finite size corrections play an important role. We find distinctly different scaling relations in the limit $N^2 P \ll 1$, which is relevant as the system size shrinks or the confinement pressure drops, and in the limit $N^2 P \gg 1$, which describes thermodynamically large ensembles of soft particles. The characteristic strains describing made and broken contacts can be rationalized via simple mean field-like scaling arguments and Poisson statistics, while log corrections or weak corrections to scaling can improve their accuracy.

Contact changes are also reflected in the mechanical response, including the shear modulus. We have shown that the ensemble-averaged differential shear modulus is unaltered by contact changes in thermodynamically large systems. This finding rationalizes the ability of effective linear response (i.e.~Hooke's law) to describe bulk mechanical properties even at finite values of the strain \cite{boschaninprep,dagoisbohy2014oscillatoryrheologynear}. However extreme value analysis suggests that the thermodynamic limit is reached more slowly than one would infer from the system size dependence of contact change strains and mechanical properties.

Finally, we have demonstrated surprising correlations in the spatiotemporal patterning of successive contact changes. These suggest the need for further study of particle scale dynamics on finite strain and time scales. Open questions include the interplay between jamming physics and microscopic irreversibility, as well as the role of viscous interactions. Both can be addressed, e.g., with simulations of oscillatory rheology \cite{dagoisbohy2014oscillatoryrheologynear,otsuki2014avalanchecontributionto}.

\appendix
\section{Creating and shearing a packing}
\label{apx:create-and-shear}
\paragraph{Boundary conditions}
We use periodic boundaries in a non-square box, where
each particle has periodic copies at
$\vec{r} = \vec{r_i} + n_x\cdot\vec{L_x} + n_y\cdot\vec{L_y}$,
where $\vec{r_i}$ is the canonical position of the particle,
$n_x$ and $n_y$ are integers and $\vec{L_x}=(L_{xx}, L_{xy})$ and $\vec{L_y}=(L_{yx}, L_{yy})$ describe the box.
The area of the unit cell is
$L^2 = L_{xx}\cdot L_{yy}$, the
Lees-Edwards shear strain is
$\alpha = L_{yx} / L$ and the
 pure shear strain in
$\delta = \frac{L_{yy} - L}{L} = \sqrt{L_{yy}/L_{xx}} - 1~$. For square cells
$\vec{L_x} = (L_{xx}, 0)$ and $\vec{L_y} = (0, L_{yy})$, and consequently $\alpha = \delta = 0$ \cite{ohern2002randompackingsfrictionless,ohern2003jammingatzero,hecke2010jammingsoftparticles}.
In contrast, here we require that the energy is at a minimum with respect to $\alpha$ and $\delta$ for the initial condition, which guarantees that we obtain $\epsilon_\textrm{all}^+$ packings where
the shear modulus is positive and the residual shear stresses are zero \cite{dagoisbohy2012softspherepackings,goodrich2013jamminginfinite}, as one expects for a physical system at rest. We keep $L_{xy} = 0$ as allowed by rotational symmetry.

\paragraph{Interactions, energy and stress}
\label{sec:interactions-stress-energy-in-static}
Our system consists of a bi-disperse mix of soft disks with repulsive harmonic interactions, using $N/2$ small particles with $R_s = 1$ and $N/2$ large particles with radius $R_l = 1.4$.
The interaction between particles is determined by their overlap $\delta_{ij} = \max(0,|\vec{r_{ij}}| - R_i - R_j)$, where $\vec{r_{ij}}$ is the center-to-center distance of the two particles: $\vec{r_{ij}} = \vec{r_i} - \vec{r_j} - n_{x,ij} \vec{L_x} - n_{y,ij} \vec{L_y}$, where $\vec{r_i}$ and $\vec{r_j}$ are the canonical particle positions, and $n_{x,ij}$ ($n_{y,ij}$) is $0$ if the closest copy of $j$ to $i$ is the canonical copy, $+1$ if it is across the right (top) boundary and $-1$ if it is across the left (bottom) boundary. Contact forces have magnitude $f_{ij} = k\delta_{ij}$, where $k=1$ is the spring constant, and result in the harmonic potential $ U_{ij} = (k/2) \delta_{ij}^2.$
The internal energy  is given by the sum of all inter-particle potentials,
$U = \sum_{i,j} U_{ij} = \sum_{i,j} (k/2)\delta_{ij}^2$.
Length scales, stresses and energies are expressed in units $R_s$,  $k$ and $k R_s^2$ respectively.
The boundary stresses are
the simple shear stress $\sigma_{yx} = \sigma_{xy},$ the deviatoric (pure shear) stress $\tau = \frac{1}{2}\left(\sigma_{xx} - \sigma_{yy}\right),$ and the volumetric stress $P_\textrm{int} = \frac{1}{2}\left(\sigma_{xx} + \sigma_{yy}\right)$, which are computed using the Born-Huang approximation \cite{simon2012rearrangementsinjammed,born1998dynamicaltheorycrystal}
$\sigma_{ab} = (1/2L^2) \sum_{i,j} \left[(\vec{r_{ij}} \cdot \hat a) (\vec{f_{ij}} \cdot \hat b) \right]
$ where $a,b\in\{x,y\}$ and the sum is over all particle pairs $i,j$.

\paragraph{Preparing a packing}
To create $\epsilon_\textrm{all}^+$ packings at given pressure $P$ between $10^{-7}$ and $10^{-2}$, we minimize the enthalpy $H = U + PL^2$.
We place our bidisperse $N$  particles within a square box with size
$L^2_\textrm{init} = \phi_\textrm{init} \left(\frac{N}{2} \pi R_s^2 + \frac{N}{2} \pi R_l^2\right),$
where $\phi_\textrm{init} \equiv 0.8$ is chosen to be far below the jamming density $\phi_J \approx 0.84$.
We use a combination of the Conjugate Gradient method \cite{shewchuk1994introductiontoconjugate} and the Fast Inertial Relaxation Engine (\fire) \cite{bitzek2006structuralrelaxationmade} algorithms. The latter is much faster, but is unstable when the overlaps between particles are large. We therefore initially relax the packing using standard Conjugate Gradient methods to resolve the largest overlaps with fixed boundaries, and minimize the energy until $|\Delta E| \leq  10^{-2}\cdot E$.
We then use the \textsc{fire} algorithm, allowing the boundaries (i.e., $L_{xx},$ $L_{yy},$ and $L_{yx}$) to deform, and relax the system until $
|\Delta H| \leq  10^{-17}\cdot H,\textrm{ and } |\sigma_{yx}| \leq  10^{-15}$.

As we will study changes in individual contacts, and in particular  probe the strain at which the first contact change takes place, we anticipate the need to study finite size effects. Moreover, we anticipate that many quantities will rescale with $N^2 P$ as has recently been found in  \cite{goodrich2012finitesizescaling,goodrich2013jamminginfinite,deen2014contactchangesnear}. We therefore
prepared ensembles of sheared systems at a range of $N$ and $P$. Most ensembles contain $100$ systems, with some ensembles containing up to $5000$ systems.
To characterize the behavior at the first contact change, we created a set of ensembles having $N$ and $P$ on a log-spaced grid, with $N=16,32,\ldots,1024$ and $P=10^{-7}, 10^{-6 \frac{5}{6}}, \ldots 10^{-2}$, and a set at intermediate $N=22,45,\ldots 724$ for $P=10^{-2}$ and $10^{-7}$. These are sheared until we find at least one contact change. % up to three contact changes. % 50..5000 depending on ensemble...
To characterize the effects of multiple contact changes (Sec.~\ref{sec:whoop-whoop-whoop}), we sheared the ensembles at $N=16$, $P=10^{-6}$, $N=1024$, $P=10^{-6}$ and $N=1024$, $P=10^{-2}$ up to $25$ contact changes.

\paragraph{Simple shear, contact changes and rattlers}
\label{sec:simple-shear-cc-method}
We perform quasistatic shear, so viscous damping is irrelevant and
only the elastic interactions between particles are taken into account. We apply shear
by distorting the unit cell as\begin{align}
\vec{L_x}(\gamma) =& \vec{L_{x}}(0),\\
\vec{L_y}(\gamma) =& \vec{L_{y}}(0) + \gamma L \cdot \hat x,
\end{align}
i.e., we change $\alpha \rightarrow \alpha + \gamma$, while keeping  $L^2$ and $\delta$ constant. We then use the \textsc{fire} algorithm to
relax the system (keeping the boundaries fixed) until
$
|\Delta H| < 10^{-13}\cdot H,
$
where we sacrifice a small error in the particle positions for simulation speed.  We found that
this is accurate enough for the detection of contact changes and to determine the stress and energy at the contact change: The details of the relaxation do not influence the detection of contact changes, and the relative error in $\sigma_{xy}$ is typically less than $10^{-6}$. Note that in this strained state, we are now no longer in an enthalpy minimum with respect to the \emph{boundary conditions}, so $\sigma_{xy} \neq 0$ and $G$ can become negative.

\section{Calculating the linear response}
\label{sec:calculate-linear-response}

In this appendix we will briefly review how, based on the initial particle positions, box size and box shape, we determine the linear response of the system. Given an applied deformation of the box, we can determine the resulting particle motion, forces and energy cost \cite{dagoisbohy2012softspherepackings,goodrich2013jamminginfinite,ellenbroek2006criticalscalingin,ohern2003jammingatzero}.

The state of the system can be described as a vector
\begin{align}
\ket{q} &= \ket{q_x, q_b} \nonumber\\
&= \ket{\{x_1\ldots x_N, y_1\ldots y_N\}, \{L_{xx}, L_{xy}, L_{yx}, L_{yy}\}}
\end{align}
where $(x_n, y_n)$ is the position of particle $n$ and the four parameters $L_{ij}$ describe the box size and shape. We only include particles that are part of the load bearing network (non-rattlers).

We then prescribe a displacement $\ket{\Delta q}$. We determine the energy in the new state $\ket{q + \Delta q}$ by expanding $U$ up to second order:
\begin{equation}
U(\ket{q + \Delta q}) = U(\ket{q}) + \braket{J_q | \Delta q} + \frac{1}{2}\braket{\Delta q | \matrixsym{H_q} | \Delta q} + O(\Delta q^3)
\end{equation}
where
\begin{equation}
\bra{J_q} = \bra{\frac{\partial U}{\partial x_1}, \cdots, \frac{\partial U}{\partial L_{yy}}}
\end{equation}
is the Jacobian and
\begin{equation}
\matrixsym{H_q}= \left(\begin{matrix}\frac{\partial^2 U}{\partial x_1 \partial x_1}  & \cdots & \frac{\partial^2 U}{\partial x_1 \partial L_{yy}} \\
\vdots & \ddots
\end{matrix} \right)
\end{equation}
the extended Hessian at $\bra{q}$ \cite{tighe2011relaxationsandrheology}. Because the initial state is at an energy minimum, the Jacobian term is zero, and the leading contribution to the energy comes from the extended Hessian.

For a given displacement, the energy cost is thus given by
\begin{equation}\Delta U = \frac{1}{2}\braket{\Delta q | \matrixsym{H_q} | \Delta q},\end{equation}
and the resulting forces on particles and boundaries by
\begin{equation}
\ket{f} = \matrixsym{H_q} \ket{\Delta q}\label{eqn:ext-hessian-force-response}.
\end{equation}

However, typically, we do not know the displacement of each particle. Instead, we wish to calculate the displacement of the particles given a change in the boundaries, i.e., find a state where, given the new boundaries, the sum of forces on each particle is zero. To find this state, we split the extended Hessian into four parts:
\begin{equation}
\matrixsym{H} = \left(\begin{matrix}
    \matrixsym{H_{xx}} & \matrixsym{H_{bx}^T} \\
    \matrixsym{H_{bx}} & \matrixsym{H_{bb}}
\end{matrix}\right)\end{equation}
where the ordinary Hessian $\matrixsym{H_{xx}}$ describes the particle-particle interactions, $\matrixsym{H_{bx}}$ the interactions between boundaries and particles, and $\matrixsym{H_{bb}}$ those between different boundaries. We can then rewrite Eq.~\ref{eqn:ext-hessian-force-response} as follows:
\begin{equation}
\left(\begin{matrix}
    \ket{\Delta f_x} \\
    \ket{\Delta f_b}
\end{matrix}\right)
=
\left(\begin{matrix}
    \matrixsym{H_{xx}} & \matrixsym{H_{bx}^T} \\
    \matrixsym{H_{bx}} & \matrixsym{H_{bb}}
\end{matrix}\right)
\left(\begin{matrix}
    \ket{\Delta q_x} \\
    \ket{\Delta q_b}
\end{matrix}\right).\label{eqn:split-hessian-force-response}
\end{equation}
where $\ket{\Delta q_x}$ and $\ket{\Delta q_b}$ are the displacements of particles and boundaries, and $\ket{\Delta f_x}$ and $\ket{\Delta f_b}$ the corresponding forces. Setting the forces on the particles to zero, we find
\begin{equation}\ket{\Delta f_x} = \matrixsym{H_{xx}}\ket{\Delta q_x} + \matrixsym{H_{bx}^T}\ket{\Delta q_b} = 0.\end{equation}
Solving for $\ket{\Delta q_x}$ gives us the particle displacement as a function of the deformation of the simulation box
\begin{equation}\ket{\Delta q_x} = - \matrixsym{H_{xx}^{-1}}\matrixsym{H_{bx}^T}\ket{\Delta q_b}.\label{eqn:displacement-given-boundary-deformation-inv}\end{equation}
Unfortunately, $\matrixsym{H_{xx}^{-1}}$ cannot be calculated due to the two zero-energy translational modes. Instead, we choose to use the Moore-Penrose pseudoinverse $\matrixsym{H_{xx}^{+}}$, which fixes the zero-energy translational modes in place \cite[\S 6.4]{strang2009introductiontolinear}:
\begin{equation}\ket{\Delta q_x} = - \matrixsym{H_{xx}^{+}}\matrixsym{H_{bx}^T}\ket{\Delta q_b}.\label{eqn:displacement-given-boundary-deformation}\end{equation}

To calculate the energy cost and the stress on the boundary, we use the full displacement vector \begin{equation}\ket{\Delta q} = \left(\begin{matrix}
    - \matrixsym{H_{xx}^{+}}\matrixsym{H_{bx}^T}\ket{\Delta q_b} \\
    \ket{\Delta q_b}
\end{matrix}\right)\end{equation} and, again using Eq.~\ref{eqn:split-hessian-force-response}, find
\begin{align}\ket{\Delta f_b} &= \matrixsym{H_{bx}}\ket{\Delta q_x} + \matrixsym{H_{bb}}\ket{\Delta q_b} \\
  &= (\matrixsym{H_{bb}} - \matrixsym{H_{bx}}\matrixsym{H_{xx}^{+}}\matrixsym{H_{bx}^T})\ket{\Delta q_b}.\end{align}
The corresponding stress can be calculated as
\begin{equation}
\ket{\Delta \sigma_b} = \ket{\frac{\Delta f_{xx}}{L_{xx}}, \frac{\Delta f_{xy}}{L_{xx}}, \frac{\Delta f_{yx}}{L_{yy}}, \frac{\Delta f_{yy}}{L_{yy}}},
\end{equation}
but in practice, it is more convenient to calculate the stress by using the Born-Huang approximation \cite{simon2012rearrangementsinjammed,born1998dynamicaltheorycrystal}, on the new particle positions $\ket{q_x'} = \ket{q_x} + \ket{\Delta q_x}$. The stress also allows us to determine the elastic modulus corresponding to a given boundary deformation
\begin{equation}c_q = \braket{\Delta\sigma_b | \Delta q_b} / \braket{\Delta q_b | \Delta q_b}.\label{eqn:elastic-modulus}\end{equation}
For the resulting energy change we use $\ket{\Delta f_x} \equiv 0$ to find
\begin{align}\Delta U &= \frac{1}{2}\braket{\Delta q_b | \Delta f_b} \\
 &= \frac{1}{2}\bra{\Delta q_b}(\matrixsym{H_{bb}} - \matrixsym{H_{bx}}\matrixsym{H_{xx}^{+}}\matrixsym{H_{bx}^T})\ket{\Delta q_b}.\label{eqn:energy-change-for-strain-LR}\end{align}

We now have all ingredients in place to calculate, for a given boundary deformation, the particle displacements, stress response and energy change from linear response.

\section{Finite size scaling of $\rho(\upar)$ and $\rho(\uperp)$}
\begin{figure}
\includegraphics[width=\columnwidth]{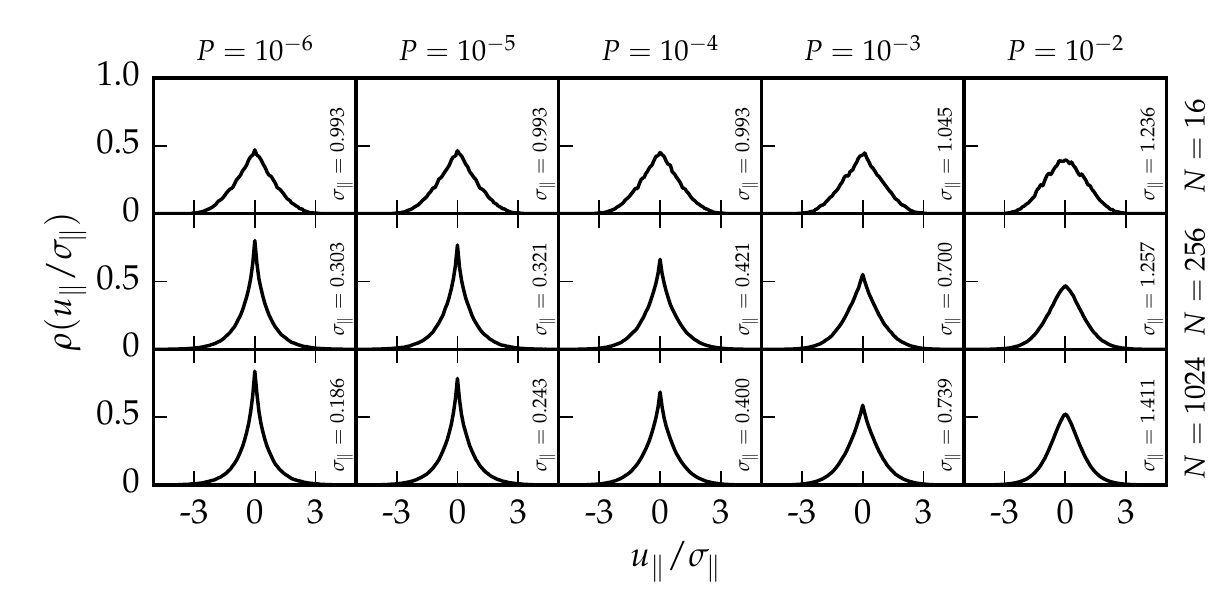}
\includegraphics[width=\columnwidth]{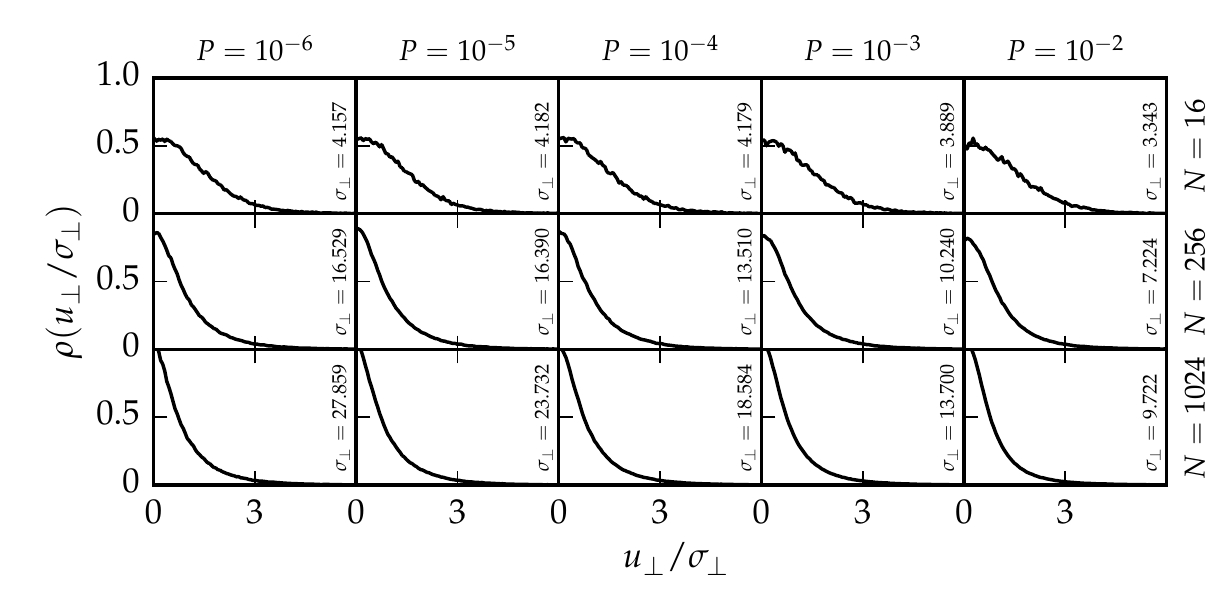}
\caption{\label{fig:upar-uperp-multi}(top) Distributions of $\upar$, rescaled by their standard deviation $\sigma_\parallel$, for ensembles with $N=16,$ $256,$ or $1024$ particles at $P=10^{-6}\ldots 10^{-1}$. $\sigma_\parallel$ is indicated in each figure. The distributions develop a sharp kink around $0$ for low pressures, and become smooth for $P \gtrapprox 10^{-2}$. There is a weak dependence on $N$, with the distribution becoming more peaked for high $N$.
(bottom) Same, for $\uperp$. Here, the distributions depend less on $N$ and $P$, although also here the distribution gains weight near $0$ for decreasing $P$.}
\end{figure}
\begin{figure}
\subfloat{\label{fig:upar-scaling}}
\includegraphics[width=\halfcolwidth]{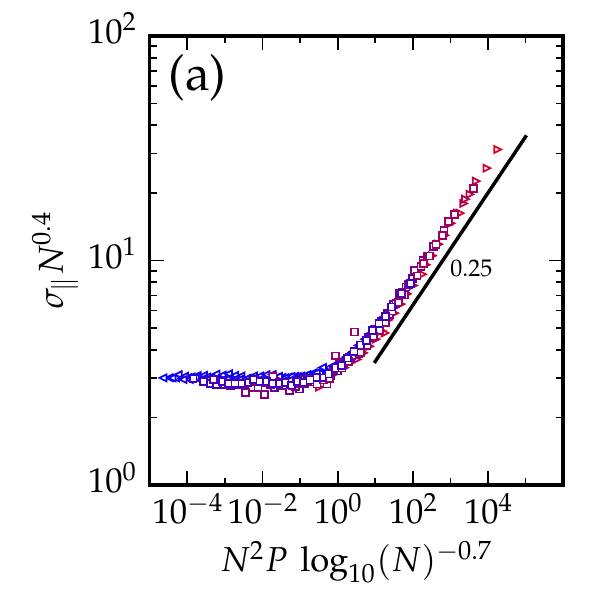}
\subfloat{\label{fig:uperp-scaling}}
\includegraphics[width=\halfcolwidth]{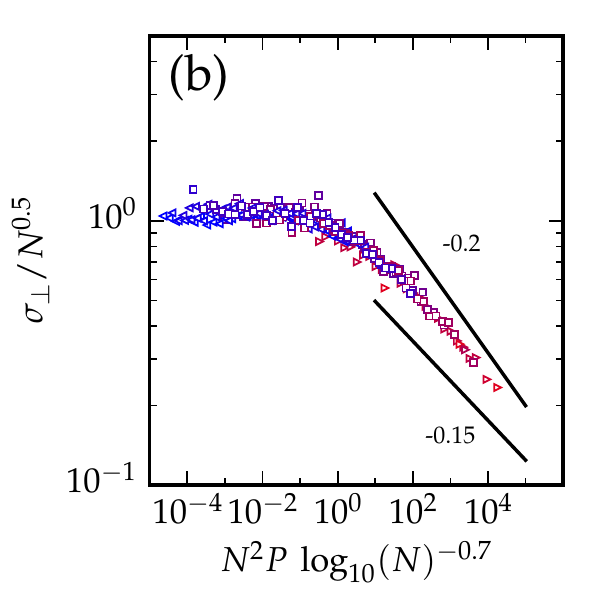}
\caption{(color online)
(a) Scaling of the standard deviation $\sigma_\parallel$ as a function of $N$ and $P$. At low $N^2P$, $\sigma_\parallel$ is independent of pressure and at high $N^2P$ we recover a scaling  to $\sigma_\parallel \sim \delta^{0.25}$, consistent with \cite{ellenbroek2009jammedfrictionlessdisks}.
(b) Same, for $\sigma_\perp$. At low $N^2P$, $\sigma_\parallel$ is independent of pressure. At high pressure we find a scaling $\sigma_\parallel \sim \delta^{-0.2\ldots-0.15}$, somewhat slower than the $\delta^{-0.25}$ found in  \cite{ellenbroek2009jammedfrictionlessdisks}.
}
\end{figure}
In this appendix, we will discuss the the distributions of $\upar$ and $\uperp$, which provide a continuum description of the interparticle motion. For each particle pair $i,j$, we split the interparticle velocity $\vec{u_{\particlepair}} = \partial \vec{x_\particlepair} / \partial\gamma$  in components parallel and perpendicular to the contact:
\begin{align}
{\upar}_{,\particlepair} = & \vec{u_{\particlepair}} \cdot \hat r_{\particlepair}~, \\
{\uperp}_{,\particlepair} = & \sqrt{u_{\particlepair}^2 - {\upar}_{,\particlepair}^2}~.
\end{align}
Using every contact in every packing in an ensemble, we then build the frequentist distributions $\rho(\upar)$ and $\rho(\uperp)$.

In the following, we will discuss the relationship between the shape and scale of these distributions and $N$ and $P$. Earlier work \cite{ellenbroek2009jammedfrictionlessdisks} has focused on Hertzian systems at intermediate to high pressure ($P^{2/3} \sim \langle\delta\rangle \geq  3 \cdot 10^{-4}$). They find the shape of the distribution does not depend on $P$, and find a simple single scaling of the overall scale with $P$.
We extend this with harmonic systems much closer to jamming ($P \sim \langle\delta\rangle \geq 10^{-7}$). At high pressures, we recover the same behavior, but close to jamming, we find \emph{(i)} the shape of the distributions depends on the pressure $P$, and \emph{(ii)} the widths of the distributions scale with $N^2P$, with two distinct scaling regimes.

\paragraph{Shape of distributions}
In Fig.~\ref{fig:upar-uperp-multi}, we plot the probability density functions of $\upar$ and $\uperp$, rescaled by their standard deviations $\sigma_\parallel$ and $\sigma_\perp$, for ensembles with different system sized and pressures. We note that, even though the different distributions cannot be collapsed with a single scale parameter, the majority of the behavior is captured in the standard deviation $\sigma$. For both distributions, we observe the distributions become increasingly peaked near $0$, and, although neither \pdf diverges, this peak appears to develop a sharp kink for small pressures. We observe the shape changes with $P$, and, for large enough $N$, is largely independent of $N$ --- $N^2P$ is not the relevant scaling parameter here. Surprisingly, this means the overabundant low values are still present for large systems at $P \approx 10^{-3}$, which would normally not be considered `close to jamming'.

\paragraph{Scaling of standard deviations}
\citet{ellenbroek2009jammedfrictionlessdisks} find the width of the distributions scale as
\begin{align}
\sigma_\parallel \sim \langle\delta\rangle^{1/4}~, \label{eqn:ellenbrok-upar} \\
\sigma_\perp \sim \langle\delta\rangle^{-1/4}~,\label{eqn:ellenbrok-uperp}
\end{align}
where $\langle\delta\rangle$ is the mean overlap between pairs of particles in contact in the ensemble.
If we assume \emph{(i)} the standard deviations will scale with $N^2P$ and \emph{(ii)} the distributions are independent of $N$ for large $N$, Eq.~\ref{eqn:ellenbrok-upar} and Eq.~\ref{eqn:ellenbrok-uperp} suggest plotting
\begin{align}
N^{0.5} \sigma_\parallel &\sim F(N^2P)~,\\
N^{-0.5} \sigma_\perp &\sim F(N^2P)~, \label{eqn:uperp-scaling-collapse}
\end{align}
should collapse our data. We note that, because the shape of the distribution varies, the choice of the scaling parameter (e.g. a percentile rather than the standard deviation) can have a rather large effect on the collapse (which can reach $\pm 0.2$ in the scaling exponent), and we therefore do not expect a perfect match.

In Fig.~\ref{fig:upar-scaling}, we find the best scaling collapse for $\sigma_\parallel$ is close but not equal to the expected scaling: we find $\sigma_\parallel \sim N^{-0.4}$ at low $N^2P$ rather than $\sigma_\parallel \sim N^{-0.5}$. Nonetheless, we suggest that the scaling is close enough to be consistent with the proposed scaling.
At low pressures, we find that $\sigma_\parallel$ only depends on $N$, and no longer depends on $P$. For $N^2P \gg 1$, we find the expected $\sigma_\parallel \sim P^{0.25}$ power law.

For $\sigma_\perp$, we find Eq.~\ref{eqn:uperp-scaling-collapse} provides a rather good collapse (Fig.~\ref{fig:uperp-scaling}). At low $N^2P$, we find $\sigma_\perp$ becomes independent of $P$, and at high $N^2P$, we find behavior similar, but different from the expected $\sigma_\perp \sim P^{-0.25}$ power law.

Surprisingly, we find both $\sigma_\perp$ and $\sigma_\parallel$ reach a pressure-independent plateau for low $N^2P$. This has important implications for the behavior close to jamming --- in contrast to what is generally assumed, $\sigma_\perp / \sigma_\parallel$ does \emph{not} diverge for low pressures, but reaches a plateau whose value diverges as $\sigma_\perp / \sigma_\parallel \sim N^{0.9}$ in the thermodynamic limit.

\section{Discussion}\label{sec:shear-alternate-scaling-models}
Finally we will discuss our findings in the light of alternative scaling models that have surfaced in the literature. Nonlinearities in jammed packings at finite temperature
were studied in
\citet{schreck2011repulsivecontactinteractions}, and these authors  find a different scaling that we attribute to their averaging over modes. Moreover, \citet{combe2000strainversusstress} and \citet{lerner2013lowenergynon} have approached the problem from a hard particle perspective, and find a scaling law very close to the behavior we find close to jamming.

\subsection{Excited eigenmodes}
\citet{schreck2011repulsivecontactinteractions} investigated contact breaking in jammed sphere packings using excited eigenmodes. They displace particles along an eigenmode:
\begin{equation}
\vec{r} =  \vec{r_0} + \sqrt{N}\delta\hat e_k,
\end{equation}
where $\vec{r_0}$ is the original state, $\vec{r}$ the excited state, $N$ the system size, $\hat e_k$ the eigenvector for eigenmode $k$, and $\delta$ the excitation amplitude. The system is then allowed to evolve at fixed energy. For small excitations $\delta$, the system oscillates around a base state, and most energy is contained in the initial eigenmode. However, for excitations larger than a critical excitation amplitude $\delta_c(k)$ there is a sharp increase in how much energy spreads into the other eigenmodes of the system.

\citeauthor{schreck2011repulsivecontactinteractions} find that $\delta_c$ is directly related to the first contact change in the system. Surprisingly, they find that contacts only \emph{break}, even for large systems ($N=1920$) at high densities ($\Delta\phi = 10^{-2}$).

For each system, $\delta_c(k)$ is calculated for every eigenmode $k$. The authors then measure the average energy
\begin{equation}
E = \langle \left(\omega_k \delta_c(k) \right)^2 \rangle_k,
\label{eqn:schreck2011-energy-calculation}
\end{equation}
where $\omega_k$ is the eigenfrequency of eigenmode $k$ and the mean is taken over all eigenmodes.

For the scaling of the energy per particle $E/N$ with the density $\Delta\phi$ and system size $N$, \citeauthor{schreck2011repulsivecontactinteractions} find a relationship
\begin{equation}
\frac{E/N}{A(\Delta\phi)\cdot(\Delta\phi)^2} \sim N^{-\beta},
\end{equation}
``where $A(\Delta\phi)$ is only weakly dependent on $\Delta\phi$ and $\beta\approx 1.7$'' \cite{schreck2011repulsivecontactinteractions}. Close to jamming ($N\Delta z = 0 \ldots 2$), they find $A(\Delta\phi)$ is constant and $\beta=1\ldots 2$ \cite{ohern20150610privcomm}. Writing this in terms of $E$, taking $A(\Delta\phi)$ as constant and using $\Delta\phi \sim P$:
\begin{equation}
E \sim N^{1-\beta} (\Delta\phi)^{2} \sim N^{1-\beta} P^{2}
\end{equation}

To compare this with our results, we note that
\begin{align}
E & \sim \sigma\gamma L^2 \sim \sigma\gamma N \sim GN \gamma^2,
\end{align}
so
\begin{equation}
\gamma \sim \sqrt{E/GN} \sim N^{-\beta/2}PG^{-1/2}.
\end{equation}
Using the known finite-size scaling of $G$ \cite{dagoisbohy2012softspherepackings}, we then find
\begin{equation}
\gamma \sim \begin{cases}
P N^{(1-\beta)/2} & (N^2P \ll 1) \\
P^{0.75} N^{-\beta/2} & (N^2P \gg 1) \end{cases}\label{eqn:corey-scaling}
\end{equation}

\begin{figure}
        \includegraphics[width=\halfcolwidth]{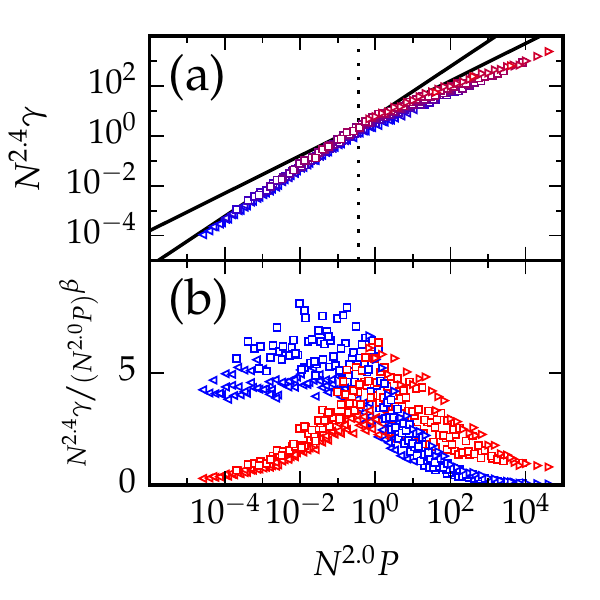}
        \includegraphics[width=\halfcolwidth]{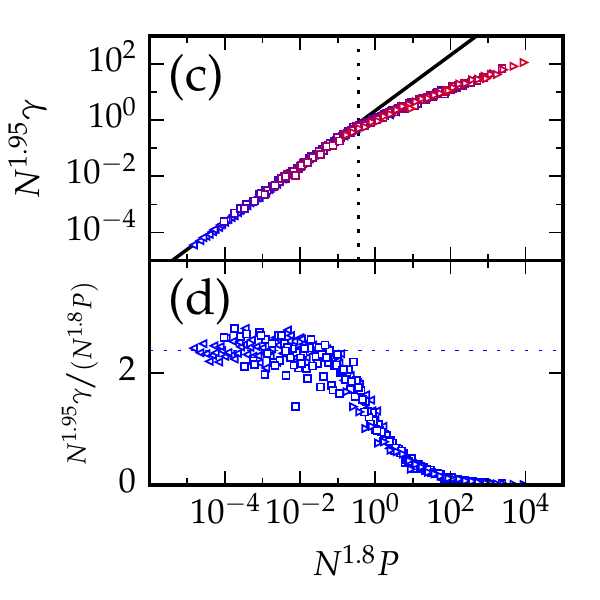}
        \subfloat{\label{fig:corey-scaling}}
        \subfloat{\label{fig:corey-residue}}
        \subfloat{\label{fig:wyart-scaling}}
        \subfloat{\label{fig:wyart-residue}}
\caption{(color online) (a) Data rescaled as in \citet{schreck2011repulsivecontactinteractions} (Eq.~\ref{eqn:corey-scaling}). Black lines indicate power laws with exponent $1$ and $0.75$.
(b) The residuals $F(x)/x^{1.0}$ (dark/blue) and $F(x)/x^{0.75}$ (light/red) do not have a plateau, indicating these power laws do not well describe the data.
(c,d) Same, but with data rescaled as in \citet{wyart201404privcomm} (Eq.~\ref{eqn:wyart-scaling}), i.e., with $q-r = 0.15$. We have chosen  $r=1.8$, as in our best collapse.}
\end{figure}

To test whether this matches the data, we plot $N^{(\beta+3)/2}\gamma$ as a function of $N^2P$ in Fig.~\ref{fig:corey-scaling}, using the published value $\beta=1.7$. We find, firstly, that the collapse is not very good. Secondly, we find the $0.75$ power law for the upper branch overestimates the actual strains. To a lesser extent, the lower branch also deviates from Eq.~\ref{eqn:corey-scaling}. This is also reflected in the residuals in Fig.~\ref{fig:corey-residue} -- neither branch collapses onto a constant value.

We expect these differences arise due to the averaging in Eq.~\ref{eqn:schreck2011-energy-calculation}, which means the energy is effectively an average over $2N$ modes within the same system. In Sec.~\ref{sec:extremal-value-scaling}, we will see that averaging over all contacts loses many of the features we found for the first contact change.

\subsection{Hard particle systems}
The question of contact breaking and plasticity has also been studied in systems of hard particles. These systems are isostatic \cite{moukarzel1998isostaticphasetransition}, which means a contact change will cause the system to unjam, and thus contact changes are directly connected to plastic events. Isostaticity also implies that the force distribution is unique, and can be derived directly from the particle positions \cite{roux2000geometricoriginmechanical}. On the other hand, because the systems are isostatic, the results can only describe the $N^2P \ll 1$ limit of soft particle systems.

\citet{combe2000strainversusstress} investigated the prevalence of and distance between strain jumps in a system under uniaxial stress-controlled compression. They found that the spacing between events is described by a exponential distribution in $\delta q (N/1024)^{1.16}$, where $\delta q$ is the relative uniaxial stress increment $\Delta\sigma/P$. This is consistent with modeling contact changes as a Poisson process.

To calculate the scaling of $\gammabk$ with $N$ and $P$ in this system, we first note that the mean stress required to break the first contact scales as
\begin{equation}
\langle\Delta\sigma\rangle \sim P\langle\Delta q\rangle \sim P/N^{1.16}.
\end{equation}
We can then calculate the $\gammabk$ using the uniaxial compression modulus $E$. Using that $K\sim 1$ and $G\sim 1/N$ near jamming, $E$ is given by \cite{thorpe1985elasticmodulitwo}
\begin{equation}
E = \frac{4}{1/K + 1/G} \sim \frac{1}{N}
\end{equation}
and the expected mean strain to break the first contact is thus given by
\begin{equation}
\gammabk \sim \langle\Delta\sigma\rangle / E \sim P/N^{0.16},\label{eqn:wyart-scaling}
\end{equation}
which is very close to the $P/N^{0.20}$ scaling we found by fitting our data to a pure power law (Eq.~\ref{eqn:scaling-fit-two-regimes}).

A theoretical argument for this power law, based on the concept of "weak" contacts that connect to local motion, and "strong" contacts that are connected to global motion, was introduced in \cite{lerner2013lowenergynon}. \citet{wyart201404privcomm} uses this to predict that the strain for the first contact change should scale as
\begin{equation}
\gamma \sim P/N^{0.15},
\end{equation}
which is close to the value found in \cite{combe2000strainversusstress}.

In Fig.~\ref{fig:wyart-scaling}, we show this scaling also provides a good match to our data -- the $0.15$ exponent can be seen as a power law correction to our initial $\gamma \sim P$ scaling near jamming, and is essentially indistinguishable from either log or $0.2$ power law corrections.

\section{Open data}
The software and data files used for this article have been published on-line, under the following identifiers:
\begin{itemize}
	\item \emph{JamBashBulk: two-dimensional packing simulation}. \doi{10.5281/zenodo.60972},
	\item \emph{Contact changes in shear-stabilized jammed packings}. \doi{10.5281/zenodo.59217}.
\end{itemize}

\bibliographystyle{apsrev4-1}
\bibliography{papers,{\jobname}Notes}

\begin{thebibliography}{61}%
\makeatletter
\providecommand \@ifxundefined [1]{%
 \@ifx{#1\undefined}
}%
\providecommand \@ifnum [1]{%
 \ifnum #1\expandafter \@firstoftwo
 \else \expandafter \@secondoftwo
 \fi
}%
\providecommand \@ifx [1]{%
 \ifx #1\expandafter \@firstoftwo
 \else \expandafter \@secondoftwo
 \fi
}%
\providecommand \natexlab [1]{#1}%
\providecommand \enquote  [1]{``#1''}%
\providecommand \bibnamefont  [1]{#1}%
\providecommand \bibfnamefont [1]{#1}%
\providecommand \citenamefont [1]{#1}%
\providecommand \href@noop [0]{\@secondoftwo}%
\providecommand \href [0]{\begingroup \@sanitize@url \@href}%
\providecommand \@href[1]{\@@startlink{#1}\@@href}%
\providecommand \@@href[1]{\endgroup#1\@@endlink}%
\providecommand \@sanitize@url [0]{\catcode `\\12\catcode `\$12\catcode
  `\&12\catcode `\#12\catcode `\^12\catcode `\_12\catcode `\%12\relax}%
\providecommand \@@startlink[1]{}%
\providecommand \@@endlink[0]{}%
\providecommand \url  [0]{\begingroup\@sanitize@url \@url }%
\providecommand \@url [1]{\endgroup\@href {#1}{\urlprefix }}%
\providecommand \urlprefix  [0]{URL }%
\providecommand \Eprint [0]{\href }%
\providecommand \doibase [0]{http://dx.doi.org/}%
\providecommand \selectlanguage [0]{\@gobble}%
\providecommand \bibinfo  [0]{\@secondoftwo}%
\providecommand \bibfield  [0]{\@secondoftwo}%
\providecommand \translation [1]{[#1]}%
\providecommand \BibitemOpen [0]{}%
\providecommand \bibitemStop [0]{}%
\providecommand \bibitemNoStop [0]{.\EOS\space}%
\providecommand \EOS [0]{\spacefactor3000\relax}%
\providecommand \BibitemShut  [1]{\csname bibitem#1\endcsname}%
\let\auto@bib@innerbib\@empty
%</preamble>
\bibitem [{\citenamefont {Goodrich}\ \emph {et~al.}(2012)\citenamefont
  {Goodrich}, \citenamefont {Liu},\ and\ \citenamefont
  {Nagel}}]{goodrich2012finitesizescaling}%
  \BibitemOpen
  \bibfield  {author} {\bibinfo {author} {\bibfnamefont {C.~P.}\ \bibnamefont
  {Goodrich}}, \bibinfo {author} {\bibfnamefont {A.~J.}\ \bibnamefont {Liu}}, \
  and\ \bibinfo {author} {\bibfnamefont {S.~R.}\ \bibnamefont {Nagel}},\ }\href
  {\doibase 10.1103/PhysRevLett.109.095704} {\bibfield  {journal} {\bibinfo
  {journal} {Phys. Rev. Lett.}\ }\textbf {\bibinfo {volume} {109}},\ \bibinfo
  {pages} {095704} (\bibinfo {year} {2012})}\BibitemShut {NoStop}%
\bibitem [{\citenamefont {Goodrich}\ \emph
  {et~al.}(2014{\natexlab{a}})\citenamefont {Goodrich}, \citenamefont
  {Dagois-Bohy}, \citenamefont {Tighe}, \citenamefont {van Hecke},
  \citenamefont {Liu},\ and\ \citenamefont
  {Nagel}}]{goodrich2013jamminginfinite}%
  \BibitemOpen
  \bibfield  {author} {\bibinfo {author} {\bibfnamefont {C.~P.}\ \bibnamefont
  {Goodrich}}, \bibinfo {author} {\bibfnamefont {S.}~\bibnamefont
  {Dagois-Bohy}}, \bibinfo {author} {\bibfnamefont {B.~P.}\ \bibnamefont
  {Tighe}}, \bibinfo {author} {\bibfnamefont {M.}~\bibnamefont {van Hecke}},
  \bibinfo {author} {\bibfnamefont {A.~J.}\ \bibnamefont {Liu}}, \ and\
  \bibinfo {author} {\bibfnamefont {S.~R.}\ \bibnamefont {Nagel}},\ }\href
  {\doibase 10.1103/PhysRevE.90.022138} {\bibfield  {journal} {\bibinfo
  {journal} {Phys. Rev. E}\ }\textbf {\bibinfo {volume} {90}},\ \bibinfo
  {pages} {022138} (\bibinfo {year} {2014}{\natexlab{a}})},\ \Eprint
  {http://arxiv.org/abs/1406.1529} {1406.1529} \BibitemShut {NoStop}%
\bibitem [{\citenamefont {Keim}\ \emph {et~al.}(2013)\citenamefont {Keim},
  \citenamefont {Paulsen},\ and\ \citenamefont
  {Nagel}}]{keim2013multipletransientmemories}%
  \BibitemOpen
  \bibfield  {author} {\bibinfo {author} {\bibfnamefont {N.~C.}\ \bibnamefont
  {Keim}}, \bibinfo {author} {\bibfnamefont {J.~D.}\ \bibnamefont {Paulsen}}, \
  and\ \bibinfo {author} {\bibfnamefont {S.~R.}\ \bibnamefont {Nagel}},\ }\href
  {\doibase 10.1103/PhysRevE.88.032306} {\bibfield  {journal} {\bibinfo
  {journal} {Phys. Rev. E}\ }\textbf {\bibinfo {volume} {88}},\ \bibinfo
  {pages} {032306} (\bibinfo {year} {2013})}\BibitemShut {NoStop}%
\bibitem [{\citenamefont {Maloney}\ and\ \citenamefont
  {Lemaître}(2006)}]{maloney2006amorphoussystemsin}%
  \BibitemOpen
  \bibfield  {author} {\bibinfo {author} {\bibfnamefont {C.~E.}\ \bibnamefont
  {Maloney}}\ and\ \bibinfo {author} {\bibfnamefont {A.}~\bibnamefont
  {Lemaître}},\ }\href {\doibase 10.1103/PhysRevE.74.016118} {\bibfield
  {journal} {\bibinfo  {journal} {Phys. Rev. E}\ }\textbf {\bibinfo {volume}
  {74}},\ \bibinfo {pages} {016118} (\bibinfo {year} {2006})}\BibitemShut
  {NoStop}%
\bibitem [{\citenamefont {Lemaître}\ and\ \citenamefont
  {Caroli}(2007)}]{lemaitre2007plasticresponsetwo}%
  \BibitemOpen
  \bibfield  {author} {\bibinfo {author} {\bibfnamefont {A.}~\bibnamefont
  {Lemaître}}\ and\ \bibinfo {author} {\bibfnamefont {C.}~\bibnamefont
  {Caroli}},\ }\href {\doibase 10.1103/PhysRevE.76.036104} {\bibfield
  {journal} {\bibinfo  {journal} {Phys. Rev. E}\ }\textbf {\bibinfo {volume}
  {76}},\ \bibinfo {pages} {036104} (\bibinfo {year} {2007})}\BibitemShut
  {NoStop}%
\bibitem [{\citenamefont {Salerno}\ \emph {et~al.}(2012)\citenamefont
  {Salerno}, \citenamefont {Maloney},\ and\ \citenamefont
  {Robbins}}]{salerno2012avalanchesinstrained}%
  \BibitemOpen
  \bibfield  {author} {\bibinfo {author} {\bibfnamefont {K.~M.}\ \bibnamefont
  {Salerno}}, \bibinfo {author} {\bibfnamefont {C.~E.}\ \bibnamefont
  {Maloney}}, \ and\ \bibinfo {author} {\bibfnamefont {M.~O.}\ \bibnamefont
  {Robbins}},\ }\href {\doibase 10.1103/PhysRevLett.109.105703} {\bibfield
  {journal} {\bibinfo  {journal} {Phys. Rev. Lett.}\ }\textbf {\bibinfo
  {volume} {109}},\ \bibinfo {pages} {105703} (\bibinfo {year}
  {2012})}\BibitemShut {NoStop}%
\bibitem [{\citenamefont {Hentschel}\ \emph {et~al.}(2010)\citenamefont
  {Hentschel}, \citenamefont {Karmakar}, \citenamefont {Lerner},\ and\
  \citenamefont {Procaccia}}]{hentschel2010sizeplasticevents}%
  \BibitemOpen
  \bibfield  {author} {\bibinfo {author} {\bibfnamefont {H.~G.~E.}\
  \bibnamefont {Hentschel}}, \bibinfo {author} {\bibfnamefont {S.}~\bibnamefont
  {Karmakar}}, \bibinfo {author} {\bibfnamefont {E.}~\bibnamefont {Lerner}}, \
  and\ \bibinfo {author} {\bibfnamefont {I.}~\bibnamefont {Procaccia}},\ }\href
  {\doibase 10.1103/PhysRevLett.104.025501} {\bibfield  {journal} {\bibinfo
  {journal} {Phys. Rev. Lett.}\ }\textbf {\bibinfo {volume} {104}},\ \bibinfo
  {pages} {025501} (\bibinfo {year} {2010})}\BibitemShut {NoStop}%
\bibitem [{\citenamefont {Chikkadi}\ \emph {et~al.}(2013)\citenamefont
  {Chikkadi}, \citenamefont {Gendelman}, \citenamefont {Ilyin}, \citenamefont
  {Ashwin}, \citenamefont {Procaccia},\ and\ \citenamefont
  {Shor}}]{chikkadi2013percolatingplasticfailure}%
  \BibitemOpen
  \bibfield  {author} {\bibinfo {author} {\bibfnamefont {V.}~\bibnamefont
  {Chikkadi}}, \bibinfo {author} {\bibfnamefont {O.}~\bibnamefont {Gendelman}},
  \bibinfo {author} {\bibfnamefont {V.}~\bibnamefont {Ilyin}}, \bibinfo
  {author} {\bibfnamefont {J.}~\bibnamefont {Ashwin}}, \bibinfo {author}
  {\bibfnamefont {I.}~\bibnamefont {Procaccia}}, \ and\ \bibinfo {author}
  {\bibfnamefont {C.~A. B.~Z.}\ \bibnamefont {Shor}},\ }\href {\doibase
  10.1209/0295-5075/110/48001} {\bibfield  {journal} {\bibinfo  {journal}
  {Europhys. Lett.}\ }\textbf {\bibinfo {volume} {110}},\ \bibinfo {pages}
  {48001} (\bibinfo {year} {2013})}\BibitemShut {NoStop}%
\bibitem [{\citenamefont {Olsson}\ and\ \citenamefont
  {Teitel}(2007)}]{olsson2007criticalscalingshear}%
  \BibitemOpen
  \bibfield  {author} {\bibinfo {author} {\bibfnamefont {P.}~\bibnamefont
  {Olsson}}\ and\ \bibinfo {author} {\bibfnamefont {S.}~\bibnamefont
  {Teitel}},\ }\href {\doibase 10.1103/PhysRevLett.99.178001} {\bibfield
  {journal} {\bibinfo  {journal} {Phys. Rev. Lett.}\ }\textbf {\bibinfo
  {volume} {99}},\ \bibinfo {pages} {178001} (\bibinfo {year}
  {2007})}\BibitemShut {NoStop}%
\bibitem [{\citenamefont {Tighe}\ \emph {et~al.}(2010)\citenamefont {Tighe},
  \citenamefont {Woldhuis}, \citenamefont {Remmers}, \citenamefont {van
  Saarloos},\ and\ \citenamefont {van Hecke}}]{tighe2010modelscalingstresses}%
  \BibitemOpen
  \bibfield  {author} {\bibinfo {author} {\bibfnamefont {B.~P.}\ \bibnamefont
  {Tighe}}, \bibinfo {author} {\bibfnamefont {E.}~\bibnamefont {Woldhuis}},
  \bibinfo {author} {\bibfnamefont {J.~J.~C.}\ \bibnamefont {Remmers}},
  \bibinfo {author} {\bibfnamefont {W.}~\bibnamefont {van Saarloos}}, \ and\
  \bibinfo {author} {\bibfnamefont {M.}~\bibnamefont {van Hecke}},\ }\href
  {\doibase 10.1103/PhysRevLett.105.088303} {\bibfield  {journal} {\bibinfo
  {journal} {Phys. Rev. Lett.}\ }\textbf {\bibinfo {volume} {105}},\ \bibinfo
  {pages} {088303} (\bibinfo {year} {2010})}\BibitemShut {NoStop}%
\bibitem [{\citenamefont {Chikkadi}\ \emph {et~al.}(2015)\citenamefont
  {Chikkadi}, \citenamefont {Woldhuis}, \citenamefont {van Hecke},\ and\
  \citenamefont {Schall}}]{chikkadi2015correlationsstrainand}%
  \BibitemOpen
  \bibfield  {author} {\bibinfo {author} {\bibfnamefont {V.}~\bibnamefont
  {Chikkadi}}, \bibinfo {author} {\bibfnamefont {E.}~\bibnamefont {Woldhuis}},
  \bibinfo {author} {\bibfnamefont {M.}~\bibnamefont {van Hecke}}, \ and\
  \bibinfo {author} {\bibfnamefont {P.}~\bibnamefont {Schall}},\ }\href
  {\doibase 10.1209/0295-5075/112/36004} {\bibfield  {journal} {\bibinfo
  {journal} {Europhys. Lett.}\ }\textbf {\bibinfo {volume} {112}},\ \bibinfo
  {pages} {36004} (\bibinfo {year} {2015})}\BibitemShut {NoStop}%
\bibitem [{\citenamefont {Woldhuis}\ \emph {et~al.}(2015)\citenamefont
  {Woldhuis}, \citenamefont {Chikkadi}, \citenamefont {van Deen}, \citenamefont
  {Schall},\ and\ \citenamefont {van Hecke}}]{woldhuis2015fluctuationsinflows}%
  \BibitemOpen
  \bibfield  {author} {\bibinfo {author} {\bibfnamefont {E.}~\bibnamefont
  {Woldhuis}}, \bibinfo {author} {\bibfnamefont {V.}~\bibnamefont {Chikkadi}},
  \bibinfo {author} {\bibfnamefont {M.~S.}\ \bibnamefont {van Deen}}, \bibinfo
  {author} {\bibfnamefont {P.}~\bibnamefont {Schall}}, \ and\ \bibinfo {author}
  {\bibfnamefont {M.}~\bibnamefont {van Hecke}},\ }\href {\doibase
  10.1039/c5sm01592h} {\bibfield  {journal} {\bibinfo  {journal} {Soft Matter}\
  }\textbf {\bibinfo {volume} {11}},\ \bibinfo {pages} {7024} (\bibinfo {year}
  {2015})}\BibitemShut {NoStop}%
\bibitem [{\citenamefont {Wyart}(2012)}]{wyart2012marginalstabilityconstrains}%
  \BibitemOpen
  \bibfield  {author} {\bibinfo {author} {\bibfnamefont {M.}~\bibnamefont
  {Wyart}},\ }\href {\doibase 10.1103/PhysRevLett.109.125502} {\bibfield
  {journal} {\bibinfo  {journal} {Phys. Rev. Lett.}\ }\textbf {\bibinfo
  {volume} {109}},\ \bibinfo {pages} {125502} (\bibinfo {year}
  {2012})}\BibitemShut {NoStop}%
\bibitem [{\citenamefont {Schreck}\ \emph {et~al.}(2011)\citenamefont
  {Schreck}, \citenamefont {Bertrand}, \citenamefont {O’Hern},\ and\
  \citenamefont {Shattuck}}]{schreck2011repulsivecontactinteractions}%
  \BibitemOpen
  \bibfield  {author} {\bibinfo {author} {\bibfnamefont {C.~F.}\ \bibnamefont
  {Schreck}}, \bibinfo {author} {\bibfnamefont {T.}~\bibnamefont {Bertrand}},
  \bibinfo {author} {\bibfnamefont {C.~S.}\ \bibnamefont {O’Hern}}, \ and\
  \bibinfo {author} {\bibfnamefont {M.~D.}\ \bibnamefont {Shattuck}},\ }\href
  {\doibase 10.1103/PhysRevLett.107.078301} {\bibfield  {journal} {\bibinfo
  {journal} {Phys. Rev. Lett.}\ }\textbf {\bibinfo {volume} {107}},\ \bibinfo
  {pages} {078301} (\bibinfo {year} {2011})}\BibitemShut {NoStop}%
\bibitem [{\citenamefont {Gómez}\ \emph
  {et~al.}(2012{\natexlab{a}})\citenamefont {Gómez}, \citenamefont {Turner},\
  and\ \citenamefont {Vitelli}}]{gomez2012uniformshockwaves}%
  \BibitemOpen
  \bibfield  {author} {\bibinfo {author} {\bibfnamefont {L.~R.}\ \bibnamefont
  {Gómez}}, \bibinfo {author} {\bibfnamefont {A.~M.}\ \bibnamefont {Turner}},
  \ and\ \bibinfo {author} {\bibfnamefont {V.}~\bibnamefont {Vitelli}},\ }\href
  {\doibase 10.1103/PhysRevE.86.041302} {\bibfield  {journal} {\bibinfo
  {journal} {Phys. Rev. E}\ }\textbf {\bibinfo {volume} {86}},\ \bibinfo
  {pages} {041302} (\bibinfo {year} {2012}{\natexlab{a}})}\BibitemShut
  {NoStop}%
\bibitem [{\citenamefont {Gómez}\ \emph
  {et~al.}(2012{\natexlab{b}})\citenamefont {Gómez}, \citenamefont {Turner},
  \citenamefont {van Hecke},\ and\ \citenamefont
  {Vitelli}}]{gomez2012shocksnearjamming}%
  \BibitemOpen
  \bibfield  {author} {\bibinfo {author} {\bibfnamefont {L.~R.}\ \bibnamefont
  {Gómez}}, \bibinfo {author} {\bibfnamefont {A.~M.}\ \bibnamefont {Turner}},
  \bibinfo {author} {\bibfnamefont {M.}~\bibnamefont {van Hecke}}, \ and\
  \bibinfo {author} {\bibfnamefont {V.}~\bibnamefont {Vitelli}},\ }\href
  {\doibase 10.1103/PhysRevLett.108.058001} {\bibfield  {journal} {\bibinfo
  {journal} {Phys. Rev. Lett.}\ }\textbf {\bibinfo {volume} {108}},\ \bibinfo
  {pages} {058001} (\bibinfo {year} {2012}{\natexlab{b}})}\BibitemShut
  {NoStop}%
\bibitem [{\citenamefont {van~den Wildenberg}\ \emph
  {et~al.}(2013)\citenamefont {van~den Wildenberg}, \citenamefont {van Loo},\
  and\ \citenamefont {van Hecke}}]{wildenberg2013shockwavesin}%
  \BibitemOpen
  \bibfield  {author} {\bibinfo {author} {\bibfnamefont {S.}~\bibnamefont
  {van~den Wildenberg}}, \bibinfo {author} {\bibfnamefont {R.}~\bibnamefont
  {van Loo}}, \ and\ \bibinfo {author} {\bibfnamefont {M.}~\bibnamefont {van
  Hecke}},\ }\href {\doibase 10.1103/PhysRevLett.111.218003} {\bibfield
  {journal} {\bibinfo  {journal} {Phys. Rev. Lett.}\ }\textbf {\bibinfo
  {volume} {111}},\ \bibinfo {pages} {218003} (\bibinfo {year}
  {2013})}\BibitemShut {NoStop}%
\bibitem [{\citenamefont {Combe}\ and\ \citenamefont
  {Roux}(2000)}]{combe2000strainversusstress}%
  \BibitemOpen
  \bibfield  {author} {\bibinfo {author} {\bibfnamefont {G.}~\bibnamefont
  {Combe}}\ and\ \bibinfo {author} {\bibfnamefont {J.-N.}\ \bibnamefont
  {Roux}},\ }\href {\doibase 10.1103/PhysRevLett.85.3628} {\bibfield  {journal}
  {\bibinfo  {journal} {Phys. Rev. Lett.}\ }\textbf {\bibinfo {volume} {85}},\
  \bibinfo {pages} {3628} (\bibinfo {year} {2000})}\BibitemShut {NoStop}%
\bibitem [{\citenamefont {Boschan}\ \emph {et~al.}(2016)\citenamefont
  {Boschan}, \citenamefont {Vågberg}, \citenamefont {Somfai},\ and\
  \citenamefont {Tighe}}]{boschaninprep}%
  \BibitemOpen
  \bibfield  {author} {\bibinfo {author} {\bibfnamefont {J.}~\bibnamefont
  {Boschan}}, \bibinfo {author} {\bibfnamefont {D.}~\bibnamefont {Vågberg}},
  \bibinfo {author} {\bibfnamefont {E.}~\bibnamefont {Somfai}}, \ and\ \bibinfo
  {author} {\bibfnamefont {B.~P.}\ \bibnamefont {Tighe}},\ }\href {\doibase
  10.1039/c6sm00536e} {\bibfield  {journal} {\bibinfo  {journal} {Soft Matter}\
  }\textbf {\bibinfo {volume} {12}},\ \bibinfo {pages} {5450} (\bibinfo {year}
  {2016})}\BibitemShut {NoStop}%
\bibitem [{\citenamefont {Bolton}\ and\ \citenamefont
  {Weaire}(1990)}]{bolton1990rigiditylosstransition}%
  \BibitemOpen
  \bibfield  {author} {\bibinfo {author} {\bibfnamefont {F.}~\bibnamefont
  {Bolton}}\ and\ \bibinfo {author} {\bibfnamefont {D.}~\bibnamefont
  {Weaire}},\ }\href {\doibase 10.1103/PhysRevLett.65.3449} {\bibfield
  {journal} {\bibinfo  {journal} {Phys. Rev. Lett.}\ }\textbf {\bibinfo
  {volume} {65}},\ \bibinfo {pages} {3449} (\bibinfo {year}
  {1990})}\BibitemShut {NoStop}%
\bibitem [{\citenamefont {Durian}(1995)}]{durian1995foammechanicsat}%
  \BibitemOpen
  \bibfield  {author} {\bibinfo {author} {\bibfnamefont {D.~J.}\ \bibnamefont
  {Durian}},\ }\href {\doibase 10.1103/PhysRevLett.75.4780} {\bibfield
  {journal} {\bibinfo  {journal} {Phys. Rev. Lett.}\ }\textbf {\bibinfo
  {volume} {75}},\ \bibinfo {pages} {4780} (\bibinfo {year}
  {1995})}\BibitemShut {NoStop}%
\bibitem [{\citenamefont {Lacasse}\ \emph {et~al.}(1996)\citenamefont
  {Lacasse}, \citenamefont {Grest}, \citenamefont {Levine}, \citenamefont
  {Mason},\ and\ \citenamefont {Weitz}}]{lacasse1996modelelasticitycompressed}%
  \BibitemOpen
  \bibfield  {author} {\bibinfo {author} {\bibfnamefont {M.-D.}\ \bibnamefont
  {Lacasse}}, \bibinfo {author} {\bibfnamefont {G.~S.}\ \bibnamefont {Grest}},
  \bibinfo {author} {\bibfnamefont {D.}~\bibnamefont {Levine}}, \bibinfo
  {author} {\bibfnamefont {T.~G.}\ \bibnamefont {Mason}}, \ and\ \bibinfo
  {author} {\bibfnamefont {D.~A.}\ \bibnamefont {Weitz}},\ }\href {\doibase
  10.1103/PhysRevLett.76.3448} {\bibfield  {journal} {\bibinfo  {journal}
  {Phys. Rev. Lett.}\ }\textbf {\bibinfo {volume} {76}},\ \bibinfo {pages}
  {3448} (\bibinfo {year} {1996})}\BibitemShut {NoStop}%
\bibitem [{\citenamefont {O'Hern}\ \emph {et~al.}(2002)\citenamefont {O'Hern},
  \citenamefont {Langer}, \citenamefont {Liu},\ and\ \citenamefont
  {Nagel}}]{ohern2002randompackingsfrictionless}%
  \BibitemOpen
  \bibfield  {author} {\bibinfo {author} {\bibfnamefont {C.~S.}\ \bibnamefont
  {O'Hern}}, \bibinfo {author} {\bibfnamefont {S.~A.}\ \bibnamefont {Langer}},
  \bibinfo {author} {\bibfnamefont {A.~J.}\ \bibnamefont {Liu}}, \ and\
  \bibinfo {author} {\bibfnamefont {S.~R.}\ \bibnamefont {Nagel}},\ }\href
  {\doibase 10.1103/PhysRevLett.88.075507} {\bibfield  {journal} {\bibinfo
  {journal} {Phys. Rev. Lett.}\ }\textbf {\bibinfo {volume} {88}},\ \bibinfo
  {pages} {075507} (\bibinfo {year} {2002})}\BibitemShut {NoStop}%
\bibitem [{\citenamefont {O'Hern}\ \emph {et~al.}(2003)\citenamefont {O'Hern},
  \citenamefont {Silbert}, \citenamefont {Liu},\ and\ \citenamefont
  {Nagel}}]{ohern2003jammingatzero}%
  \BibitemOpen
  \bibfield  {author} {\bibinfo {author} {\bibfnamefont {C.~S.}\ \bibnamefont
  {O'Hern}}, \bibinfo {author} {\bibfnamefont {L.~E.}\ \bibnamefont {Silbert}},
  \bibinfo {author} {\bibfnamefont {A.~J.}\ \bibnamefont {Liu}}, \ and\
  \bibinfo {author} {\bibfnamefont {S.~R.}\ \bibnamefont {Nagel}},\ }\href
  {\doibase 10.1103/PhysRevE.68.011306} {\bibfield  {journal} {\bibinfo
  {journal} {Phys. Rev. E}\ }\textbf {\bibinfo {volume} {68}},\ \bibinfo
  {pages} {011306} (\bibinfo {year} {2003})}\BibitemShut {NoStop}%
\bibitem [{\citenamefont {Wyart}(2005)}]{wyart2005rigidityamorphoussolids}%
  \BibitemOpen
  \bibfield  {author} {\bibinfo {author} {\bibfnamefont {M.}~\bibnamefont
  {Wyart}},\ }\href {\doibase doi:10.1051/anphys:2006003} {\bibfield  {journal}
  {\bibinfo  {journal} {Ann. Phys. Fr.}\ }\textbf {\bibinfo {volume} {30}},\
  \bibinfo {pages} {1} (\bibinfo {year} {2005})}\BibitemShut {NoStop}%
\bibitem [{\citenamefont {Wyart}\ \emph {et~al.}(2008)\citenamefont {Wyart},
  \citenamefont {Liang}, \citenamefont {Kabla},\ and\ \citenamefont
  {Mahadevan}}]{wyart2008elasticityfloppyand}%
  \BibitemOpen
  \bibfield  {author} {\bibinfo {author} {\bibfnamefont {M.}~\bibnamefont
  {Wyart}}, \bibinfo {author} {\bibfnamefont {H.}~\bibnamefont {Liang}},
  \bibinfo {author} {\bibfnamefont {A.}~\bibnamefont {Kabla}}, \ and\ \bibinfo
  {author} {\bibfnamefont {L.}~\bibnamefont {Mahadevan}},\ }\href {\doibase
  10.1103/PhysRevLett.101.215501} {\bibfield  {journal} {\bibinfo  {journal}
  {Phys. Rev. Lett.}\ }\textbf {\bibinfo {volume} {101}},\ \bibinfo {pages}
  {215501} (\bibinfo {year} {2008})}\BibitemShut {NoStop}%
\bibitem [{\citenamefont {Katgert}\ and\ \citenamefont {van
  Hecke}(2010)}]{katgert2010jammingandgeometry}%
  \BibitemOpen
  \bibfield  {author} {\bibinfo {author} {\bibfnamefont {G.}~\bibnamefont
  {Katgert}}\ and\ \bibinfo {author} {\bibfnamefont {M.}~\bibnamefont {van
  Hecke}},\ }\href {\doibase 10.1209/0295-5075/92/34002} {\bibfield  {journal}
  {\bibinfo  {journal} {Europhys. Lett.}\ }\textbf {\bibinfo {volume} {92}},\
  \bibinfo {pages} {34002} (\bibinfo {year} {2010})}\BibitemShut {NoStop}%
\bibitem [{\citenamefont {Tighe}(2011)}]{tighe2011relaxationsandrheology}%
  \BibitemOpen
  \bibfield  {author} {\bibinfo {author} {\bibfnamefont {B.~P.}\ \bibnamefont
  {Tighe}},\ }\href {\doibase 10.1103/PhysRevLett.107.158303} {\bibfield
  {journal} {\bibinfo  {journal} {Phys. Rev. Lett.}\ }\textbf {\bibinfo
  {volume} {107}},\ \bibinfo {pages} {158303} (\bibinfo {year}
  {2011})}\BibitemShut {NoStop}%
\bibitem [{\citenamefont {Wyart}\ \emph {et~al.}(2005)\citenamefont {Wyart},
  \citenamefont {Nagel},\ and\ \citenamefont
  {Witten}}]{wyart2005geometricoriginexcess}%
  \BibitemOpen
  \bibfield  {author} {\bibinfo {author} {\bibfnamefont {M.}~\bibnamefont
  {Wyart}}, \bibinfo {author} {\bibfnamefont {S.~R.}\ \bibnamefont {Nagel}}, \
  and\ \bibinfo {author} {\bibfnamefont {T.~A.}\ \bibnamefont {Witten}},\
  }\href {\doibase 10.1209/epl/i2005-10245-5} {\bibfield  {journal} {\bibinfo
  {journal} {Europhys. Lett.}\ }\textbf {\bibinfo {volume} {72}},\ \bibinfo
  {pages} {486} (\bibinfo {year} {2005})}\BibitemShut {NoStop}%
\bibitem [{\citenamefont {Silbert}\ \emph {et~al.}(2005)\citenamefont
  {Silbert}, \citenamefont {Liu},\ and\ \citenamefont
  {Nagel}}]{silbert2005vibrationsanddiverging}%
  \BibitemOpen
  \bibfield  {author} {\bibinfo {author} {\bibfnamefont {L.~E.}\ \bibnamefont
  {Silbert}}, \bibinfo {author} {\bibfnamefont {A.~J.}\ \bibnamefont {Liu}}, \
  and\ \bibinfo {author} {\bibfnamefont {S.~R.}\ \bibnamefont {Nagel}},\ }\href
  {\doibase 10.1103/PhysRevLett.95.098301} {\bibfield  {journal} {\bibinfo
  {journal} {Phys. Rev. Lett.}\ }\textbf {\bibinfo {volume} {95}},\ \bibinfo
  {pages} {098301} (\bibinfo {year} {2005})}\BibitemShut {NoStop}%
\bibitem [{\citenamefont {Ellenbroek}\ \emph {et~al.}(2006)\citenamefont
  {Ellenbroek}, \citenamefont {Somfai}, \citenamefont {van Hecke},\ and\
  \citenamefont {van Saarloos}}]{ellenbroek2006criticalscalingin}%
  \BibitemOpen
  \bibfield  {author} {\bibinfo {author} {\bibfnamefont {W.~G.}\ \bibnamefont
  {Ellenbroek}}, \bibinfo {author} {\bibfnamefont {E.}~\bibnamefont {Somfai}},
  \bibinfo {author} {\bibfnamefont {M.}~\bibnamefont {van Hecke}}, \ and\
  \bibinfo {author} {\bibfnamefont {W.}~\bibnamefont {van Saarloos}},\ }\href
  {\doibase 10.1103/PhysRevLett.97.258001} {\bibfield  {journal} {\bibinfo
  {journal} {Phys. Rev. Lett.}\ }\textbf {\bibinfo {volume} {97}},\ \bibinfo
  {pages} {258001} (\bibinfo {year} {2006})}\BibitemShut {NoStop}%
\bibitem [{\citenamefont {Ellenbroek}\ \emph {et~al.}(2009)\citenamefont
  {Ellenbroek}, \citenamefont {van Hecke},\ and\ \citenamefont {van
  Saarloos}}]{ellenbroek2009jammedfrictionlessdisks}%
  \BibitemOpen
  \bibfield  {author} {\bibinfo {author} {\bibfnamefont {W.~G.}\ \bibnamefont
  {Ellenbroek}}, \bibinfo {author} {\bibfnamefont {M.}~\bibnamefont {van
  Hecke}}, \ and\ \bibinfo {author} {\bibfnamefont {W.}~\bibnamefont {van
  Saarloos}},\ }\href {\doibase 10.1103/PhysRevE.80.061307} {\bibfield
  {journal} {\bibinfo  {journal} {Phys. Rev. E}\ }\textbf {\bibinfo {volume}
  {80}},\ \bibinfo {pages} {061307} (\bibinfo {year} {2009})}\BibitemShut
  {NoStop}%
\bibitem [{\citenamefont {Dagois-Bohy}\ \emph {et~al.}(2012)\citenamefont
  {Dagois-Bohy}, \citenamefont {Tighe}, \citenamefont {Simon}, \citenamefont
  {Henkes},\ and\ \citenamefont {van
  Hecke}}]{dagoisbohy2012softspherepackings}%
  \BibitemOpen
  \bibfield  {author} {\bibinfo {author} {\bibfnamefont {S.}~\bibnamefont
  {Dagois-Bohy}}, \bibinfo {author} {\bibfnamefont {B.~P.}\ \bibnamefont
  {Tighe}}, \bibinfo {author} {\bibfnamefont {J.}~\bibnamefont {Simon}},
  \bibinfo {author} {\bibfnamefont {S.}~\bibnamefont {Henkes}}, \ and\ \bibinfo
  {author} {\bibfnamefont {M.}~\bibnamefont {van Hecke}},\ }\href {\doibase
  10.1103/PhysRevLett.109.095703} {\bibfield  {journal} {\bibinfo  {journal}
  {Phys. Rev. Lett.}\ }\textbf {\bibinfo {volume} {109}},\ \bibinfo {pages}
  {095703} (\bibinfo {year} {2012})}\BibitemShut {NoStop}%
\bibitem [{\citenamefont {Goodrich}\ \emph
  {et~al.}(2014{\natexlab{b}})\citenamefont {Goodrich}, \citenamefont {Liu},\
  and\ \citenamefont {Nagel}}]{goodrich2013commentrepulsivecontact}%
  \BibitemOpen
  \bibfield  {author} {\bibinfo {author} {\bibfnamefont {C.~P.}\ \bibnamefont
  {Goodrich}}, \bibinfo {author} {\bibfnamefont {A.~J.}\ \bibnamefont {Liu}}, \
  and\ \bibinfo {author} {\bibfnamefont {S.~R.}\ \bibnamefont {Nagel}},\ }\href
  {\doibase 10.1103/PhysRevLett.112.049801} {\bibfield  {journal} {\bibinfo
  {journal} {Phys. Rev. Lett.}\ }\textbf {\bibinfo {volume} {112}},\ \bibinfo
  {pages} {049801} (\bibinfo {year} {2014}{\natexlab{b}})}\BibitemShut
  {NoStop}%
\bibitem [{\citenamefont {Schreck}\ \emph {et~al.}(2013)\citenamefont
  {Schreck}, \citenamefont {Bertrand}, \citenamefont {O'Hern},\ and\
  \citenamefont {Shattuck}}]{schreck2013responsetocomment}%
  \BibitemOpen
  \bibfield  {author} {\bibinfo {author} {\bibfnamefont {C.~F.}\ \bibnamefont
  {Schreck}}, \bibinfo {author} {\bibfnamefont {T.}~\bibnamefont {Bertrand}},
  \bibinfo {author} {\bibfnamefont {C.~S.}\ \bibnamefont {O'Hern}}, \ and\
  \bibinfo {author} {\bibfnamefont {M.~D.}\ \bibnamefont {Shattuck}},\
  }\href@noop {} {\enquote {\bibinfo {title} {Response to comment on 'repulsive
  contact interactions make jammed particulate systems inherently
  nonharmonic'},}\ } (\bibinfo {year} {2013}),\ \Eprint
  {http://arxiv.org/abs/1306.1961} {arXiv:1306.1961 [cond-mat.soft]}
  \BibitemShut {NoStop}%
\bibitem [{\citenamefont {{Lerner}}\ \emph {et~al.}(2013)\citenamefont
  {{Lerner}}, \citenamefont {{D{\"u}ring}},\ and\ \citenamefont
  {{Wyart}}}]{lerner2013lowenergynon}%
  \BibitemOpen
  \bibfield  {author} {\bibinfo {author} {\bibfnamefont {E.}~\bibnamefont
  {{Lerner}}}, \bibinfo {author} {\bibfnamefont {G.}~\bibnamefont
  {{D{\"u}ring}}}, \ and\ \bibinfo {author} {\bibfnamefont {M.}~\bibnamefont
  {{Wyart}}},\ }\href {\doibase 10.1039/c3sm50515d} {\bibfield  {journal}
  {\bibinfo  {journal} {Soft Matter}\ }\textbf {\bibinfo {volume} {9}},\
  \bibinfo {pages} {8252} (\bibinfo {year} {2013})}\BibitemShut {NoStop}%
\bibitem [{\citenamefont {van Deen}\ \emph {et~al.}(2014)\citenamefont {van
  Deen}, \citenamefont {Simon}, \citenamefont {Zeravcic}, \citenamefont
  {Dagois-Bohy}, \citenamefont {Tighe},\ and\ \citenamefont {van
  Hecke}}]{deen2014contactchangesnear}%
  \BibitemOpen
  \bibfield  {author} {\bibinfo {author} {\bibfnamefont {M.~S.}\ \bibnamefont
  {van Deen}}, \bibinfo {author} {\bibfnamefont {J.}~\bibnamefont {Simon}},
  \bibinfo {author} {\bibfnamefont {Z.}~\bibnamefont {Zeravcic}}, \bibinfo
  {author} {\bibfnamefont {S.}~\bibnamefont {Dagois-Bohy}}, \bibinfo {author}
  {\bibfnamefont {B.~P.}\ \bibnamefont {Tighe}}, \ and\ \bibinfo {author}
  {\bibfnamefont {M.}~\bibnamefont {van Hecke}},\ }\href {\doibase
  10.1103/PhysRevE.90.020202} {\bibfield  {journal} {\bibinfo  {journal} {Phys.
  Rev. E}\ }\textbf {\bibinfo {volume} {90}},\ \bibinfo {pages} {020202}
  (\bibinfo {year} {2014})}\BibitemShut {NoStop}%
\bibitem [{\citenamefont {Manning}\ and\ \citenamefont
  {Liu}(2011)}]{manning2011vibrationalmodesidentify}%
  \BibitemOpen
  \bibfield  {author} {\bibinfo {author} {\bibfnamefont {M.~L.}\ \bibnamefont
  {Manning}}\ and\ \bibinfo {author} {\bibfnamefont {A.~J.}\ \bibnamefont
  {Liu}},\ }\href {\doibase 10.1103/PhysRevLett.107.108302} {\bibfield
  {journal} {\bibinfo  {journal} {Phys. Rev. Lett.}\ }\textbf {\bibinfo
  {volume} {107}},\ \bibinfo {pages} {108302} (\bibinfo {year}
  {2011})}\BibitemShut {NoStop}%
\bibitem [{\citenamefont {Karmakar}\ \emph {et~al.}(2010)\citenamefont
  {Karmakar}, \citenamefont {Lemaître}, \citenamefont {Lerner},\ and\
  \citenamefont {Procaccia}}]{karmakar2010predictingplasticflow}%
  \BibitemOpen
  \bibfield  {author} {\bibinfo {author} {\bibfnamefont {S.}~\bibnamefont
  {Karmakar}}, \bibinfo {author} {\bibfnamefont {A.}~\bibnamefont {Lemaître}},
  \bibinfo {author} {\bibfnamefont {E.}~\bibnamefont {Lerner}}, \ and\ \bibinfo
  {author} {\bibfnamefont {I.}~\bibnamefont {Procaccia}},\ }\href {\doibase
  10.1103/PhysRevLett.104.215502} {\bibfield  {journal} {\bibinfo  {journal}
  {Phys. Rev. Lett.}\ }\textbf {\bibinfo {volume} {104}},\ \bibinfo {pages}
  {215502} (\bibinfo {year} {2010})}\BibitemShut {NoStop}%
\bibitem [{\citenamefont {Binder}\ \emph {et~al.}(1985)\citenamefont {Binder},
  \citenamefont {Nauenberg}, \citenamefont {Privman},\ and\ \citenamefont
  {Young}}]{binder1985finitesizetests}%
  \BibitemOpen
  \bibfield  {author} {\bibinfo {author} {\bibfnamefont {K.}~\bibnamefont
  {Binder}}, \bibinfo {author} {\bibfnamefont {M.}~\bibnamefont {Nauenberg}},
  \bibinfo {author} {\bibfnamefont {V.}~\bibnamefont {Privman}}, \ and\
  \bibinfo {author} {\bibfnamefont {A.~P.}\ \bibnamefont {Young}},\ }\href
  {\doibase 10.1103/PhysRevB.31.1498} {\bibfield  {journal} {\bibinfo
  {journal} {Phys. Rev. B}\ }\textbf {\bibinfo {volume} {31}},\ \bibinfo
  {pages} {1498} (\bibinfo {year} {1985})}\BibitemShut {NoStop}%
\bibitem [{\citenamefont {Ikeda}\ \emph {et~al.}(2013)\citenamefont {Ikeda},
  \citenamefont {Berthier},\ and\ \citenamefont
  {Biroli}}]{ikeda2013dynamiccriticalityat}%
  \BibitemOpen
  \bibfield  {author} {\bibinfo {author} {\bibfnamefont {A.}~\bibnamefont
  {Ikeda}}, \bibinfo {author} {\bibfnamefont {L.}~\bibnamefont {Berthier}}, \
  and\ \bibinfo {author} {\bibfnamefont {G.}~\bibnamefont {Biroli}},\ }\href
  {\doibase 10.1063/1.4769251} {\bibfield  {journal} {\bibinfo  {journal} {J.
  Chem. Phys}\ }\textbf {\bibinfo {volume} {138}},\ \bibinfo {pages} {12A507}
  (\bibinfo {year} {2013})}\BibitemShut {NoStop}%
\bibitem [{\citenamefont {Wang}\ and\ \citenamefont
  {Xu}(2013)}]{wang2013criticalscalingin}%
  \BibitemOpen
  \bibfield  {author} {\bibinfo {author} {\bibfnamefont {L.}~\bibnamefont
  {Wang}}\ and\ \bibinfo {author} {\bibfnamefont {N.}~\bibnamefont {Xu}},\
  }\href {\doibase 10.1039/c2sm27148f} {\bibfield  {journal} {\bibinfo
  {journal} {Soft Matter}\ }\textbf {\bibinfo {volume} {9}},\ \bibinfo {pages}
  {2475} (\bibinfo {year} {2013})}\BibitemShut {NoStop}%
\bibitem [{\citenamefont {Goodrich}\ \emph
  {et~al.}(2014{\natexlab{c}})\citenamefont {Goodrich}, \citenamefont {Liu},\
  and\ \citenamefont {Nagel}}]{goodrich2014whendojammed}%
  \BibitemOpen
  \bibfield  {author} {\bibinfo {author} {\bibfnamefont {C.~P.}\ \bibnamefont
  {Goodrich}}, \bibinfo {author} {\bibfnamefont {A.~J.}\ \bibnamefont {Liu}}, \
  and\ \bibinfo {author} {\bibfnamefont {S.~R.}\ \bibnamefont {Nagel}},\ }\href
  {\doibase 10.1103/PhysRevE.90.022201} {\bibfield  {journal} {\bibinfo
  {journal} {Phys. Rev. E}\ }\textbf {\bibinfo {volume} {90}},\ \bibinfo
  {pages} {022201} (\bibinfo {year} {2014}{\natexlab{c}})}\BibitemShut
  {NoStop}%
\bibitem [{\citenamefont {Donev}\ \emph {et~al.}(2004)\citenamefont {Donev},
  \citenamefont {Torquato}, \citenamefont {Stillinger},\ and\ \citenamefont
  {Connelly}}]{donev2004commentjammingat}%
  \BibitemOpen
  \bibfield  {author} {\bibinfo {author} {\bibfnamefont {A.}~\bibnamefont
  {Donev}}, \bibinfo {author} {\bibfnamefont {S.}~\bibnamefont {Torquato}},
  \bibinfo {author} {\bibfnamefont {F.~H.}\ \bibnamefont {Stillinger}}, \ and\
  \bibinfo {author} {\bibfnamefont {R.}~\bibnamefont {Connelly}},\ }\href
  {\doibase 10.1103/PhysRevE.70.043301} {\bibfield  {journal} {\bibinfo
  {journal} {Phys. Rev. E}\ }\textbf {\bibinfo {volume} {70}},\ \bibinfo
  {pages} {043301} (\bibinfo {year} {2004})}\BibitemShut {NoStop}%
\bibitem [{\citenamefont {Croarkin}\ and\ \citenamefont
  {Tobias}(2014)}]{NIST2014ehandbookstatistical}%
  \BibitemOpen
  \bibinfo {editor} {\bibfnamefont {C.}~\bibnamefont {Croarkin}}\ and\ \bibinfo
  {editor} {\bibfnamefont {P.}~\bibnamefont {Tobias}},\ eds.,\ \href
  {http://www.itl.nist.gov/div898/handbook/} {\emph {\bibinfo {title}
  {e-Handbook of Statistical Methods}}}\ (\bibinfo  {publisher}
  {NIST/SEMATECH},\ \bibinfo {year} {2014})\BibitemShut {NoStop}%
\bibitem [{\citenamefont {Goodrich}\ \emph
  {et~al.}(2015{\natexlab{a}})\citenamefont {Goodrich}, \citenamefont {Liu},\
  and\ \citenamefont {Sethna}}]{goodrich2015scalingtheoryjamming}%
  \BibitemOpen
  \bibfield  {author} {\bibinfo {author} {\bibfnamefont {C.~P.}\ \bibnamefont
  {Goodrich}}, \bibinfo {author} {\bibfnamefont {A.~J.}\ \bibnamefont {Liu}}, \
  and\ \bibinfo {author} {\bibfnamefont {J.~P.}\ \bibnamefont {Sethna}},\
  }\href@noop {} {\enquote {\bibinfo {title} {Scaling theory for the jamming
  transition},}\ } (\bibinfo {year} {2015}{\natexlab{a}}),\ \Eprint
  {http://arxiv.org/abs/1510.03469} {arXiv:1510.03469} \BibitemShut {NoStop}%
\bibitem [{\citenamefont {Goodrich}\ \emph
  {et~al.}(2015{\natexlab{b}})\citenamefont {Goodrich}, \citenamefont {Liu},\
  and\ \citenamefont {Nagel}}]{goodrich2015principleindependentbond}%
  \BibitemOpen
  \bibfield  {author} {\bibinfo {author} {\bibfnamefont {C.~P.}\ \bibnamefont
  {Goodrich}}, \bibinfo {author} {\bibfnamefont {A.~J.}\ \bibnamefont {Liu}}, \
  and\ \bibinfo {author} {\bibfnamefont {S.~R.}\ \bibnamefont {Nagel}},\ }\href
  {\doibase 10.1103/PhysRevLett.114.225501} {\bibfield  {journal} {\bibinfo
  {journal} {Phys. Rev. Lett.}\ }\textbf {\bibinfo {volume} {114}},\ \bibinfo
  {pages} {225501} (\bibinfo {year} {2015}{\natexlab{b}})}\BibitemShut
  {NoStop}%
\bibitem [{\citenamefont {Dagois-Bohy}\ \emph {et~al.}()\citenamefont
  {Dagois-Bohy}, \citenamefont {Somfai}, \citenamefont {Tighe},\ and\
  \citenamefont {van Hecke}}]{dagoisbohy2014oscillatoryrheologynear}%
  \BibitemOpen
  \bibfield  {author} {\bibinfo {author} {\bibfnamefont {S.}~\bibnamefont
  {Dagois-Bohy}}, \bibinfo {author} {\bibfnamefont {E.}~\bibnamefont {Somfai}},
  \bibinfo {author} {\bibfnamefont {B.~P.}\ \bibnamefont {Tighe}}, \ and\
  \bibinfo {author} {\bibfnamefont {M.}~\bibnamefont {van Hecke}},\ }\href@noop
  {} {\enquote {\bibinfo {title} {Oscillatory rheology near jamming},}\
  }\bibinfo {note} {In preparation.}\BibitemShut {Stop}%
\bibitem [{\citenamefont {Lerner}\ \emph {et~al.}(2012)\citenamefont {Lerner},
  \citenamefont {Düring},\ and\ \citenamefont
  {Wyart}}]{lerner2012towardmicroscopicdescription}%
  \BibitemOpen
  \bibfield  {author} {\bibinfo {author} {\bibfnamefont {E.}~\bibnamefont
  {Lerner}}, \bibinfo {author} {\bibfnamefont {G.}~\bibnamefont {Düring}}, \
  and\ \bibinfo {author} {\bibfnamefont {M.}~\bibnamefont {Wyart}},\ }\href
  {\doibase 10.1209/0295-5075/99/58003} {\bibfield  {journal} {\bibinfo
  {journal} {Europhys. Lett.}\ }\textbf {\bibinfo {volume} {99}},\ \bibinfo
  {pages} {58003} (\bibinfo {year} {2012})}\BibitemShut {NoStop}%
\bibitem [{\citenamefont {Otsuki}\ and\ \citenamefont
  {Hayakawa}(2014)}]{otsuki2014avalanchecontributionto}%
  \BibitemOpen
  \bibfield  {author} {\bibinfo {author} {\bibfnamefont {M.}~\bibnamefont
  {Otsuki}}\ and\ \bibinfo {author} {\bibfnamefont {H.}~\bibnamefont
  {Hayakawa}},\ }\href {\doibase 10.1103/PhysRevE.90.042202} {\bibfield
  {journal} {\bibinfo  {journal} {Phys. Rev. E}\ }\textbf {\bibinfo {volume}
  {90}},\ \bibinfo {pages} {042202} (\bibinfo {year} {2014})}\BibitemShut
  {NoStop}%
\bibitem [{\citenamefont {van Hecke}(2010)}]{hecke2010jammingsoftparticles}%
  \BibitemOpen
  \bibfield  {author} {\bibinfo {author} {\bibfnamefont {M.}~\bibnamefont {van
  Hecke}},\ }\href {\doibase 10.1088/0953-8984/22/3/033101} {\bibfield
  {journal} {\bibinfo  {journal} {J. Phys.: Condens. Matter}\ }\textbf
  {\bibinfo {volume} {22}},\ \bibinfo {pages} {033101} (\bibinfo {year}
  {2010})}\BibitemShut {NoStop}%
\bibitem [{\citenamefont {Simon}(2012)}]{simon2012rearrangementsinjammed}%
  \BibitemOpen
  \bibfield  {author} {\bibinfo {author} {\bibfnamefont {J.}~\bibnamefont
  {Simon}},\ }\emph {\bibinfo {title} {Rearrangements in Jammed Two-Dimensional
  Packings of Spherical Particles}},\ \href@noop {} {Master's thesis},\
  \bibinfo  {school} {Leiden University} (\bibinfo {year} {2012})\BibitemShut
  {NoStop}%
\bibitem [{\citenamefont {Born}\ and\ \citenamefont
  {Huang}(1998)}]{born1998dynamicaltheorycrystal}%
  \BibitemOpen
  \bibfield  {author} {\bibinfo {author} {\bibfnamefont {M.}~\bibnamefont
  {Born}}\ and\ \bibinfo {author} {\bibfnamefont {K.}~\bibnamefont {Huang}},\
  }\href@noop {} {\emph {\bibinfo {title} {Dynamical Theory of Crystal
  Lattices}}}\ (\bibinfo  {publisher} {Clarendon Press},\ \bibinfo {year}
  {1998})\BibitemShut {NoStop}%
\bibitem [{\citenamefont
  {Shewchuk}(1994)}]{shewchuk1994introductiontoconjugate}%
  \BibitemOpen
  \bibfield  {author} {\bibinfo {author} {\bibfnamefont {J.}~\bibnamefont
  {Shewchuk}},\ }\href {https://www.cs.cmu.edu/~jrs/jrspapers.html} {\emph
  {\bibinfo {title} {An introduction to the conjugate gradient method without
  the agonizing pain}}},\ \bibinfo {type} {Tech. Rep.}\ (\bibinfo
  {institution} {School of Computer Science, Carnegie Mellon University,
  Pittsburgh, PA},\ \bibinfo {year} {1994})\BibitemShut {NoStop}%
\bibitem [{\citenamefont {Bitzek}\ \emph {et~al.}(2006)\citenamefont {Bitzek},
  \citenamefont {Koskinen}, \citenamefont {Gähler}, \citenamefont {Moseler},\
  and\ \citenamefont {Gumbsch}}]{bitzek2006structuralrelaxationmade}%
  \BibitemOpen
  \bibfield  {author} {\bibinfo {author} {\bibfnamefont {E.}~\bibnamefont
  {Bitzek}}, \bibinfo {author} {\bibfnamefont {P.}~\bibnamefont {Koskinen}},
  \bibinfo {author} {\bibfnamefont {F.}~\bibnamefont {Gähler}}, \bibinfo
  {author} {\bibfnamefont {M.}~\bibnamefont {Moseler}}, \ and\ \bibinfo
  {author} {\bibfnamefont {P.}~\bibnamefont {Gumbsch}},\ }\href {\doibase
  10.1103/PhysRevLett.97.170201} {\bibfield  {journal} {\bibinfo  {journal}
  {Phys. Rev. Lett.}\ }\textbf {\bibinfo {volume} {97}},\ \bibinfo {pages}
  {170201} (\bibinfo {year} {2006})}\BibitemShut {NoStop}%
\bibitem [{\citenamefont {Strang}(2009)}]{strang2009introductiontolinear}%
  \BibitemOpen
  \bibfield  {author} {\bibinfo {author} {\bibfnamefont {G.}~\bibnamefont
  {Strang}},\ }\href@noop {} {\emph {\bibinfo {title} {Introduction to Linear
  Algebra}}},\ \bibinfo {edition} {4th}\ ed.\ (\bibinfo  {publisher} {Wellesley
  Cambridge Press},\ \bibinfo {year} {2009})\BibitemShut {NoStop}%
\bibitem [{\citenamefont {O’Hern}\ and\ \citenamefont
  {Wu}(2015)}]{ohern20150610privcomm}%
  \BibitemOpen
  \bibfield  {author} {\bibinfo {author} {\bibfnamefont {C.~S.}\ \bibnamefont
  {O’Hern}}\ and\ \bibinfo {author} {\bibfnamefont {Q.}~\bibnamefont {Wu}},\
  }\href@noop {} {}\bibinfo {howpublished} {Private communication} (\bibinfo
  {year} {2015})\BibitemShut {NoStop}%
\bibitem [{\citenamefont {Wyart}(2014)}]{wyart201404privcomm}%
  \BibitemOpen
  \bibfield  {author} {\bibinfo {author} {\bibfnamefont {M.}~\bibnamefont
  {Wyart}},\ }\href@noop {} {}\bibinfo {howpublished} {Private communication}
  (\bibinfo {year} {2014})\BibitemShut {NoStop}%
\bibitem [{\citenamefont
  {Moukarzel}(1998)}]{moukarzel1998isostaticphasetransition}%
  \BibitemOpen
  \bibfield  {author} {\bibinfo {author} {\bibfnamefont {C.}~\bibnamefont
  {Moukarzel}},\ }\href {\doibase 10.1103/PhysRevLett.81.1634} {\bibfield
  {journal} {\bibinfo  {journal} {Phys. Rev. Lett.}\ }\textbf {\bibinfo
  {volume} {81}},\ \bibinfo {pages} {1634} (\bibinfo {year}
  {1998})}\BibitemShut {NoStop}%
\bibitem [{\citenamefont {Roux}(2000)}]{roux2000geometricoriginmechanical}%
  \BibitemOpen
  \bibfield  {author} {\bibinfo {author} {\bibfnamefont {J.-N.}\ \bibnamefont
  {Roux}},\ }\href {\doibase 10.1103/PhysRevE.61.6802} {\bibfield  {journal}
  {\bibinfo  {journal} {Phys. Rev. E}\ }\textbf {\bibinfo {volume} {61}},\
  \bibinfo {pages} {6802} (\bibinfo {year} {2000})}\BibitemShut {NoStop}%
\bibitem [{\citenamefont {Thorpe}(1985)}]{thorpe1985elasticmodulitwo}%
  \BibitemOpen
  \bibfield  {author} {\bibinfo {author} {\bibfnamefont {M.~F.}\ \bibnamefont
  {Thorpe}},\ }\href {\doibase 10.1121/1.391966} {\bibfield  {journal}
  {\bibinfo  {journal} {J. Acoust. Soc. Am.}\ }\textbf {\bibinfo {volume}
  {77}},\ \bibinfo {pages} {1674} (\bibinfo {year} {1985})}\BibitemShut
  {NoStop}%
\end{thebibliography}%

\end{document}